\documentclass[11pt]{cernrepNum}
\usepackage{graphicx}
\usepackage{here}
\usepackage{color}
\usepackage{amsmath}
\usepackage{longtable}
\def\etal{{\it et al.}}

\def\meg{\mu^+ \rightarrow e^+ + \gamma}
\def\megs{\mu \rightarrow e \, \gamma}
\def\rdecay{\mu^+ \rightarrow  e^+ \nu \overline{\nu} \gamma}
\def\Bmeg{\rm{B}_{\megs}}

\begin{document}
\title{PHYSICS WITH LOW-ENERGY MUONS AT A NEUTRINO FACTORY COMPLEX}
\author{J.~\"Ayst\"o$^1$, A.~Baldini$^2$, A.~Blondel$^3$,
A.~de~Gouv\^ea$^4$, 
J.~Ellis$^4$, W.~Fetscher$^5$, G.F.~Giudice$^4$, K.~Jungmann$^6$, 
S.~Lola$^4$, V.~Palladino$^7$, K.~Tobe$^4$, A.~Vacchi$^8$, 
A.~van~der~Schaaf$^9$, K.~Zuber$^{10}$}
\institute{$^1$Jyvaskyla University, Finland.\\
$^2$INFN Pisa, Italy.\\
$^3$Geneva University, Switzerland.\\
$^4$CERN, Geneva, Switzerland.\\
$^5$ETH, Z\"urich, Switzerland.\\
$^6$KVI, Groningen, The Netherlands.\\
$^7$Universit\`a Federico II and INFN  Napoli, Italy.\\
$^8$INFN Trieste, Italy.\\
$^9$Physik-Institut der Universit\"at Z\"urich, Switzerland.\\
$^{10}$Dortmund University, Germany.}
\maketitle

\begin{abstract}
The physics potential of an intense source of low-energy muons is
studied. Such a source is a necessary stage towards building the neutrino 
factories and muon colliders which are being considered at present. The 
CERN Neutrino Factory could deliver muon beams with intensities 3--4 orders 
of magnitude higher than
available now, with large freedom in the choice of the time structure.
Low-energy muon physics contributes to many fields of basic research,
including rare muon decays,  i.e., decays that do not conserve
muon number, measurements of fundamental constants, the muon anomalous magnetic
moment, determination of the Lorentz structure of the weak interaction, 
QED tests, CPT tests, proton and nuclear charge distributions (even for 
short-lived isotopes), and condensed matter physics.
In studying the experimental programme, we analyse the present limitations,
list the requirements on the new muon beams, and describe some
ideas on how to implement these beam lines in a CERN neutrino factory complex.
\end{abstract}
 
\centerline{\hfill    CERN-TH/2001--231}
\newpage
\tableofcontents
\newpage
\section{INTRODUCTION}

Ever since the discovery of the muon the study of its properties has
contributed to a deeper understanding of Nature at the smallest distance
scale. Muon physics played a fundamental role in establishing the V--A
structure of weak interactions and the validity of quantum electrodynamics.  
Moreover, muon physics has not yet exhausted its potential and, indeed,
may provide crucial information regarding one of the most fundamental quests
in modern physics: the structure of the theory which lies beyond the
Standard Model of particle physics. 

Muon storage rings are currently, and seriously, being considered as options
for future endeavors of accelerator laboratories such as CERN and Fermilab. 
The primary aim of the machine presently considered is the study of neutrino
properties, hence they are also referred to as neutrino factories (NUFACT). 
They are a first step towards a still more ambitious goal: muon colliders,
capable of `precision' Higgs physics and the exploration of  the highest energy
frontier.

Neutrino factories are also an ideal place to study muon properties, since
they provide, necessarily, muon fluxes which are orders of magnitude larger
than what can be obtained at present. It is, therefore, imperative to 
understand how to take full advantage of these intense muon beams in order
to significantly improve on the reach of stopped-muon experiments.
This is the aim of this part of the report. 

While precise measurements of the muon lifetime and Michel parameters
provide tests for the theory of weak interactions and its possible extensions,
one of the main interests in  muon physics lies in the search for processes
that violate muon number. The discovery of decays such as
$\mu^+ \to e^+ \gamma$ and $\mu^+\to e^+e^-e^+$ or of $\mu^-$--$e^-$
conversion in nuclei would be an indisputable proof of the existence of new
dynamics beyond the Standard Model.

Global symmetries (like individual lepton numbers), as opposed to local
symmetries, are considered not to be based on fundamental principles and are
expected to be violated by gravitational effects, in the strong regime, and,
more generally, by higher-dimensional effective operators which describe local
interactions originating from some unknown high-energy dynamics. Baryon number
conservation is another example of an Abelian global symmetry of the
Standard Model, which can be broken by new-physics effects.

Atmospheric and solar neutrino experiments have provided strong
evidence for neutrino oscillations. This implies  violation of individual
lepton numbers ($L_i$) and, most likely, of total lepton number ($L$), which
is a first indication of physics beyond the Standard Model.
Current neutrino data indicate values of the neutrino masses corresponding
to non-renormalizable interactions at a scale $M\sim 10^{9-14}$~GeV. New
lepton-number violating dynamics at the scale $M$ cannot yield observable 
rates for rare muon processes, since the corresponding effects are suppressed
by $(m_{\mu}/M)^4$. The observation of muon-number violation in muon decays
would thus require new physics beyond the one responsible for neutrino masses.
Theoretically, however, there is no reason why $L_i$ and $L$ would be
broken at the same energy scale. Indeed, in many frameworks, such as
supersymmetry, the $L_i$ breaking scale can be close to the weak scale.
In this case, muon processes with $L_\mu$ violation would occur with rates
close to the current experimental bounds.

It is also very important to stress that the information which can be
extracted from the study of rare muon processes is, in many cases, not
accessible to high-energy colliders. Take supersymmetry as an example. 
While the LHC can significantly probe slepton masses, it cannot compete
with muon-decay experiments in constraining the slepton mixing angles.

In Section~\ref{sec:2.1} we discuss `normal' muon decay and its importance to
fundamental physics. In Section~3 we address muon-number violating processes,
first discussing the theoretical expectations and then addressing the
issue of how to improve on current experimental bounds given a larger
muon flux. In particular we discuss which elements of the experimental
set-ups need to be improved or drastically modified. In Section~4 we address
the measurements of fundamental properties of the muon, and in Section~5 we
discuss bound muon systems (muonic atoms and condensed matter systems).
In Section~6 we summarize the beam requirements imposed by the various
experiments and discuss how these could be met in a neutrino factory
complex. Our conclusions are given in Section~7.

\section{NORMAL MUON DECAY}\label{sec:2.1}
\subsection{Theoretical background}\label{subsec:2.1.1}

All measurements in normal muon decay, $\mu^{-} \to e^{-}
\overline{\nu}_{e} \nu_{\mu}$, and its inverse, $\nu_{\mu} e^{-}
\to \mu^{-} \nu_{e}$, are successfully described by
the `V--A' interaction, which is a particular case of the local,
derivative-free, lepton-number-conserving, four-fermion interaction
\cite{Michel_50}.  The `V--A' form and the nature of the neutrinos
($\overline{\nu}_{e}$ and $\nu_{e}$) have been determined by
experiment \cite{FGJ_86,Langacker_89}.

The observables in muon decay (energy spectra, polarizations and angular
distributions) and in inverse muon decay (the reaction cross section) at
energies well below $m_{W}c^{2}$ may be parametrized in terms of the
dimensionless coupling constants $g_{\varepsilon \mu}^{\gamma}$ and the Fermi
coupling constant $G_{\mathrm{F}}$. The matrix element is 
\begin{equation}
{\mathcal M} = \frac{4 G_{\mathrm{F}}}{\sqrt{2}}\,
\sum_{\substack{\gamma = \mathrm{S, V, T} \\ \varepsilon, \mu = \mathrm{R, L}
}}g^{\gamma}_{\varepsilon \mu} \langle \overline{e}_{\varepsilon}
|\Gamma^{\gamma}|(\nu_{e})_{n}\rangle \langle (\overline{\nu}_{\mu})_{m}|
\Gamma_{\gamma}|\mu_{\mu}\rangle \;.\label{eq:wfetscher:1}
\end{equation}
We use here the notation of Fetscher \etal, \cite{FGJ_86,Fetscher_00}
who in turn use the sign conventions and definitions of Scheck
\cite{Scheck_83}. Here $\gamma = \mathrm{S,V,T}$ indicate a (Lorentz) scalar,
vector, or tensor interaction, and the chirality of the electron or muon
(right- or left-handed) is labelled by $\varepsilon, \mu = \mathrm{R, L}$.
The chiralities $n$ and $m$ of the $\nu_{e}$ and the $\overline{\nu}_{\mu}$ are
determined by given values of $\gamma$, $\varepsilon$ and $\mu$. The 10 complex
amplitudes $g_{\varepsilon \mu}^{\gamma}$ and $G_{\mathrm{F}}$ constitute 19
independent parameters to be determined by experiment.  The `V--A' interaction
corresponds to $g_{\mathrm{LL}}^{\mathrm{V}} = 1$, with all other amplitudes
being 0.

With the deduction \emph{from experiments} that the interaction is
predominantly of the  vector type and left-handed [$g_{\mathrm{LL}}^{\mathrm{V}} >
0.96~(90~\%\mathrm{CL}$)], there remain several separate routes of
investigation of normal muon decay, which will be discussed in the following.

\subsection{Muon lifetime measurements}\label{subsec:2.1.3}

The measurement of the muon lifetime yields the most precise determination of
the Fermi coupling constant $G_{\mathrm{F}}$, which is presently known with a
relative precision of $9 \times 10^{-6}$. Improving this measurement is
certainly an interesting goal~\cite{Marciano_00}, since $G_{\mathrm{F}}$ is one
of the fundamental parameters of the Standard Model. A clean beam pulse structure with
very good suppression of particles between pulses is indispensable. Presently three
experiments are in progress, two of which are at PSI \cite{mu_lifetime_PSI} and
one is located at RAL \cite{mu_lifetime_RAL}. An improvement in the
precision of $\tau_{\mu}$ by
about a factor of 20 can be expected. 
An additional order of magnitude could be gained at a neutrino
factory
primarily from increased muon flux with the major systematics being
pile-up and detector timing stability.

There are two caveats, however:
\begin{itemize}
\item
Reducing the error on $G_{\mathrm{F}}$ by precise measurements of the muon
lifetime would unfortunately not improve the electroweak fits, because the
error on the dimensionless input $G_{\mathrm{F}} M_{Z}^{2}$ is dominated by
the uncertainty on $M_{Z}^{2}$, which is now $48 \times 10^{-6}$ $[M_{Z} =
(91\,188.2 \pm 2.2]$~MeV \cite{pdg}).
\item
$G_{\mathrm{F}}$ is commonly determined assuming exclusively V-A interactions.
A somewhat more general formula has been given by Greub {\it et al.}
\cite{Greub_94}:
\begin{align}
G_{\mathrm{F}}^{2} = & \frac{192 \pi^{3} \hbar}{ \tau_{\mu}
m^{5}_{\mu}} \left[ 1 + \frac{\alpha}{ 2 \pi}
\left(\pi^{2} -\frac{25}{4}\right)\right]  \left[ 1 - \frac{3}{5}
\left(\frac{m_{\mu}}{m_{W}}\right)^{2}\right] \nonumber \\[5mm]
    &  \times \left[1 - 4 \eta  \frac{m_{e}}{m_{\mu}} - 4 \lambda 
\frac{m_{\nu_{\mu}}}{m_{\mu}} + 8 \left(\frac{m_{e}}{m_{\mu}}\right)^{2} + 8
\left(\frac{m_{\nu_{\mu}}}{m_{\mu}}\right)^{2}\right] , 
\label{eq:wfetscher:2}
\end{align}
\begin{equation}
                        \eta = \frac{1}{2} {\rm Re} \left[
                        g^{\mathrm{V}}_{\mathrm{LL}} 
                        g^{\mathrm{S}*}_{\mathrm{RR}} + 
                        g^{\mathrm{V}}_{\mathrm{RR}}
                        g^{\mathrm{S}*}_{\mathrm{LL}} +
                        g^{\mathrm{V}}_{\mathrm{LR}} (
                        g^{\mathrm{S}*}_{\mathrm{RL}} +
                        g^{\mathrm{T}*}_{\mathrm{RL}} ) +
                        g^{\mathrm{V}}_{\mathrm{RL}} (
                        g^{\mathrm{S}*}_{\mathrm{LR}} +
                        g^{\mathrm{T}*}_{\mathrm{LR}} ) \right] ,
                        \label{eq:wfetscher:3}
\end{equation}
\begin{equation}
                        \lambda = \frac{1}{2} {\rm Re} \left[
                        g^{\mathrm{S}}_{\mathrm{LL}} 
                        g^{\mathrm{S}*}_{\mathrm{LR}} + 
                        g^{\mathrm{S}}_{\mathrm{RR}}
                        g^{\mathrm{S}*}_{\mathrm{RL}} -2
                        g^{\mathrm{V}}_{\mathrm{RR}} 
                        g^{\mathrm{V}*}_{\mathrm{RL}} -2
                        g^{\mathrm{V}}_{\mathrm{LL}} 
                        g^{\mathrm{V}*}_{\mathrm{LR}}  \right] .
\end{equation}

Here, besides the muon lifetime and the muon mass, radiative corrections to
first order and mass terms are included. Most important is the muon decay
parameter $\eta$ which is 0 in the SM. If we assume that only one additional
interaction contributes to muon decay, 
then $\eta \simeq \frac{1}{2}{\rm Re} g^{\mathrm{S}}_{\mathrm{RR}}$, where
$g^{\mathrm{S}}_{\mathrm{RR}}$
corresponds to a scalar coupling with right-handed charged leptons. Including
the experimental value of $\eta = (-7 \pm 13) \times 10^{-3}$
\cite{Burkard_85} the error on $G_{\mathrm{F}}$ increases by a factor of 20.
\end{itemize}

\subsection{Precision measurement of the Michel parameters}\label{subsec:2.1.4}

The measurement of individual decay parameters alone generally does not give
conclusive information about the decay interaction owing to the many different
couplings and interference terms. An example is the spectrum Michel parameter
$\varrho$. A precise measurement yielding the V--A value of 3/4 by no means
establishes the V--A interaction. In fact, \emph{any} interaction consisting of
an arbitrary combination of $g_{\mathrm{LL}}^{\mathrm{S}}$,
$g_{\mathrm{LR}}^{\mathrm{S}}$, $g_{\mathrm{RL}}^{\mathrm{S}}$,
$g_{\mathrm{RR}}^{\mathrm{S}}$, $g_{\mathrm{RR}}^{\mathrm{V}}$ and
$g_{\mathrm{LL}}^{\mathrm{V}}$ will yield exactly $\varrho = 3/4$ 
\cite{Fetscher_90}. This can be seen if we write $\varrho$ in the form
\cite{Gerber_87}:
\begin{equation}
        \varrho - \textstyle{\frac{3}{4}} = - \textstyle{\frac{3}{4}} 
        \displaystyle \left\{|g_{\mathrm{LR}}^{\mathrm{V}}|^{2} +
        |g_{\mathrm{RL}}^{\mathrm{V}}|^{2}\right\} + 
        2 \left(|g_{\mathrm{LR}}^{\mathrm{T}}|^{2} +
        |g_{\mathrm{RL}}^{\mathrm{T}}|^{2} \right) + Re \left( 
        g_{\mathrm{LR}}^{\mathrm{S}}g_{\mathrm{LR}}^{\mathrm{T*}} + 
        g_{\mathrm{RL}}^{\mathrm{S}}g_{\mathrm{RL}}^{\mathrm{T*}}
        \right)\;.
        \label{eq:wfetscher:5}
\end{equation}
For $\varrho = \textstyle{\frac{3}{4}}$ and $g_{\mathrm{RL}}^{\mathrm{T}} =
g_{\mathrm{LR}}^{\mathrm{T}} = 0$ (no tensor interaction) one finds
$g_{\mathrm{RL}}^{\mathrm{V}} = g_{\mathrm{LR}}^{\mathrm{V}} = 0$, with all of
the remaining six couplings being arbitrary! On the other hand, any deviation
from the canonical value certainly would signify new physics. Tree-level new
physics contributions to the Michel parameters occur in supersymmetric theories
with $R$-parity violation or theories with left--right symmetric gauge groups.
For instance, the $R$-parity violating interactions $\lambda_{311}
L_{\mathrm{L}}^{(3)} L_{\mathrm{L}}^{(1)} {\bar E}_{\mathrm{R}}^{(1)}
+\lambda_{322} L_{\mathrm{L}}^{(3)} L_{\mathrm{L}}^{(2)} {\bar
E}_{\mathrm{R}}^{(2)}$ (where the index denotes the lepton generation)
give the following contributions~\cite{RPV_all}
\begin{equation}
        \Delta \rho = \frac{3 \epsilon^{2}}{16},~~ \Delta \eta
        =\frac{\epsilon}{2},~~ \Delta \xi = -\frac{\epsilon^{2}}{4},~~
        \Delta \delta =0,~~~~ \epsilon \equiv \frac{\lambda_{311}
        \lambda_{322}}{4\sqrt{2} G_{\mathrm{F}}{\tilde
       m}^{2}_{e_{\mathrm{L}}^{(3)}}}\;.
        \label{eq:wfetscher:6}
\end{equation}
For a left--right model, one finds
\begin{equation}
        \Delta \rho =-\frac{3}{2} \vartheta_{W_{\mathrm{R}}}^{2},~~\Delta
        \xi = -2 \vartheta_{W_{\mathrm{R}}}^{2} - 2 \left(
        \frac{M_{W_{1}}}{M_{W_{2}}}\right)^{4},
        \label{eq:wfetscher:7}
\end{equation}
where $\vartheta_{W_{\mathrm{R}}}$ is the
$W_{\mathrm{L}}$--$W_{\mathrm{R}}$ mixing angle, and $M_{W_{1}}$
($M_{W_{2}}$) is the mass of the mainly left (right) charged gauge
boson.  Measurements of $\varrho$ and $\xi$ with a precision of
$10^{-4}$ can probe $W_{\mathrm{R}}$ masses of about 1~TeV (in the most
unfavourable case $\vartheta_{W_{\mathrm{R}}} = 0$) and values of the
$R$-parity violating couplings $\lambda_{311} \approx \lambda_{322} 
\approx 0.2$
(for a slepton mass of 200~GeV).  These tests are competitive with
direct searches at high-energy colliders.

There exist also observables which yield valuable information even if they
assume their canonical values, all of which are related to the spin variables
of the muon and the electron:
\begin{itemize}
\item
A measurement of the decay asymmetry yields the parameters $\delta$ and
$P_{\mu}\xi$. Especially interesting is the combination
$P_{\mu}\xi\delta/\varrho$, which has been measured at TRIUMF \cite{Jodidio_86}
with a precision of $\approx 3 \times 10^{-3}$. A new, ambitious experiment
measuring $\varrho$, $\delta$ and $P_{\mu}\xi$ is currently under construction
at TRIUMF \cite{Gill_93}.
\item
A measurement of the longitudinal polarisation of the decay electrons
$P_{\mathrm{L}}$ consistent with 1  yields limits for all five couplings where
the electrons are right-handed. This is a difficult experiment due to the lack
of highly polarised electron targets used as analysers. The present precision 
is
$\Delta P_{\mathrm{L}} = 45 \times 10^{-3}$.
\item
The angular dependence of the longitudinal polarisation 
of decay positrons at the endpoint energy
is currently being  measured
at PSI by the Louvain-la-Neuve--PSI--ETH Z\"urich Collaboration
\cite{Prieels_97}. This yields the parameter $\xi''$ which is sensitive to
the right-handed vector and the tensor currents.
\item
A measurement of the transverse polarization of the decay positrons requires a
highly polarised pulsed muon beam. From the energy dependence of the component
$P_{\mathrm{T}_{1}}$ one can deduce the low-energy decay parameter $\eta$ which
is needed for a model-independent value of the Fermi coupling constant. The
second component $P_{\mathrm{T}_{2}}$, which is transverse to the positron
momentum and the muon polarisation, is non-invariant under time reversal. A
second generation experiment performed at PSI by the ETH Z\"urich--Cracow--PSI
Collaboration has just finished data-taking \cite{Bodek_00,Barnett_00}. The
expected experimental errors are $\Delta P_{\mathrm{T}_{1}} =
\Delta P_{\mathrm{T}_{2}} = 5 \times 10^{-3}$.
\end{itemize}

\subsection{Experimental prospects}\label{subsec:2.1.5}

As mentioned above, the precision on the muon lifetime can presumably be 
increased over the ongoing measurements by one order of magnitude. 
Improvement in measurements of the decay
parameters seems more difficult. Most ambitious is the TRIUMF project which
will soon start taking data and thereby improve the parameters $\varrho$ 
(positron energy spectrum), $P_{\mu} \xi$ and $\delta$ (decay asymmetry) by
more than one order of magnitude. The limits on most other observables are not
given by the muon rates which usually are high enough already ($\approx 3
\times 10^{8}$~s$^{-1}$ at the $\mu$E1 beam at PSI, for example), but rather by
effects like positron depolarisation in matter or by the small available
polarisation ($< 7 \%$) of the electron targets used as analysers. The
measurement of the transverse positron polarisation might be improved with a
smaller phase space (lateral beam dimension of a few millimetres or better). This 
experiment needs a \emph{pulsed} beam with high polarisation. 

\section{SEARCH FOR MUON NUMBER VIOLATION}

\subsection{Theoretical considerations}
\label{LNV}

In the Standard Model (SM), muon number is exactly conserved. When neutrino
masses are added and neutrino oscillations take place, muon-number violating
processes involving charged leptons become possible as well. However, because
of the smallness of neutrino masses, the rates for these processes are
unobservable \cite{meg_SM,shrock}; for instance
\begin{equation}
B(\mu \to e \gamma )=\frac{3\alpha}{32 \pi}\sum_i \left|
V_{\mu i}^*V_{ei} \frac{m_{\nu_i}^2}{M_W^2}\right|^2 \sim
10^{-60} \left|\frac{V_{\mu i}^*V_{ei}}{10^{-2}}\right|^2
\left( \frac{m_{\nu_i}}{10^{-2}~{\rm eV}}\right)^4\, .
\end{equation}
The observation of muon number violation in charged muon decay would,
therefore, serve as an unambiguous sign of new physics and indeed, a number of
SM extensions may be probed sensitively by the study of rare muon decays. Here
we will concentrate on supersymmetric models and models with extra dimensions,
but it should be pointed out that
various other SM extensions also predict observable rates for the rare $\mu$
decays, like models with new $\rm Z^\prime$ gauge bosons~\cite{zprime},
leptoquarks~\cite{lepto_quarks} or Lorentz-invariance 
violation~\cite{lorentz,Kostelecki_00}.
For a recent review
on muon number violation, see Ref.~\cite{Kuno:1999jp}.

We first present a model-independent formalism for studying rare muon decays
and for comparing the rates of the different muon-number-violating channels.
Then, we discuss predictions for the branching ratios of rare muon processes
in extensions of the SM with low-energy supersymmetry, in the context of the
seesaw mechanism and grand unification, both with and without the conservation
of $R$ parity. Finally, we present the expectations in models with extra 
dimensions. 

\subsubsection{Model-independent analysis of rare muon processes}

Although a purely model-independent analysis based on effective operators
cannot make any prediction for the absolute rate of rare muon processes, it
can be very useful in determining the relative rates. We will compare the
rates for $\mu^+\to e^+\gamma$, $\mu^+\to e^+e^-e^+$, and $\mu^-$--$e^-$
conversion. In a large class of models, the dominant source of individual
lepton number violation comes from a flavour non-diagonal magnetic-moment
transition. Let us therefore consider the effective operator
\begin{equation}
{\cal L}=\frac{m_\mu}{\Lambda^2} {\bar \mu}_R \sigma^{\mu\nu}e_L
F_{\mu\nu}+{\rm h.c.} \label{magmom}
\end{equation}
This interaction leads to the following results for the branching ratios of
$\mu^+\to e^+\gamma$ ($B(\mu\to e\gamma)$) and $\mu^+\to e^+e^-e^+$ 
($B(\mu\to 3e)$), and for the rate of $\mu^-$--$e^-$ conversion in
nuclei normalised to the nuclear capture rate ($B(\mu N\to eN)$):
\begin{eqnarray}
B(\mu \to e\gamma)&=& \frac{3(4\pi)^2}{G_F^2\Lambda^4}\;, \label{meg}\\
\frac{B(\mu \to 3e)}{B(\mu \to e\gamma)}&=&\frac{\alpha}{3\pi}
\left( \ln\frac{m_\mu^2}{m_e^2}-\frac{11}{4}\right) =6\times 10^{-3}\;,
\label{m3e}\\
\frac{B(\mu N\to eN)}{B(\mu \to e\gamma)}&=& 10^{12}~B(A,Z)~
\frac{2G_F^2m_\mu^4}{(4\pi)^3\alpha}=2\times 10^{-3}~B(A,Z)\;.\label{m4e}
\end{eqnarray}
Here $B(A,Z)$ is an effective nuclear coefficient which is of order 1 for
elements heavier than aluminium ~\cite{Cza}. The logarithm in Eq.~(\ref{m3e})
is an enhancement factor for $B(\mu \to 3e)$, which is a consequence of the
collinear divergence of the electron-positron pair in the $m_e\to 0$ limit.
Nevertheless, because of the smaller phase space and extra power of $\alpha$,
$B(\mu \to 3e)$ and $B(\mu N\to eN)$ turn out to be suppressed with
respect to $B(\mu \to e\gamma)$ by factors of $6\times 10^{-3}$
and $(2$--$4)\times 10^{-3}$, respectively. 
This should be compared with the experimental sensitivities for
the different processes, discussed in the next subsections.

Next, let us include an effective four-fermion operator which violates
individual lepton number 
\begin{equation}
{\cal L}=\frac{1}{\Lambda_F^2}{\bar \mu}_L\gamma^\mu e_L {\bar f}_L
\gamma_\mu f_L +{\rm h.c.}\;, \label{op4}
\end{equation}
where $f$ is a generic quark or lepton. The choice of the operator in
Eq.~(\ref{op4}) is made for concreteness, and our results do not depend
significantly on the specific chiral structure of the operator. First we
consider the case in which $f$ is neither an electron nor a light quark, and
therefore $\mu^+\to e^+e^-e^+$ and $\mu^-$--$e^-$ conversion occur only at the
loop level. Comparing the $\mu^+ \to e^+ \gamma$ rate in Eq.~(\ref{meg}) with
the contributions from the four-fermion operator to $B(\mu \to 3e)$ and
$B(\mu N\to eN)$, we find
\begin{eqnarray}
\frac{B(\mu \to 3e)}{B(\mu \to e\gamma)}&=&\frac{8\alpha^2N_f^2}{9(4\pi)^4}
\left(\frac{\Lambda}{\Lambda_F}\right)^4\left[ \ln \frac{{\rm max}
(m_f^2,m_\mu^2)}{M_F^2}\right]^2\;, \label{prim}\\
\frac{B(\mu N\to eN)}{B(\mu \to e\gamma)}&=& 10^{12}~B(A,Z)~
\frac{32G_F^2m_\mu^4N_f^2}{9(4\pi)^6}
\left(\frac{\Lambda}{\Lambda_F}\right)^4\left[ \ln \frac{{\rm max}
(m_f^2,m_\mu^2)}{M_F^2}\right]^2\;. \label{prim2}
\end{eqnarray}
Here $N_f$ is the number of colours of the fermion $f$ and $M_F$ is the 
heavy-particle mass generating the effective operators (typically $M_F$ is
much smaller than $\Lambda$ or $\Lambda_F$ because of loop factors and mixing
angles). The logarithms in Eqs.~(\ref{prim}) and (\ref{prim2}) correspond to
the anomalous dimension mixing of the operator in Eq.~(\ref{op4}) with the
four-fermion operator generating the relevant rare muon
process~\cite{Raidal:1998hq}. If $\Lambda \sim \Lambda_F$, then the
contributions from the four-fermion operator are irrelevant, since the ratios
in Eqs.~(\ref{m3e}) and (\ref{m4e}) are larger than those in Eqs.~(\ref{prim})
and (\ref{prim2}). More interesting is the case in which the four-fermion 
operator in Eq.~(\ref{op4}) is generated at tree level, while the
magnetic-moment transition in Eq.~(\ref{magmom}) is generated only at one loop,
as in models with $R$-parity violation \cite{RPV_all} or with leptoquarks
\cite{lepto_quarks}. In this case, we expect
\begin{equation}
\left(\frac{\Lambda}{\Lambda_F}\right)^4\simeq \frac{(4\pi)^3}{\alpha}\;.
\label{parg}
\end{equation}
If Eq.~(\ref{parg}) holds and if we take $M_F\simeq 1$~TeV, then the ratios in 
Eqs.~(\ref{prim}) and (\ref{prim2}) become of order unity, so the
different rare muon processes have comparable rates.

Alternatively, if the fermion $f$ in Eq.~(\ref{op4}) is an electron (or a
light quark), the effective operator can mediate $\mu \to 3e$ (or
$\mu^-$--$e^-$ conversion) at tree-level, and the corresponding process can
dominate over the others~\cite{RPV}. For instance, we obtain
\begin{equation}
\frac{B(\mu \to 3e)}{B(\mu \to e\gamma)}=
\frac{1}{12(4\pi)^2}\left(\frac{\Lambda}{\Lambda_F}\right)^4\;,
\end{equation}
for the case $f=e$.

\begin {figure}[htb]
\centerline{
\parbox{0.5\textwidth}{\includegraphics[height=7.5cm]{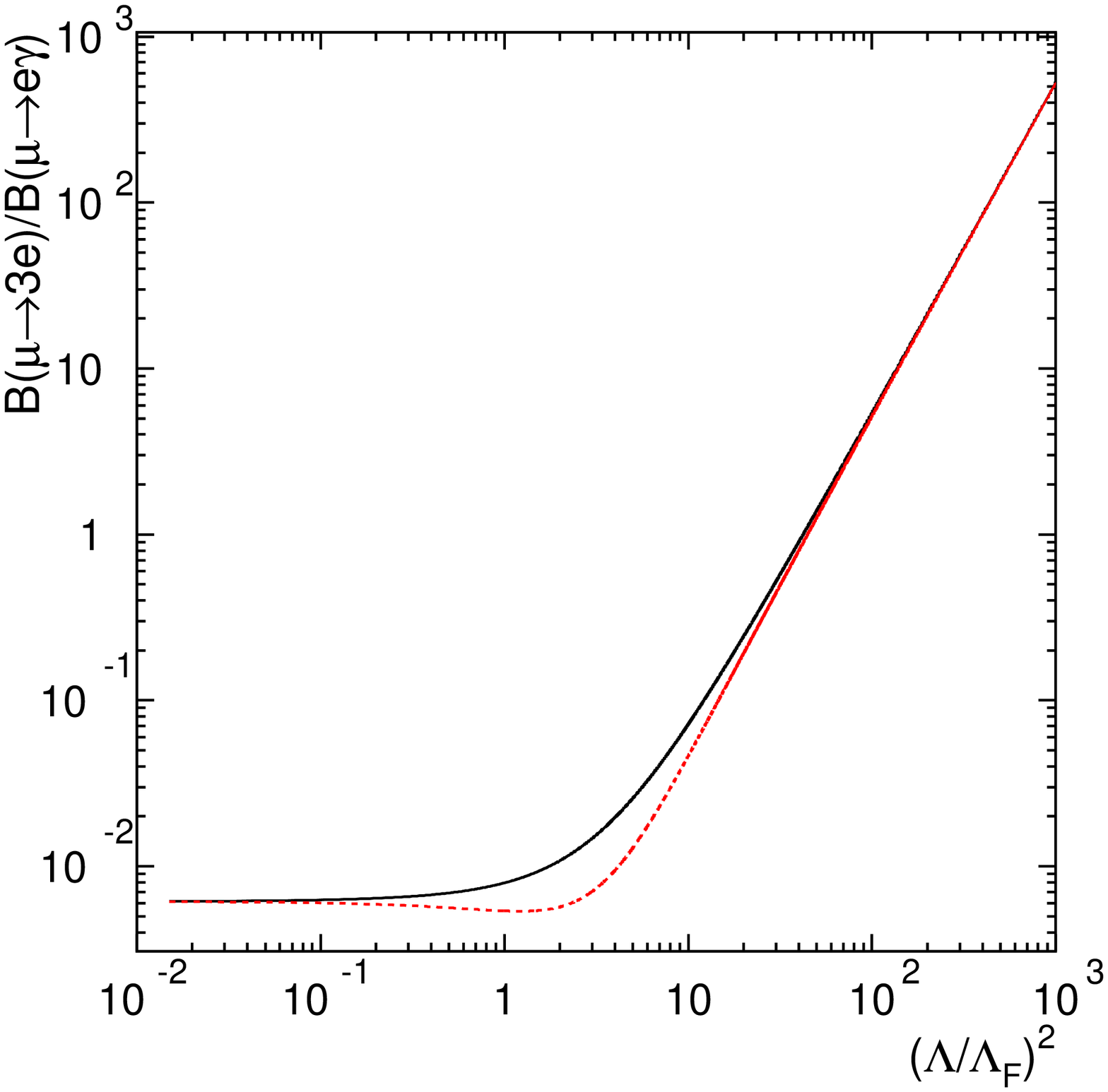}}
\parbox{0.5\textwidth}{\includegraphics[height=7.5cm]{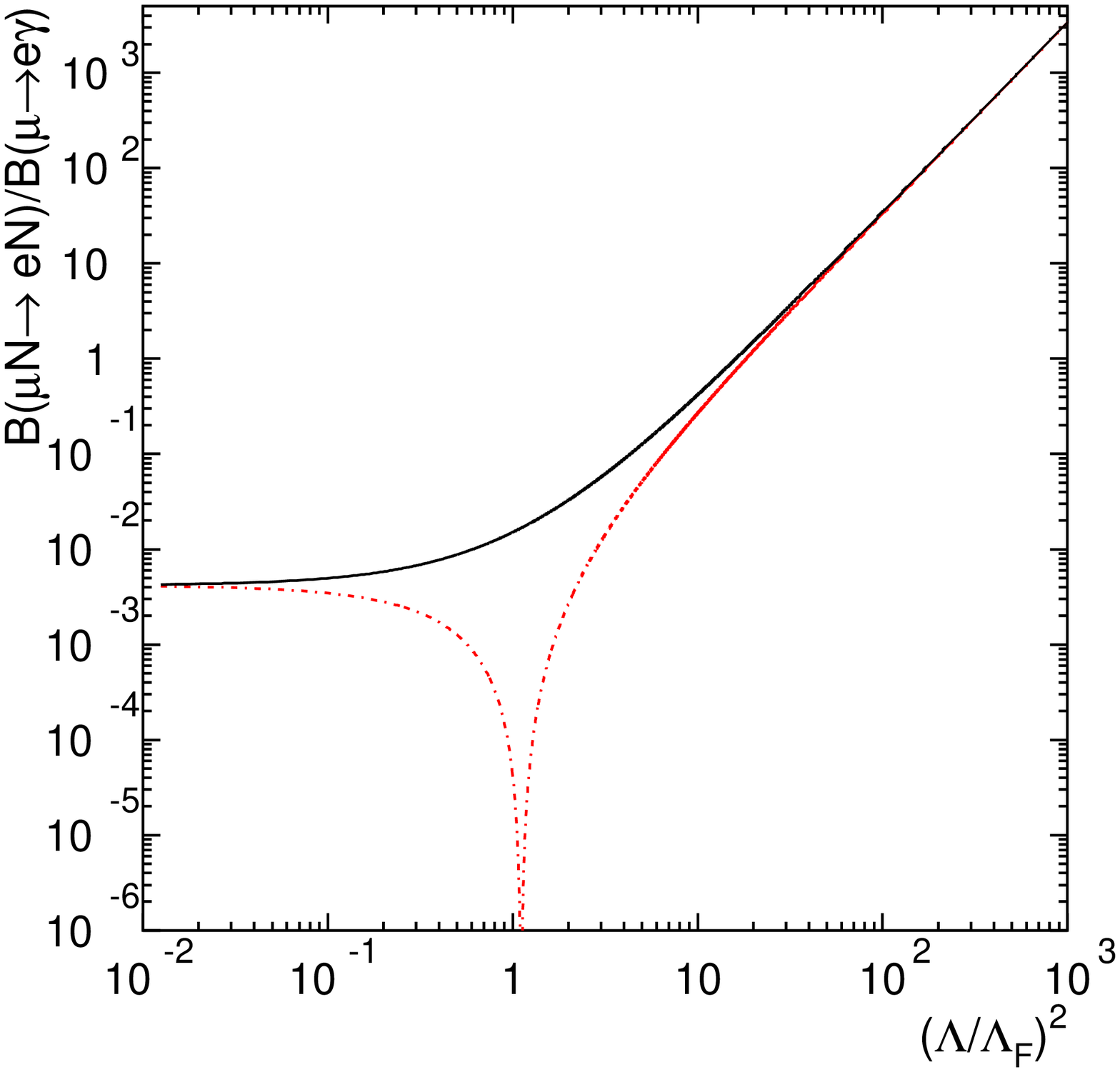}}}
\caption{Branching ratios normalised to $B(\mu\to e\gamma)$ 
as a function of the ratio of the couplings of effective dimension-5 and 
dimension-6 operators (see text), for $\mu^+\to e^+e^-e^+$ (left)
and $\mu^- N\to e^-N$ conversion in $^{48}Ti$. The solid (dashed)
curves apply when the two operators interfere constructively (destructively).}
\label{ratio_of_bs}
\end {figure}

Figure~\ref{ratio_of_bs} summarizes the behaviour of the ratio of branching
ratios as a function of the relative strength of the effective operators in
Eq.~(\ref{magmom}) and Eq.~(\ref{op4}), when the fermion $f$ in Eq.~(\ref{op4})
is an electron, as in the left part of Fig.~\ref{ratio_of_bs}, or a
combination of first generation quarks, as in the right part of
Fig.~\ref{ratio_of_bs}. It can easily be seen from the plots that when the
magnetic moment operator dominates ($\Lambda^2\ll\Lambda_F^2$) the ratio of
branching ratios saturates at several times $10^{-3}$, while it grows like
$(\Lambda^2/\Lambda_F^2)^2$ when the four-fermion operators are dominant
($\Lambda^2\gg\Lambda_F^2$). Interference effects are largest when
$\Lambda^2\sim\Lambda_F^2$, as expected.   

In conclusion, the various rare muon processes are all potentially very
interesting. In the event of a positive experimental signal for muon-number
violation, a comparison between searches in the different channels and the use
of the effective-operator approach discussed here will allow us to quickly
identify the correct class of models.

The study of rare muon decays also allows for the possibility of observing
`weak' time-reversal invariance violation (T violation)\footnote{It should
be kept in mind that the study of CP violation is not practical due to the 
experimental difficulty of stopping negative muons in matter without absorbing 
them.}. While one cannot construct experimentally accessible T-violating
observables in $\mu^+\to e^+\gamma$ decay, in 
$\mu^+\to e^+e^-e^+$ this is a possibility if the muon is polarized
\cite{cpv_in_mueee,shrock}. In this case, one may test whether the angular
distribution of the decay electrons depends on the T-odd Lorentz 
invariant $\propto \vec{P}_{\mu}\cdot (\vec{p}_{e1}\times\vec{p}_{e2})$ 
(here $p_{ei}$, $i=1,2$ are the momenta of the two positrons in the $\mu^+$ 
rest frame and $\vec{P}_{\mu}$ is the muon polarisation). In particular, a
T-odd asymmetry can be defined (see Refs.~\cite{cpv_in_mueee,OOS} for details).
In a number of theoretical models (see Refs.~\cite{OOS,RPV,raidal} 
and references therein)
`large' values (up to 24\% \cite{RPV}) for this T-odd asymmetry can be
obtained. Unfortunately, in these cases the branching ratio for
$\mu^+\to e^+\gamma$ is significantly larger than that for
$\mu^+\to e^+e^-e^+$.

It should also be noted that, again when the decaying muons are
polarised,
a number of P-odd observables in $\mu^+\to e^+\gamma$ and
$\mu^+\to e^+e^-e^+$ can be defined. The measurement of these
P-odd asymmetries can play a significant role when it comes to 
distinguishing different classes of models that yield significant flavour
violation in the charged lepton sector (see~Refs.~\cite{OOS,RPV,raidal} 
and references
therein).

\subsubsection{Rare muon processes in supersymmetry}
Supersymmetric extensions
of the SM provide a very promising way of rendering the
hierarchy of physical mass scales more natural.
Moreover, low-energy supersymmetry often leads to 
large sources of individual lepton number violation.
It provides a framework for computing physics observables in
a
controlled manner as a function of a well-defined set of parameters.
Those, in turn, can thus be constrained by the experimental limits.
Whereas lepton-flavour number violation in rare $\mu$ decays may well have
the same source as neutrino oscillations, the rates generically are no
longer
suppressed by powers of neutrino masses: 
It is worth recalling that, for large values of the ratio of the  two Higgs 
vacuum expectation values $\tan\beta$, the coefficient
of the non-diagonal
magnetic-moment operator in Eq.~(\ref{magmom}) grows
linearly with $\tan\beta$ since, at the loop level, the chiral flip 
can be generated by the Higgs with the larger vacuum expectation value,
and thus
\begin{equation}
B(\mu \to e \gamma )\propto \tan^2\beta \, .
\end{equation}

We distinguish two different supersymmetric sources of lepton-flavour number
violation.

\vskip0.3cm
$\bullet$ {\it Flavour non-diagonal soft terms}

It is well known \cite{nilles} that a mismatch in flavour space between
the 
lepton and 
slepton mass matrices generates tree-level transitions between different
leptonic generations, both in charged and neutral currents. For instance,
if we indicate the mixing angle between the first two generations of
sleptons
schematically by $\theta_{{\tilde e}{\tilde \mu}}$, we obtain
\begin{equation}
B(\mu \to e \gamma )\simeq\frac{\alpha^3\pi
\theta_{{\tilde e}{\tilde \mu}}^2}{G_F^2 {\tilde m}^4}\;,
\label{megs}
\end{equation}
where ${\tilde m}$ is a typical supersymmetric mass. In general,  
$\theta_{{\tilde e}{\tilde \mu}}$ 
receives contributions from the flavour mismatches of left- and right-handed
sleptons and sneutrinos.
Complete formulae
for the rates of rare muon processes with the functional dependence on
the different supersymmetric parameters can be found in Ref.~\cite{hisano}.
As is apparent from
Eq.~(\ref{megs}), the rates for rare muon processes can be large in 
supersymmetric models, since $\theta_{{\tilde e}{\tilde \mu}}$
is not necessarily suppressed by powers of neutrino masses.

For generic values of the soft supersymmetry-breaking mass scale $\tilde
m
\sim 1$~TeV, the experimental limits on rare muon decays
transform in very stringent upper limits on $\theta_{{\tilde e}{\tilde
\mu}}$.
Therefore, it is common to invoke some universality condition or flavour
symmetry that implies $\theta_{{\tilde e}{\tilde \mu}}=0$
at the scale at which the soft terms are generated. If this scale is
sufficiently low, as in gauge-mediated supersymmetry, then loop
corrections are small and rare muon processes are unobservable
\cite{gmsb}. 
However, if the scale of supersymmetry breaking is
large, as in supergravity models, then at the quantum level the soft
terms 
are corrected by any high-energy flavour-violating interactions, 
and we generally expect more sizeable slepton mixing angles~\cite{HKR}. 

For quantitative estimates we consider two examples for flavour-violating
interactions at high energy. Both are commonly present in supergravity
models.
Additional sources of flavour violations should be expected and
therefore our estimates should be viewed as conservative lower bounds.

The first mechanism comes from renormalisable Dirac neutrino Yukawa
couplings
$y_\nu$ at energies larger than $M_R$ \cite{hisano,nu_r,nu_r_2,EGLLN}, the
scale
of the right-handed neutrino masses. Here we are assuming that the
smallness of the neutrino masses results from the seesaw mechanism
\cite{seesaw}.
In the basis in which the charged-lepton Yukawa coupling matrix is
diagonal,
loop corrections to the slepton mass matrix are not diagonal in flavour
space:
\begin{equation}
\left( m_{{\tilde \ell}_L}^2\right)_{ij}
\simeq -\frac{3m_0^2+A_0^2}{8\pi^2}
\left( y_\nu \right)^*_{ki} \left( y_\nu \right)_{kj} \ln
\frac{M_{Pl}}{M_R}\;.
\label{sleppe}
\end{equation}
Here $m_0$ and $A_0$ are the universal soft supersymmetry-breaking scalar
mass and
trilinear term, respectively, and we have assumed that all three right-handed
neutrinos have the same mass $M_R$. The contribution in Eq.~(\ref{sleppe})
induces
flavour-violating mixing angles for both sneutrinos and charged sleptons,
and therefore rare muon processes are generated by loop diagrams
involving
charginos and neutralinos.
 
 \begin {figure}[htb]
 \parbox{0.5\textwidth}{\includegraphics[height=7.5cm,angle=-0]{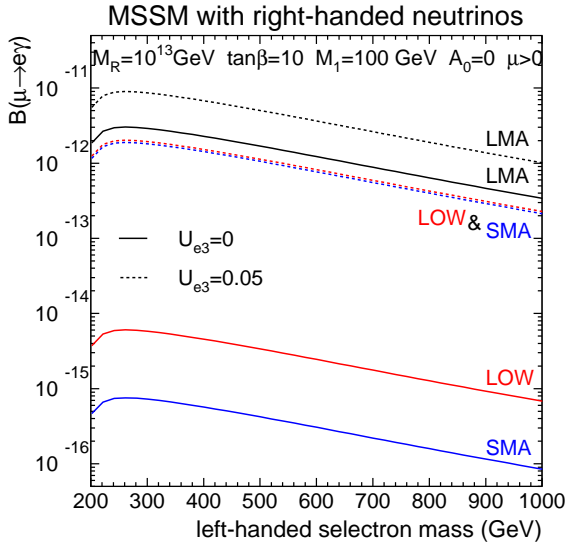}}
 \parbox{0.5\textwidth}{\caption{B($\mu\to e\gamma$) as a function of
the left-handed selectron mass for SUSY models with heavy right-handed 
neutrinos. 
The assumption here is that the right-handed neutrino mass matrix is 
proportional to the identity matrix with $M_R=10^{13}$~GeV, and 
that all the neutrino mixing comes from the Yukawa couplings. LMA, SMA,
and LOW 
refer to different solutions to the solar neutrino puzzle (see 
Ref.~\cite{neutrino_fits}).
 \label{SUSY_nu_r}}} 
 \end {figure}

The experimental information on neutrino oscillation parameters
\cite{neutrino_fits}, 
is not sufficient to reconstruct the complete structure of the relevant
mass matrices.
Therefore, it is necessary to rely on specific ans\"atze for them.  
Figure~\ref{SUSY_nu_r} depicts the value of $B(\mu\to e\gamma)$ as a 
function of the left-handed selectron mass, assuming that the
right-handed
neutrino mass matrix is proportional to the identity and that
$M_R=10^{13}$~GeV. Results are shown for
three different solutions to the solar neutrino puzzle and different
values of the $U_{e3}$ element of the neutrino mixing matrix 
(see Ref.~\cite{neutrino_fits}), assuming maximal
$\nu_{\mu}\leftrightarrow
\nu_{\tau}$ mixing as the solution to the atmospheric neutrino puzzle.   
Although the prediction for $B(\mu \to e \gamma)$ can vary significantly,
depending on differences in the assumed texture of neutrino masses, 
supergravity models in which neutrino
masses are obtained via the seesaw mechanism can induce $\mu \to e 
\gamma$ at rates close to the current experimental 
sensitivity. 
This is especially true if the solution to the solar neutrino puzzle is
indeed
the large mixing angle solution or if the $U_{e3}$ element is nonzero.
It should be noted, however, that 
because the magnitudes of neutrino Dirac Yukawa couplings are related to
the right-handed neutrino mass scale $M_R$ in seesaw models,
$B(\mu \to e \gamma)$ is very sensitive to the
right-handed neutrino scale, roughly
\begin{equation}
B(\mu \to e \gamma )\propto M_R^2[\ln(M_{Pl}/M_R)]^2 \;.
\end{equation}
As can be seen in Fig.~\ref{SUSY_nu_r}, the prediction for $B(\mu \to e
\gamma)$ also depends quite sensitively on the assumed value of $U_{e3}$. This
is a reflection of the general sensitivity to assumptions on the textures
of the charged-lepton and neutrino mass matrices.

 \begin {figure}[htb]
\parbox{0.5\textwidth}{\includegraphics[height=5.5cm,angle=-0]{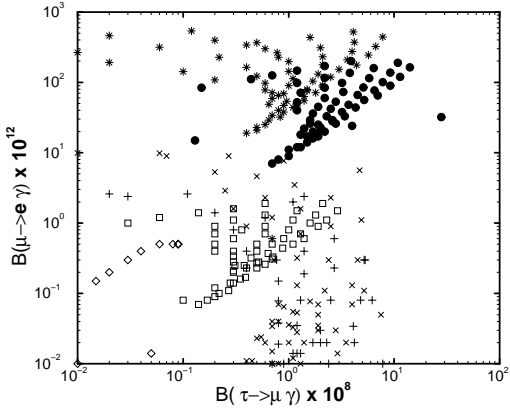}}
 \parbox{0.5\textwidth}{\caption{Scatter plot of values
of $B(\mu \to e\gamma)$ and $B(\tau \to \mu \gamma)$ in a sampling of
models of lepton masses and mixing with textures generated by $U(1)$
flavour symmetries. There is considerable spread in the predictions,
but it seems that $\mu \to e \gamma$ decay may offer better prospects
for discovering new physics than does $\tau \to \mu \gamma$, within
this class of models~\cite{EGLLN}. Note that current experiments already
constrain $B(\mu \to e \gamma)<1.2\times 10^{-11}$ and 
$B(\tau \to \mu \gamma)<1.1\times 10^{-6}$.
 \label{SUSY_mutau}}}
 \end {figure}

Figure~\ref{SUSY_mutau}
displays values of $B(\mu \to e \gamma)$ and $B(\tau \to \mu \gamma)$
found~\cite{EGLLN} in a sampling of specific 
models whose masses and mixing angles
have different textures generated by $U(1)$
flavour symmetries. We see that the predictions span a considerable range,
but, in this class of models, values of $B(\mu \to e \gamma)$ close
to the present experimental limit tend to be associated with 
values of $B(\tau \to
\mu \gamma)$ that are considerably below the present experimental
upper limit $B(\tau \to \mu \gamma) \sim 10^{-6}$, suggesting that
$B(\mu \to e \gamma)$ may be a more promising place to search for new
physics. On the other hand, improving the sensitivity to 
$B(\tau \to \mu \gamma)$ provides very useful information when it comes
to disentangling different neutrino mass models~\cite{nu_r_2}.

\begin {figure}[htb]
\parbox{0.5\textwidth}{\includegraphics[height=5.5cm,angle=-0]{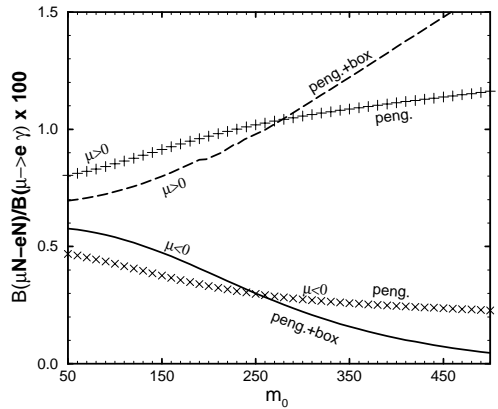}}
 \parbox{0.5\textwidth}{\caption{The ratio of the rates for $\mu$--$e$ conversion
and $\mu \to e \gamma$,
 as a function of $m_0$ for $\mu >0$ and $\mu <0$ ($\tan \beta
=3$)~\cite{EGLLN}.
 The penguin contribution, which includes the magnetic-moment piece, is
plotted separately from
 the sum of the penguin and box diagrams. The results are in the range
 expected from Eq.~(\ref{m4e}).
 \label{SUSY_meco}}}
 \end {figure}

The decay $\mu^+ \to e^+ e^- e^+$ is usually dominated by an
intermediate photon, and therefore Eq.~(\ref{m3e}) holds rather
generally. This dominance is less certain in
$\mu^-$--$e^-$ conversion, so there may be some deviations from
Eq.~(\ref{m4e}). One particular example~\cite{EGLLN} is shown in
Fig.~\ref{SUSY_meco},
where the penguin-diagram contribution (which includes the
magnetic-moment part) is plotted separately
from the sum of penguin and box diagrams. As in Fig.~\ref{ratio_of_bs},
we
see that
Eq.~(\ref{m4e}) holds generically to within some numerical factor of
order
unity.

\begin {figure}[htb]
\parbox{0.4\textwidth}{\includegraphics[height=5.5cm,angle=-0]{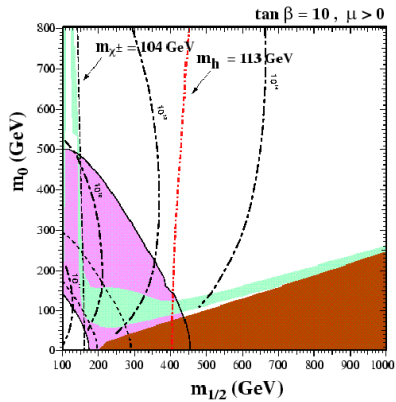}}
 \parbox{0.6\textwidth}{\caption{The contours $B(\mu \to e \gamma)=
10^{-11},10^{-12}, 10^{-13}$ and
$10^{-14}$, for a specific neutrino mass texture described
in Ref.~\cite{Carvalho:2001ex}, in the gaugino ($m_{1/2}$) versus scalar mass
($m_0$) plane, assuming $\tan\beta =10$, $A_0=0$, and $\mu>0$. The
regions
allowed by the E821 measurement of $a_\mu$ at the 2-$\sigma$
level~\cite{Brown_01} are shaded (pink) and bounded by solid black lines,
with dashed lines indicating the 1-$\sigma$ ranges. The dark (brick-red)
shaded regions are excluded because the LSP is the charged stau and the
light (turquoise) shaded regions are those with $0.1<\Omega_\chi h^2<0.3$
that are preferred by cosmology.  Also shown are the contours
$m_h=113$~GeV (assuming $m_t = 175$~GeV) and $m_{\chi^+}=104$~GeV. 
\label{SUSYcorr}}}
\end {figure}

The possible discrepancy between the Standard Model and the recent
measurement of the muon anomalous magnetic moment $\delta a_\mu \equiv
(g_\mu-2)$~\cite{Brown_01} puts a new perspective on the searches for
$\mu
\to e \gamma$ and related processes. Quite generally, the
discrepancy, if confirmed, would be evidence for new physics at the TeV
scale that might be detectable in other processes involving 
muons~\cite{Czarnecki_00_a}.
Within
any specific model that incorporates lepton-flavour violation as inferred
from neutrino-oscillation experiments, the magnitude of $\delta a_\mu$
may
be
used to normalise the prediction for $B(\mu \to e \gamma)$. This
correlation is very close if $\delta a_\mu$ is due to
supersymmetry~\cite{HT,Carvalho:2001ex,attri}, since $\mu \to e \gamma$
decay is
dominated by the magnetic-dipole-moment type operator in
Eq.~(\ref{magmom}), and the supersymmetric contribution $\delta
a_\mu^{\rm
SUSY}$ to the muon anomalous magnetic moment is induced by similar
diagrams mediated by sleptons, neutralinos and charginos. If the
left-handed sleptons have lepton-flavour-violating soft masses, the
correlation is particularly strong, since the dominant contributions to
both $\mu
\to e \gamma$ and $\delta a_\mu^{\rm SUSY}$ are from the
left-handed slepton--chargino loop diagrams: 
\begin{eqnarray}
\delta a_\mu^{\rm SUSY} &\simeq& 
\frac{5\alpha_2^2 + \alpha_Y^2}{48 \pi} \frac{m_\mu^2}{m^2_{\rm SUSY}}
\tan\beta,
\\
B(\mu \to e \gamma) &\simeq& \frac{\pi}{75}
\alpha (\alpha_2+\frac{5}{4} \alpha_Y)^2 (G_{\rm F}^2 m^4_{\rm SUSY})^{-1}
\tan^2 \beta \left(\frac{(m^2_{\tilde{\ell}_L})_{12}}{m^2_{\rm SUSY}} 
\right)^2,
\\
&=& 3 \times 10^{-5} \left(\frac{\delta a_\mu^{\rm SUSY}}{10^{-9}}
\right)^2
\left(\frac{(m^2_{\tilde{\ell}_L})_{12}}{m^2_{\rm SUSY}} \right)^2.
\end{eqnarray}
Here we have assumed, for illustration, that all the supersymmetric mass
parameters are equal to a common mass, $m_{\rm SUSY}$.  Assuming that the
2.6-sigma deviation recently observed in the Brookhaven E821
experiment~\cite{Brown_01} is due to low energy SUSY, ({\it
i.e.}\/$\delta
a_\mu^{\rm SUSY} \sim 10^{-9})$, we can use the present experimental
upper
limit $B(\mu \to e \gamma) < 1.2 \times 10^{-11}$ to obtain a
stringent constraint on the lepton-flavour-violating mass insertion: 
\begin{eqnarray}
\frac{(m^2_{\tilde{\ell}_L})_{12}}{m^2_{\rm SUSY}}
\leq 6\times 10^{-4} 
\left(\frac{\delta a_\mu^{\rm SUSY}}{10^{-9}} \right)^{-1}
\left(\frac{B(\mu \to e \gamma)}{1.2 \times 10^{-11}}
\right)^{\frac{1}{2}}.
\label{limit_g2}
\end{eqnarray}
As can be seen from Eq.~(\ref{sleppe}), for example, 
large neutrino Yukawa couplings could easily induce 
an observable effect in the near future.
The correlation between $B(\mu \to e \gamma)$
and $\delta a_\mu^{\rm SUSY}$ in the case of a specific choice of 
neutrino mass texture~\cite{Carvalho:2001ex},
is illustrated in Fig.~\ref{SUSYcorr}. We see that, at least within the
context of this model, $\mu \to e \gamma$ decay may well be
observable in the next round of experiments, and the same is true for
anomalous $\mu \to e$ conversion on nuclei. 
The establishment of the anomaly in $g_\mu-2$, together with
improvements of the present experimental limits on rare muon
decay processes, will probe severely supersymmetric models.

\begin {figure}[htb]
\centerline{
 \parbox{0.5\textwidth}{\includegraphics[height=7.5cm,angle=0]{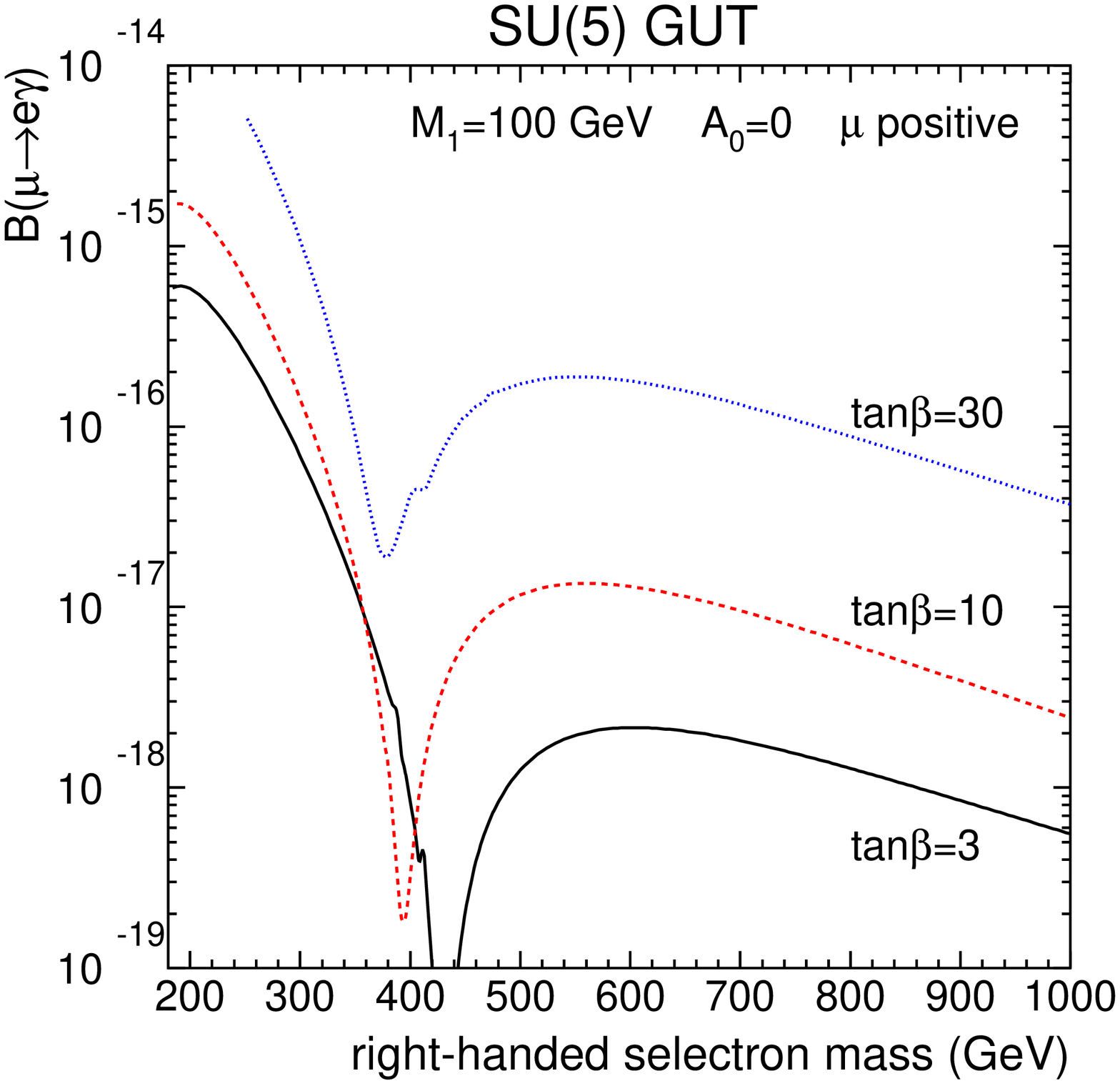}}
 \parbox{0.5\textwidth}{\includegraphics[height=7.5cm,angle=0]{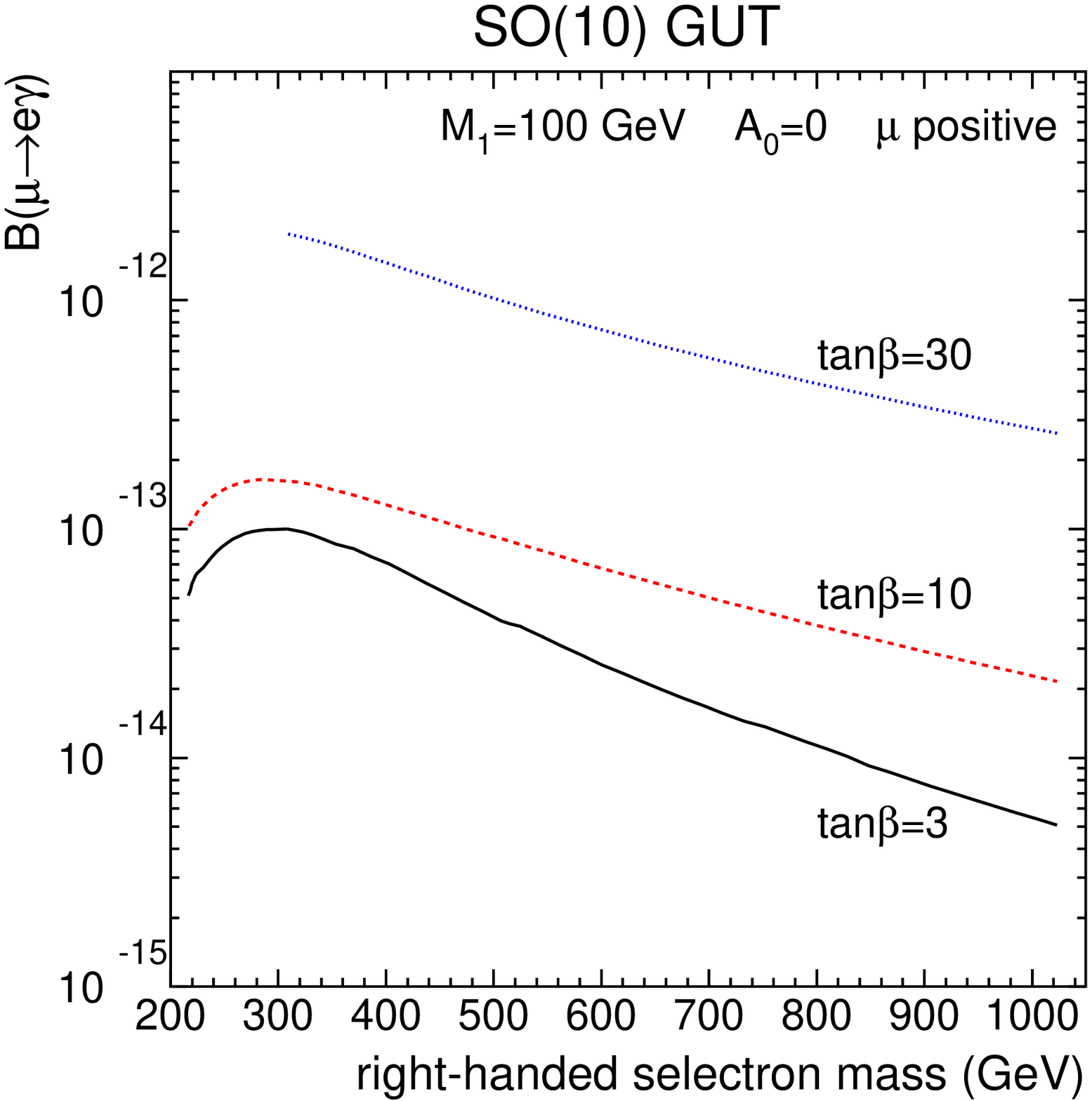}}}
 \caption{$B(\mu\to e\gamma)$ as a function of the 
right-handed selectron
mass in supersymmetric models with SU(5) (left) \cite{SU(5)} and SO(10)
(right) 
\cite{meg_SUSY_so10} 
grand unification.}
\label{SUSY_GUTs} 
\end {figure}

The second mechanism we investigate is present when the 
minimal supersymmetric extension of the SM is embedded into
a grand unified theory (GUT). 
In this case, leptons and quarks belong to the same group representation,
and the large top Yukawa coupling $y_t$ will split the mass of
the third-generation slepton from the other 
two~\cite{meg_SUSY,meg_SUSY_so10}. 
In SU(5) models~\cite{SU(5),SU(5)_2}, the 
up-type quarks are in the same representation as the right-handed charged
leptons, and therefore we find
\begin{equation}
\left( m_{\tilde \ell_R}^2\right)_{33}\simeq
-\frac{3(3m_0^2+A_0^2)}{8\pi^2}
y_t^2 \ln \frac{M_{Pl}}{M_{\rm GUT}}\;.
\label{sleppe2}
\end{equation}
Once the charged-lepton mass matrix is diagonalised, Eq.~(\ref{sleppe2})
induces slepton mixing. The left part of 
Fig.~\ref{SUSY_GUTs} depicts $B(\mu\to e\gamma)$ as a function
of the right-handed selectron mass for different values of $\tan\beta$ in
the
case of an SU(5) GUT. 
In this case, since the mixing is present only 
in the right-handed lepton sector, the positron in the anti-muon decay is
left-handed, $\mu^+ \to e_L^+ \gamma$.
This decay could be distinguished from $\mu^+ \to e_R^+ \gamma$
by studying the angular distribution of the positron in
polarised $\mu^+ \to e^+ \gamma$ decays~\cite{Kuno2,pol_mu}.

In the case of SU(5) models with massless neutrinos, 
the rates for $\mu \to e \gamma$ are
typically much smaller than the present experimental bound.
This is because chargino loop diagrams are absent and neutralino diagrams
suffer from a partial cancellation. Note that the position of the dip in
the curves
of Fig.~\ref{SUSY_GUTs} (corresponding to the exact cancellation) depends
on the value of the gaugino mass $M_1$.
However, if higher-dimensional
operators
are added to models with SU(5) symmetry, $B(\mu\to e\gamma)$
may be enhanced (see, for example, Ref.~\cite{SU(5)_2}). Such operators may be
required in order to obtain realistic mass relations for first 
and second generation quarks and leptons. Also note that the values of
$B(\mu\to e\gamma)$ depends very strongly on the precise value of the top
Yukawa coupling at the GUT scale, which is not well determined by the 
measurement of the top quark mass, because of its quasi-fixed-point structure. 

The situation changes in SO(10) models\footnote{Other unification
scenarios have also been explored. See, for example, Ref.~\cite{other_susy}.} 
\cite{meg_SUSY_so10,SO(10)}, 
where all SM fermions belong to the same irreducible representation, and
therefore the top Yukawa effect splits not only $\tilde \ell_R$, but
also $\tilde \ell_L$. Furthermore, since the loop diagram
involves the exchange of a third-generation slepton, the necessary chiral
flip can occur along the $\tilde \tau$ line, and $B(\mu \to e \gamma )$
is enhanced by a factor $m_\tau^2/m_\mu^2$. Figure~\ref{SUSY_GUTs}(right)
depicts
$B(\mu\to e\gamma)$ as a function
of the right-handed selectron mass for different values of $\tan\beta$ in
the
case of an SO(10) GUT. 

\vskip0.3cm
\newpage
$\bullet$ {\it R-parity violation}

In supersymmetric models where $R$ parity is not imposed as a discrete 
symmetry, 
one encounters renormalisable interactions that violate total lepton
number 
and individual lepton numbers. At the scale of the muon mass, these
interactions 
correspond to effective four-fermion operators of the kind discussed in 
Section~3.1.1.
Therefore, depending on the field content of the operator, we can have
situations in which $\mu^+\to e^+e^-e^+$ or $\mu^-$--$e^-$ conversion 
become the dominant rare muon process. For instance, if a model predicts
the
simultaneous presence of the two $R$-parity-violating
interactions $\lambda_{131}L_L^{(1)}L_L^{(3)}E_R^{c(1)}$ and
$\lambda_{231}L_L^{(2)}L_L^{(3)}E_R^{c(1)}$, then $\mu^+\to e^+e^-e^+$
occurs at the tree level with a branching ratio
\begin{equation}
B(\mu \to 3 e)= \frac{(\lambda_{131}\lambda_{231})^2}{32G_F^2m_{{\tilde 
\nu}_\tau}^4} = \left( \frac{\lambda_{131}\lambda_{231}}{10^{-6}}
\right)^2 \left( \frac{200~{\rm GeV}}{m_{{\tilde 
\nu}_\tau}}\right)^4 10^{-13}\;.
\end{equation}
Present constraints from rare muon processes provide
very stringent upper limits on numerous products of  $R$-parity-violating
couplings, some of which are listed in Table~\ref{rpv_constr}. At a
neutrino factory complex, these bounds could be improved by two orders of
magnitude if $\mu \to e \gamma$ and $\mu \to eee$ searches are not
successful
and by three orders of magnitude if $\mu$--$e$ conversion in nuclei is
not observed.

\begin{table}[hbt]
\caption{Upper limits on products of $R$-parity-violating couplings
from the current limits for muon-number-violating processes, taking
slepton masses equal to 100~GeV and squark masses equal to 300~GeV. 
The values in parentheses indicate the sensitivity which could be
achieved at a neutrino factory complex, i.e., resulting from improvements
in
experimental sensitivities to rare $\mu$ decays by 4-6 orders of
magnitude.
The coupling constants $\lambda$ and $\lambda^\prime$ refer to the
interactions
$W_{R\hspace*{-1.5mm}\slash}\supset \lambda_{ijk}L^iL^jE^k + 
\lambda'_{ijk}L^iQ^jD^k$, where $i,j,k$ are generation indices.
Tree-level constraints are indicated by [tree].}
\label{rpv_constr}
\begin{center}
\begin{tabular}{|c|l|l|l|} \hline
 & $\mu \to e\gamma$ & $\mu \to eee$ 
& $\mu \to e$~conversion in nuclei              \\ \hline
$|\lambda_{131}\lambda_{231}|$ 
& $2.3\times 10^{-4}$ $(2\times 10^{-6})$&$6.7\times 10^{-7}$
$(7\times 10^{-9}) [{\rm tree}]$
&$1.1\times 10^{-5}$ $(1\times 10^{-8})$        \\
$|\lambda_{132}\lambda_{232}|$ 
& $2.3\times 10^{-4}$ $(2\times 10^{-6} )$& $7.1\times 10^{-5}$ 
$(7\times 10^{-7} )$ 
&$1.3\times 10^{-5}$ $(2\times 10^{-8})$        \\
$|\lambda_{133}\lambda_{233}|$ & 
$2.3\times 10^{-4}$ $(2\times 10^{-6} )$& $1.2\times 10^{-4}$ $(1\times
10^{-6})$ & $2.3\times 10^{-5}$ $(3\times 10^{-8})$     \\
$|\lambda_{121}\lambda_{122}|$ 
& $8.2\times 10^{-5}$ $(7\times 10^{-7} )$ &$6.7\times 10^{-7}$ 
$(7\times 10^{-9}) 
[{\rm tree}]$ & $ 6.1\times 10^{-6}$ $(8\times 10^{-9})$        \\
$|\lambda_{131}\lambda_{132}|$ 
& $8.2\times 10^{-5}$ $(7\times 10^{-7} )$ & $6.7\times 10^{-7}$ 
$(7\times 10^{-9}) [{\rm tree}]$ &$7.6 \times 10^{-6}$ $(1\times 10^{-8})$\\
$|\lambda_{231}\lambda_{232}|$ 
& $8.2\times 10^{-5}$ $(7\times 10^{-7} )$ & $4.5\times 10^{-5}$ 
$(5\times 10^{-7})$ &$8.3\times 10^{-6}$ $(1\times 10^{-8})$    \\
$|\lambda '_{111}\lambda '_{211}|$ 
& $6.8\times 10^{-4}$ $(6\times 10^{-6})$ & $1.3\times 10^{-4}$ $(1\times
10^{-6})$&      $5.4\times 10^{-6}$ $(7\times 10^{-9})~[{\rm tree}]$    \\
$|\lambda '_{112}\lambda '_{212}|$ 
& $6.8\times 10^{-4}$ $(6\times 10^{-6})$ & $1.4\times 10^{-4}$ $(1\times
10^{-6})$ &     $3.9\times 10^{-7}$ $(5\times 10^{-10})~[{\rm tree}]$   \\
$|\lambda '_{113}\lambda '_{213}|$ 
& $6.8\times 10^{-4}$ $(6\times 10^{-6})$ & $1.6\times 10^{-4}$ $(2\times
10^{-6})$ &     $3.9\times 10^{-7}$ $(5\times 10^{-10})~[{\rm tree}]$   \\
$|\lambda '_{121}\lambda '_{221}|$ 
& $6.8\times 10^{-4}$ $(6\times 10^{-6})$ & $2.0\times 10^{-4}$ $(2\times
10^{-6})$ &     $3.6\times 10^{-7}$ $(5\times 10^{-10})~[{\rm tree}]$   \\
$|\lambda '_{122}\lambda '_{222}|$ 
& $6.8\times 10^{-4}$ $(6\times 10^{-6})$ & $2.3\times 10^{-4}$ $(2\times
10^{-6} )$&     $4.3\times 10^{-5}$ $(6\times 10^{-8})$ \\
$|\lambda '_{123}\lambda '_{223}|$ 
& $6.9\times 10^{-4}$ $(6\times 10^{-6})$ & $2.9\times 10^{-4}$ $(3\times
10^{-6})$ &     $5.4\times 10^{-5}$ $(7\times 10^{-8})$ \\ \hline
\end{tabular}
\end{center}
\end{table}

\subsubsection{Rare muon processes in models with extra dimensions}

Recently, theories with extra spatial dimensions have attracted a lot of
attention as a possible solution to the problem of the large hierarchy
between the Planck and Fermi mass scales. The hypothesis is that the universe
possesses $1+n$ ($n>3$) space-time dimensions, while the Standard
Model particles are constrained to live on a 1+3-dimensional subspace. 
Gravity, which is described by the geometry and
therefore propagates in
the full space, appears very weak to us either because its strength is
diluted in a large compactified extra-dimensional space~\cite{add} or because it is
localised away from us in spaces with non-factorizable geometries~\cite{rs}. Since
these  theories assume that
gravity becomes strongly coupled at an energy scale $M_D$ comparable to the 
electroweak scale, the possibility of explaining the smallness of the 
neutrino masses via the classical see-saw mechanism is lost. Nevertheless,
the small neutrino masses could now have an explanation based on geometrical
arguments, similar to those that led to a justification of the small ratio
$M_W/M_{Pl}$. For this to happen, one needs to assume the existence of
right-handed neutrinos which, like the graviton, also propagate
in the extra dimensions. If this is the case, their
Yukawa interactions with the SM left-handed neutrinos are effectively suppressed by
large geometrical factors~\cite{neuextr}, and therefore the SM neutrinos obtain,
after electroweak symmetry breaking, a 
very small Dirac mass. When neutrino family mixing is included,
one finds that the Kaluza--Klein modes of the right-handed neutrino mediate,
at the loop level, flavour transitions in the charged 
sector~\cite{violextr,dGGST}. 

Particularly interesting are `minimal models' in which all flavour transitions 
are described by only two free parameters, besides the observable neutrino 
masses and mixing angles. These two parameters are the `fundamental' 
cut-off scale $\Lambda$ (expected to be of the order of the weak scale, if 
these models are motivated by the hierarchy problem) and a dimensionless coefficient 
$\epsilon$, which is currently constrained to be approximately less than $10^{-2}$
and whose expected value depends strongly on the neutrino Yukawa couplings
and on details of the extra dimensional model,
such as the number of extra dimensions $\delta$ in which the right-handed neutrinos
propagate (see Ref.~\cite{dGGST} for details). 

Under these conditions, the branching ratio for  
$\mu\rightarrow e\gamma$ is
\begin{equation}
B(\mu \rightarrow e \gamma)  =  \frac{3 \alpha}{8\pi} \epsilon^2
\left|U_{e2} U_{\mu 2}^* \frac{\Delta m^2_{\rm sun}}{\Delta m^2_{\rm atm}}
+U_{e3} U_{\mu 3}^* 
\right|^2\;.
\end{equation} 
Here the unitary matrix $U$ describes the neutrino mixing angles and 
$\Delta m^2_{\rm sun , atm}$ are the neutrino mass-squared differences relevant to
solar and atmospheric oscillation, for a hierarchical neutrino mass spectrum.  

\begin{figure}[htb]
\centerline{\includegraphics[width=0.65\textwidth]{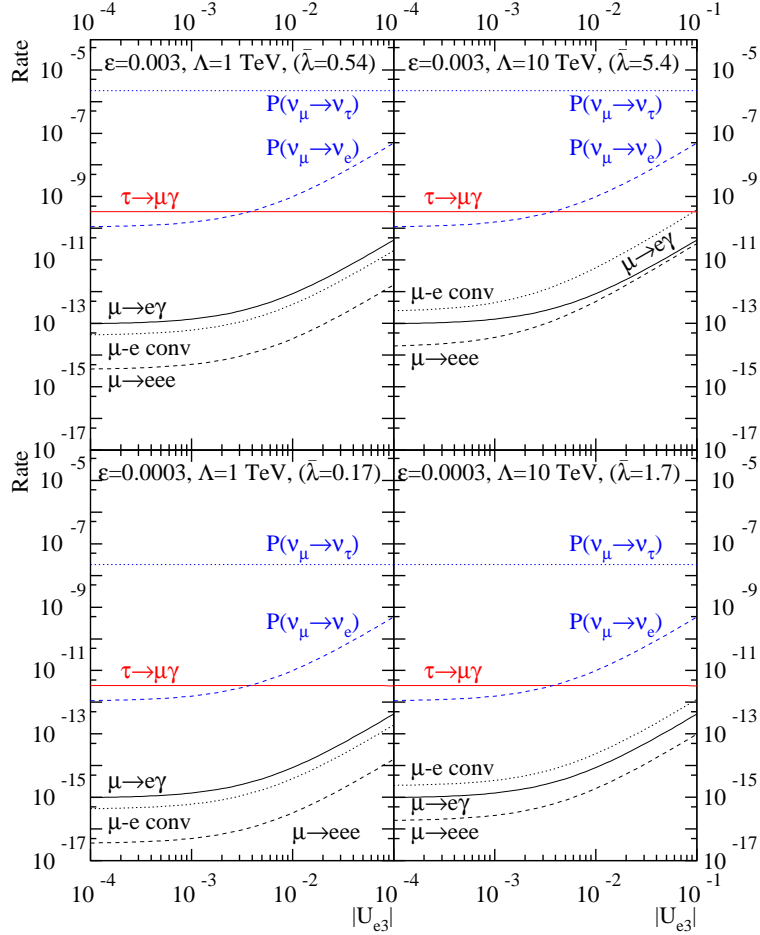}}
\caption{Branching ratios of $\mu  \to e \gamma$, $\mu \to
3e$ and $\tau \to \mu 
\gamma$, 
normalized rate of $\mu$--$e$ conversion in $^{27}$Al, 
and transition probabilities for $\nu_\mu \to \nu_\tau$ and
$\nu_\mu \to \nu_e$ at very short baseline as  functions of $|U_{e3}|$.
We have chosen the parameters $\delta=5$,
$\epsilon=0.003$ (top) or 0.0003 (bottom), and $\Lambda=1$~TeV (left) or 10~TeV (right). 
We assumed $\Delta m^2_{\rm sun}/\Delta m^2_{\rm atm}=10^{-2}$, $|U_{\mu3}/U_{\tau3}|^2=1$,
$|U_{e2}/U_{e1}|^2=2/3$ (i.e.\ maximal mixing in the atmospheric sector and the LMA
solution to the solar neutrino puzzle), no
CP-violation in the neutrino mixing matrix, and
hierarchical neutrino masses ($m_1^2\ll m_2^2\ll m_3^2$). See Ref.~\cite{dGGST} for
details.
\label{fig:br}}
\end {figure}

Unlike $\ell_i\rightarrow \ell_j\gamma$ decays, the rates for
$\mu\rightarrow eee$ and $\mu\rightarrow e$ conversion in nuclei
are quite dependent on the unknown ultraviolet details 
of the models. 
Nevertheless, these rates can be predicted as a function of
$\epsilon$ and $\Lambda$
(see Refs.~\cite{violextr,dGGST} for complete 
expressions). It should be noted that the 
rates for $\mu\rightarrow eee$ and $\mu\rightarrow e$ conversion in 
nuclei can be significantly enhanced with respect to the rate for 
$\mu\rightarrow e\gamma$ in some regions of the $\epsilon, \Lambda$ parameter
space, similarly to the case of SUSY models with $R$-parity violation.
Figure~\ref{fig:br} shows the rates for the different 
flavour-violating lepton processes,  
 as functions of $|U_{e3}|$, assuming 
$\Delta m^2_{\rm sun}/\Delta m^2_{\rm atm}=10^{-2}$, 
$|U_{\mu3}/U_{\tau3}|^2=1$,
$|U_{e2}/U_{e1}|^2=2/3$
(as suggested by atmospheric and LMA solar oscillations),
 and no CP violation in the neutrino mixing matrix,
for different values of $\epsilon$ and $\Lambda$. We also show the  
conversion probabilities for $\nu_\mu \to \nu_e$ and $\nu_\mu \to \nu_\tau$
at very short baseline experiments. This allows a comparison of the discovery reach
for different facets of a neutrino factory, namely, experiments with stopped muons and
experiments with an intense neutrino beam.

We emphasize that the
one-loop effects  giving rise to rare muon and tau processes that violate 
individual lepton flavour number  are cut-off dependent 
and can only be qualitatively estimated. Nonetheless, in minimal models, their 
flavour structure is
directly related to the physical neutrino mass matrix, and one is capable of 
predicting and relating the rates of rare charged and neutral lepton processes 
in terms of observable neutrino oscillation parameters. This is in sharp contrast to
other cases of physics beyond the SM. In SUSY models, for instance,
the rates for rare muon processes are perturbatively calculable, but their
relations with neutrino oscillations parameters are indirect and strongly model-dependent,
as described in the previous subsection. 

 \clearpage
 \begin{figure}[htb]
\centerline{\includegraphics[width=10cm]{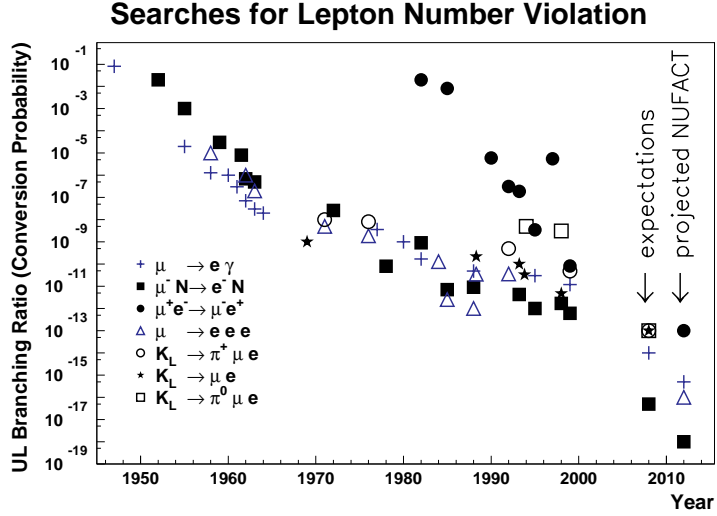}}
 \caption{Historical development of the 90\% C.L. upper limits (UL) on branching
ratios
 respectively conversion probabilities
 of muon-number violating processes which involve muons and kaons.
 Sensitivities expected for planned searches are indicated in the year
 2008 (see also Ref.~\cite{Kuno:1999jp}). The projections for
 a neutrino factory (NUFACT) are also shown.
 \label{fig:history}}
 \end{figure}

 \subsection{Experimental prospects}
 The experimental sensitivities achieved during the past decades in tests
 of muon number conservation are illustrated in  Fig.~\ref{fig:history}. 
 Generally the tests were limited up to now by the intensities of the available
 $\mu$ and K beams, but in some cases detector limitations have played a role
 as well.

 All recent results with $\mu^+$ beams were obtained with `surface' muon
 beams~\cite{surface}, i.e. beams of muons originating in the decay of
 $\pi^+$'s that stopped in the pion production target. These beams offer the
 highest muon stop densities that can be obtained at present, allowing  for the
 low-mass experimental targets that are required for the ultimate resolution
 in positron momentum and emission angle or the efficient production of
 muonium in vacuum.

 In this subsection we study the question of  how far experimental searches
 could benefit from muon beam intensities which are 2--3 orders of magnitude
 higher than presently available. We do so by analysing the present state of
 the art, assuming modest improvements in detector technology.

 \subsubsection{$\megs$}
 
 During the past 25 years the experimental sensitivity to this decay mode
 was raised by two orders of magnitude (see Table~\ref{tab:prevexp}).
 \begin{table}[htbp]
 \caption{90\% C.L. upper limit on the branching ratio for $\mu \to e
\gamma$
 obtained by previous experiments at meson factories}
 \label{tab:prevexp}
 \begin{center}
 \begin{tabular}{llrr}
 \hline 
 Experiment or Lab.     & Year\hspace*{3cm} & Limit\hspace*{5mm} & Ref.
\\
 \hline
 SIN (presently PSI)    & 1977 & $1  \times 10^{-9}$\phantom{0}    &\cite{SIN}
\\
 TRIUMF                 & 1977 & $3.6 \times 10^{-9}$\phantom{0}   &\cite{TRIUMF}
\\
 LANL                   & 1979 & $1.7 \times 10^{-10}$  &\cite{LANL}
\\
 Crystal Box            & 1986 & $4.9 \times 10^{-11}$
&\cite{Crystalbox}      \\
 MEGA                   & 1999 & $1.2 \times 10^{-11}$  &\cite{MEGA}
\\
 \hline
 \end{tabular}
 \end{center}
 \end{table}
 The most sensitive search was performed recently by the MEGA
Collaboration,
 establishing an upper (90\% C.L.) limit on $\Bmeg$ of $1.2 \times
10^{-11}$~\cite{MEGA}.
 An experiment~\cite{PSIprop}, which aims at a single-event sensitivity
of $\sim 10^{-14}$ 
 was approved by the PSI scientific committee in July 1999. Measurements
should start in 
 2003 and data taking should go on for two or three years. This
experiment will
 use a surface muon beam that reaches an intensity around
$5\times10^8\,\mu^+/s$. As a next
 step it seems reasonable to consider experiments aiming at a sensitivity
of $10^{-15}$
 or better. As we shall see below, it is not at all obvious how to reach
such levels of
 sensitivity without running into the background of accidental $e\gamma$
coincidences.

\newpage

 {\bf Experimental sensitivity}

 \smallskip
 In order to understand how the sensitivity\footnote{In this document, 
sensitivity is defined as the 90\% CL upper limit in the absence of a
signal.} to the $\megs$ decay is related to the
 experimental resolutions, let us consider an experiment in which muons
are stopped
 in a thin (typically 10 ${\rm mg/cm^2}$ ) target; photons and positrons
are detected
 by a pair spectrometer or an electromagnetic calorimeter and a magnetic
spectrometer, respectively.
 The expected number $N_s$ of observed $\megs$ decays can be written as:
 \begin{equation}
 N_s = R_{\mu} T \cfrac{\Omega}{4\pi} \, 
 \epsilon_e \epsilon_{\gamma} \epsilon_{cut} \, \Bmeg \;,
 \label{eqn1}
 \end{equation}
 where $R_{\mu}$ is the muon stop rate, $T$ is the total measuring time,
$\Omega$ is
 the detector solid angle (we assume identical values for the photon and
the positron
 detectors), $\epsilon_e$ and  $\epsilon_{\gamma}$ are the positron and
photon detection
 efficiencies, $\epsilon_{cut}$ is the efficiency of the selection cuts.
Selection cuts
 can be applied on the reconstructed positron energy ($E_e$), photon
energy ( $E_\gamma$ ),
 opening angle ($\theta_{e \gamma}$) and relative timing ($t_{e
\gamma}$). Considering
 selection windows (indicated with a $\Delta$ in front of the relevant
quantity) covering
 90 \% of the signal (full window = 1.4 FWHM for Gaussian distributions)
we have
 $\epsilon_{cut} = 0.9^4 \simeq 0.66$.
 
\begin{figure}[htb]
\centerline{\includegraphics[width=7cm]{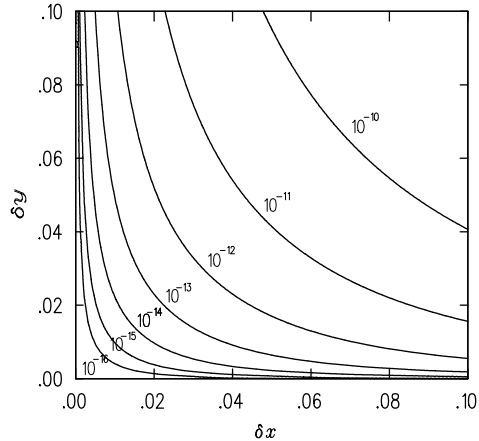}}
 \caption{Prompt background from $\rdecay$ as a function of the
 selection windows on the positron ($x = 2E_{e^+}/m_\mu$) and photon
 ($y = 2E_\gamma/m_\mu$) energies. We have defined
 $\delta x =1-x$ and $\delta y =1-y$ (results from Ref.~\cite{LOI}).
 \label{fig:pict1}}
 \end{figure}

 The level of prompt background from $\rdecay$ depends on the selection 
 windows on the energies and opening angle of the $e\gamma$ pair. The
background
 level predicted with no angular cut is shown in Fig.~\ref{fig:pict1}.
Whereas
 these estimates do not take into account details of the detector
response
 functions, they are accurate enough for our purpose. 
 It appears that detector technology is sufficiently advanced to keep
prompt background at a
 negligible level

 In the proposed PSI experiment (with $R_{\mu} \simeq 10^8 \mu/$s) the
background is
 dominated by accidental coincidences of a positron from the normal muon
decay and a
 photon which may originate in the decay $\rdecay$ or may be produced by
an ${e}^+$ through
 external bremsstrahlung or annihilation in flight. In a DC beam the
number of accidental
 coincidences is given by:
 \begin{equation}
 N_b = {R_{\mu}}^2 f_e \epsilon_e f_\gamma \epsilon_\gamma   
 (\cfrac{\Omega}{4\pi})^2 \, \pi \cfrac{\Delta \theta_{e
\gamma}^2}{\Omega}
 \, 2\Delta t \, T\;,
 \label{eqn2}
 \end{equation}
 where $f_e$ ($f_\gamma$) is the $e^+$ ($\gamma$) yield per stopped muon
within the 
 selection window and $\Delta t$ is the cut applied on the $e^+ - \gamma$
time difference.
 For a non DC beam $N_b$ must multiplied by the inverse of the duty
cycle.
 
\begin{figure}[htb]
\centerline{\includegraphics[width=7cm]{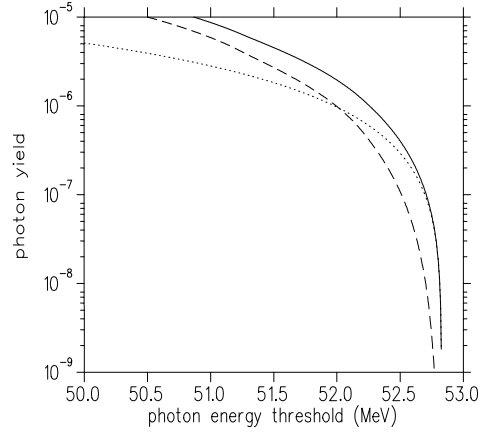}}
 \caption{Photon yield vs. energy threshold.
 Dotted: annihilation in flight in a $50 \, {\rm mg/cm^2}$ target.
 Dashed: radiative decay.
  Solid: sum of the two.
 \label{fig:pict2}}
 \end{figure}

 An evaluation of $f_e$ and $f_{\gamma}$ is given in Refs.~\cite{LOI,Kuno1}.
 Figure~\ref{fig:pict2} shows $f_{\gamma}$ as a function of the energy
threshold
 calculated for a $50 \, {\rm mg/cm^2}$ target.

 By taking into account the Michel spectrum for the positron and only
 the radiative muon decay for the photon, one finds near the endpoint 
 ($E_e \, {\rm and} \, E_\mu \simeq m_\mu/2$):
 \begin{equation}
 f_e \propto \frac{\Delta E_e}{E_e} \hspace{1cm} f_\gamma \propto
 (\frac{\Delta E_\gamma}{E_\gamma})^2\;.
 \label{eqn4}
 \end{equation}
 From Eqs.~(\ref{eqn1}) and (\ref{eqn2}) the signal-to-background ratio
is
 \begin{equation}
 \cfrac{N_s}{N_b} = \cfrac{\epsilon_{cut} \Bmeg} {R_{\mu} (f_e f_{\gamma}
 \cfrac{\Delta \theta_{e \gamma}^2}{4} \; 2\Delta t)}\;.
 \label{eqn3}
  \end{equation}

 The detector resolutions quoted in the PSI proposal are shown in
Table~\ref{tab:resolutions}
 together with those of previous experiments.
 \begin{table}[htbp]
 \caption{Detector resolutions, muon stop rates and accelerator duty
 cycles of the various $\megs$ searches.}
 \label{tab:resolutions}
 \begin{center}
 \begin{tabular}{lcccccc}
 \hline 
 Exp./Lab. & $\Delta E_e/E_e$ & $\Delta E_\gamma/E_\gamma$
 & $\Delta t_{e \gamma}$ & $\Delta \theta_{e \gamma}$ & Instantaneous
stop rate & Duty cycle \\
 &\% FWHM       &\% FWHM   &(ns)  &(mrad)  & (1/s) &(\%)  \\
 \hline
 SIN (now PSI)  &8.7    &9.3    &1.4    &-      &$(4\ldots 6)\times 10^5$
&100            \\
 TRIUMF         &10     &8.7    &6.7    &-      &$2 \times 10^5$
&100            \\
 LANL           &8.8    &8      &1.9    &37     &$2.4 \times 10^6$
&6.4            \\
 Crystal Box    &8      &8      &1.3    &87     &$4 \times 10^5$
&$(6 \ldots 9)$     \\
 MEGA           &1.2    &4.5    &1.6    &17     &$2.5 \times 10^8$
&$(6 \ldots 7)$     \\
 PSI            &0.7    &1.4    &0.15   &12     &$10^8$
&100            \\
 \hline
 \end{tabular}
 \end{center}
 \end{table}
 For this experiment a solid angle $\Omega / 4\pi$ of about 10\% and a
total measuring
 time $T$ of almost a year are being considered. It turns out that one
background
 event is expected and one $\megs$ event for $B_{\megs}=10^{-14}$.

 Since the accidental background rises quadratically with the muon stop
rate, it will be
 even more problematic in future experiments using a higher beam
intensity.
 An experiment with $R_\mu = 10^{10} \mu/{\rm s}$ and all the other
quantities of
 Eq.~(\ref{eqn1}) unchanged would yield one $\megs$ event for $B_{\megs}
= 10^{-16}$.
 However, the accidental background would increase to $10^4$ events. It
is obvious
 that better detector resolutions so as to reduce the product
 \begin{equation}
 \frac{\Delta E_e}{E_e} \;
 (\frac{\Delta E_\gamma}{E_\gamma})^2  
 \Delta \theta_{e \gamma}^2 \, \Delta t
 \label{neqn}
 \end{equation}
 and/or improved experimental concepts are required.

 \bigskip

 {\bf Ideas for background reduction}

 \smallskip
 We report here on some possibilities discussed in the study group.

 \bigskip
 {\it Detector improvements}

 \smallskip
 The safest improvement can be obtained in the positron momentum
 resolution where a substantial factor ($\sim 10$) can be gained
 by utilising a beta-spectrometer~\cite{LOI} which may
 reach resolutions of the order of 0.1\%. As an additional advantage
 such a spectrometer would result in reduced rates in the tracking
 detectors as compared to the broad-band systems used so far.

 The resolution in $e^+- \gamma$ opening angle is determined by
 positron multiple scattering in the stopping target, so another
 possible improvement could be the reduction of the target thickness.
 The use of very intense muon beams would give the possibility to
 obtain a $10^{10} \mu/$s stop rate with much thinner targets.
 A reduction by one order of magnitude would reduce
 $\Delta \theta_{e \gamma}$ by a factor of 3, corresponding to
 an order of magnitude less accidental background, see Eq.~(\ref{neqn}). 
 Photon background from positron annihilation would be reduced by
 the same factor. Since the decay $\rdecay$ would then dominate the
 photon background, one could envisage reducing even this background by
 vetoing on the accompanying positron.

 \bigskip
 {\it Target subdivision}

 \smallskip
 Another way to diminish the accidental background could be the use of
 a row of very thin targets. The requirement for positrons and photons to
 originate in the same target would reduce the background in
 proportion to the number of targets. One might think of it as a series
of
 identical $\megs$ set-ups sharing the same beam. However, this
 configuration poses several complications:

 \begin{itemize}
 \item 
 the photon detector must have sufficient directional selectivity
 to distinguish between two adjacent targets;
 \item
 it is not obvious how to combine the scheme with the idea of a
high-resolution
 positron spectrometer.
 \end{itemize} 

 \bigskip
 {\it Muon polarization}

 \smallskip
 The use of the muon polarization for reducing the accidental background
has
 been suggested in the past~\cite{Kuno1,Kuno2}. In muon decay, positrons
and
 photons at energies close to the endpoint follow a $(1 + P_\mu
\cos\theta)$
 angular distribution (where $P_\mu$ is the degree of muon polarization).
 Configurations with back-to-back $e^+$--$\gamma$ pairs are therefore
suppressed
 with respect to the unpolarized case. If one requires a reasonable solid
angle
 subtended by the detectors ($\Omega/4\pi \simeq 0.1$ ) the rejection
factor
 turns out to be approximately 5. It should be noted, however, that in some
 models $\megs$ would be suppressed as well.

\subsubsection{$\mu^+ \to e^+e^+e^-$}
{}From an experimental point of view the decay $\mu\to 3e$ offers some 
important advantages compared to the more familiar $\mu \to e \gamma$
discussed in the previous section. Since the final state contains
only charged particles there is no need for an electromagnetic
calorimeter with its limited performance in terms of energy and
directional resolution, rate capability, and event definition in
general. Apart from the obvious constraints on relative timing and
total energy and momentum, as can be applied in $\mu \to e \gamma$
searches as well, there are powerful additional constraints on vertex
quality and location to suppress the accidental background. Of major
concern, however, are the high rates in the tracking system which has
to stand the load of the full muon decay spectrum.

The present experimental limit, $B(\mu \to 3e) < 1 \times 10^{-12}$
\cite{SNDR88}, was published in 1988. Since no new proposals
exist for this decay mode we shall analyse the prospects of an improved
experiment with this SINDRUM experiment as a point of reference. A
detailed description of the experiment may be found in
Ref.~\cite{SNDR85}.

Data were taken during a period of six months in the years 1984 and 1986 when a
25~MeV/$c$
subsurface beam was brought to rest in a hollow double-cone target at a
rate of $6 \times 10^6\, \mu^+$~s$^{-1}$. The target geometry is
described
in  Table\,\ref{SNDR_target}.
\begin{table}[hbt]
\begin{center}
\caption{\label{SNDR_target} SINDRUM I target for $\mu \to 3e$}
\begin{tabular}{l|rl}
\hline
Shape          &hollow double-cone        \\
Material       &foam                    \\
Length         &220mm                         \\
Diameter       &58mm                             \\
Mass/cm$^2$    &11mg/cm$^2$                      \\
Total mass     &2.4 g                              \\
\hline
\end{tabular}
\end {center}
\end {table}
The spectrometer acceptance for $\mu \to 3e$ was 24\% of 4$\pi$ sr (for
a constant transition-matrix element) so the only place for a
significant improvement in sensitivity would be the beam intensity.
SINDRUM I is a solenoidal spectrometer with a relatively low magnetic
field of 0.33~T corresponding to a transverse-momentum threshold
around 18~MeV/$c$ for particles crossing the tracking system. This
system consists of five cylindrical MWPCs concentric with the beam
axis. Three-dimensional space points are found by measuring the charges
induced on cathode strips oriented $\pm 45^\circ$ relative to the sense
wires. Gating times were typically 50~ns.

\begin {figure}[htb]
\centerline{\includegraphics[height=70mm]{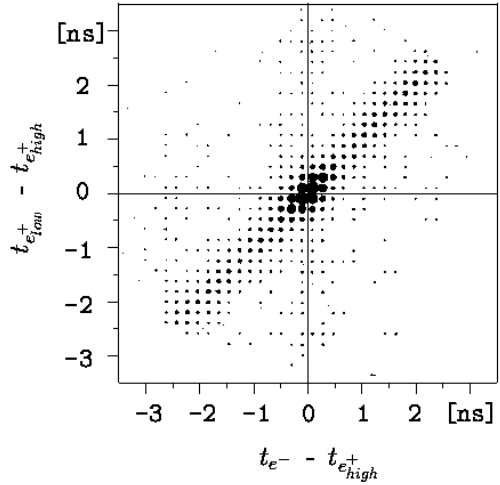}}
\caption{Relative timing of $e^+e^+e^-$
events. The two positrons are labelled according to the invariant mass
when combined with the electron. One notices a contribution of
correlated triples in the centre of the distribution. These events
are mainly $\mu \to 3e \nu \overline{\nu}$ decays. The concentration
of events along the diagonal is due to low-invariant-mass $e^+e^-$
pairs in accidental coincidence with a positron originating in the
decay of a second muon. The $e^+e^-$ pairs are predominantly due to
Bhabha scattering in the target.
\label{SNDR_dt}}
\end {figure}
Figure~\ref{SNDR_dt} shows the time distribution of the recorded
$e^+e^+e^-$ triples. Apart from a prompt contribution of correlated
triples one notices a dominant contribution from accidental
coincidences involving low-invariant-mass $e^+e^-$ pairs. These are explained by
Bhabha scattering of positrons
from normal muon decay $\mu \to e \nu \overline{\nu}$. The accidental
background thus scales with the target mass, but it is not obvious
how to reduce this mass  significantly from the 11~mg/cm$^2$
shown in Table\,\ref{SNDR_target}.

\begin {figure}[htb]
\centerline{\includegraphics[height=60mm]{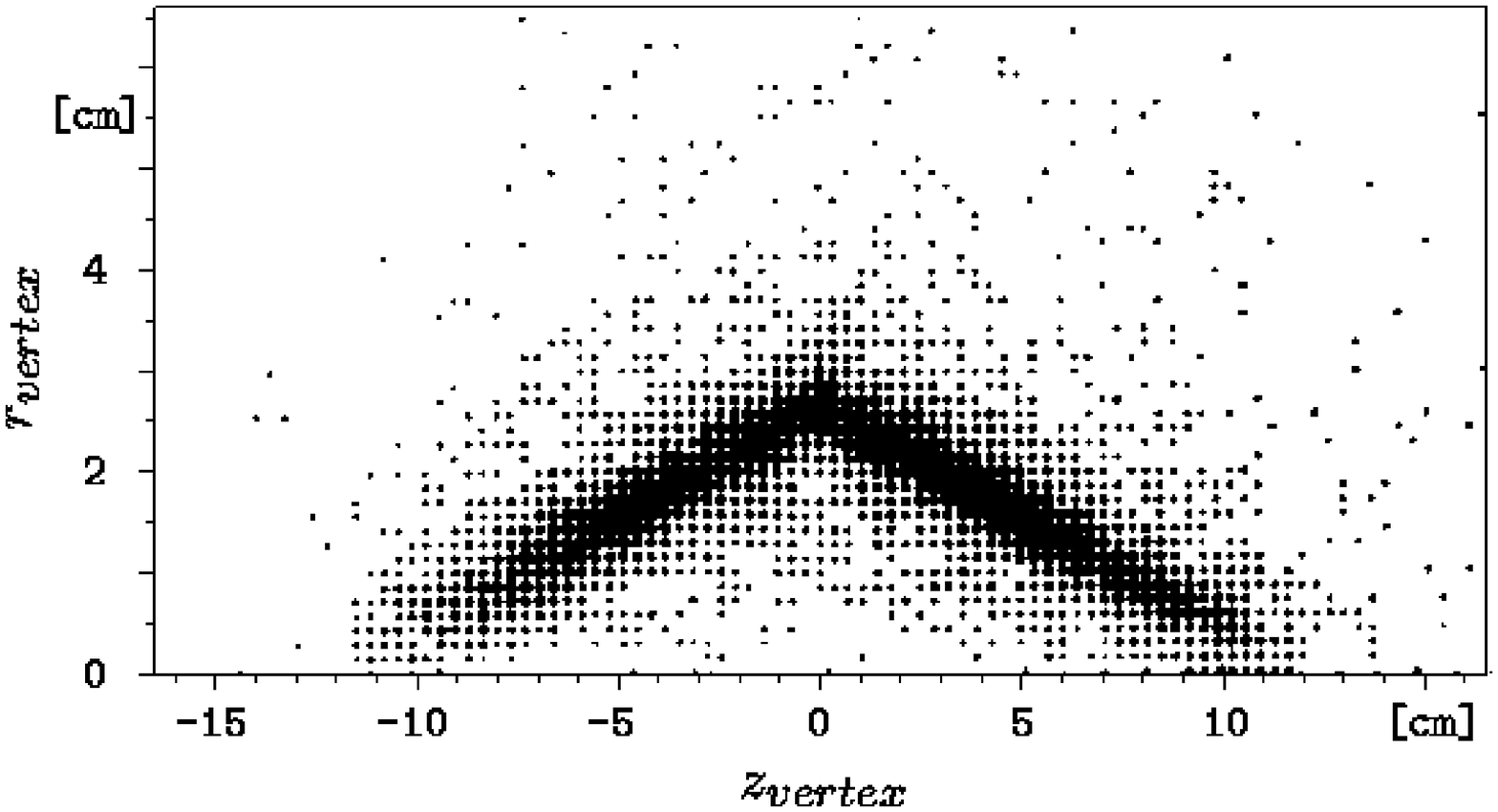}}
\caption{Spatial distribution of the vertex fitted to
prompt $e^+e^+e^-$ triples. One clearly notices the double-cone target.
\label{SNDR_vtx}}
\end {figure}

Figure~\ref{SNDR_vtx} shows the vertex distribution of prompt events.
One should keep in mind that most of the uncorrelated triples contain
$e^+e^-$ pairs coming from the target and their vertex distribution
will thus follow the target contour as well. This 1-fold accidental
background is
suppressed by the ratio of the vertex resolution (couple of mm$^2$ ?)
and the target area. There is no reason, other than the cost of  the
detection system, not to choose a much larger target than that  given in
Table~\ref{SNDR_target}. Such an  increase might also help to reduce the
load on the tracking detectors.

\begin {figure}[htb]
\centerline{
\includegraphics[height=5.5cm]{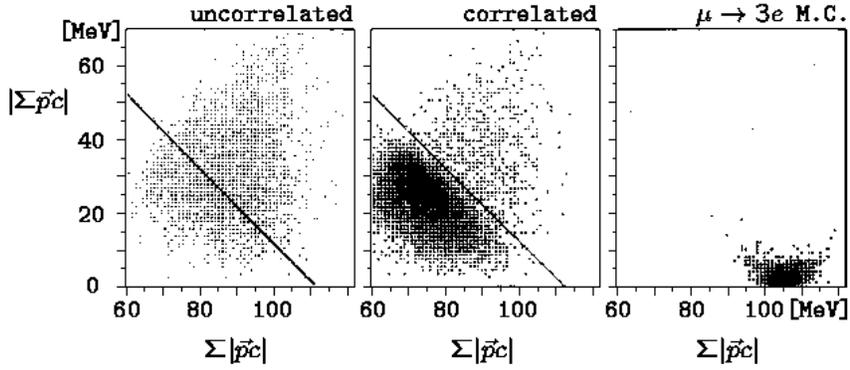}}
\caption{Total momentum versus total energy for three event
classes discussed in the text. The line shows the kinematic limit
(within resolution) defined by
$\Sigma |\vec{p}c| + |\Sigma \vec{p}c| \leq m_{\mu}c^2$ for any
muon decay. The enhancement in the distribution of correlated triples
below this limit is due to the decay $\mu \to 3e \nu \overline{\nu}$.
\label{SNDR_ep}}
\end {figure}

For completeness, Fig.~\ref{SNDR_ep} shows the distribution of total
momentum versus total energy for three classes of events, (i)
uncorrelated $e^+e^+e^-$  triples, (ii) correlated  $e^+e^+e^-$
triples, and (iii) simulated $\mu \to 3e$ decays. The distinction
between uncorrelated and correlated triples has been made on the
basis of relative timing and vertex as discussed above.

 What would  a $\mu \to 3e$ set-up look like that would aim at a single-event
sensitivity around $10^{-16}$, i.e., would make use of a beam rate around
$10^{10}$~$\mu^+$/s? The SINDRUM I measurement was background-free at the level of
$10^{-12}$ with a beam of $0.6  \times 10^{7}$~$\mu^+$/s. Taking into account that
background would have set in at $10^{-13}$, the increased stop rate would raise the
background level to $\approx 10^{-10}$, so six orders of magnitude
in background reduction would have to be achieved. Increasing the
target size and improving the tracking resolution should bring
two orders of magnitude from the vertex requirement alone. Since
the dominant sources of background are accidental coincidences
between two decay positrons (one of which undergoes Bhabha
scattering) the background rate scales with the momentum resolution
squared. Assuming an improvement by one order of magnitude, i.e.,
from the $\approx 10\%$ FWHM obtained by SINDRUM I to $\approx 1\%$
for a new search, one would gain two orders of magnitude from the
constraint on total energy alone. The remaining factor 100 would result
from the test on the collinearity of the $e^+$ and the $e^+e^-$ pair.

As mentioned in Ref.~\cite{SNDR85} a dramatic suppression of background
could be achieved by requiring a minimal opening angle (typically
30$^\circ$) for both $e^+e^-$ combinations. Depending on the mechanism for
$\mu \to 3e$, such a cut might, however, lead to a strong loss in
$\mu \to 3e$ sensitivity as well.

Whereas background levels may be under control, the question remains
whether detector concepts can be developed that work at the high beam
rates proposed. A large modularity will be required to solve problems
of pattern recognition. Also the trigger for data readout may turn
out to be  a great challenge.

\subsubsection{$\mu e$ conversion} 
Neutrinoless $\mu^- \to e^-$ conversion in muonic atoms,
$\mu^-(A,Z) \to e^-(A,Z)$ with $A$ mass number and $Z$ atomic number,
offers some of the best tests of lepton flavour conservation (LFC). For
conversions leaving the nucleus in its ground state the nucleons act
coherently, which boosts the conversion probability relative to the
rate of the dominant process of nuclear muon capture. 

Assuming LFC violation one may wonder how the conversion probability
$B_{\mu e}$ varies as a function of $Z$ and $A$, and with what
probability the nucleus stays in its ground state. Earlier
calculations \cite{Wei59,Kos89,Kos90} predicted a steady rise of the
branching ratio until $Z \approx 30$, from where on it was expected
to drop again. For this reason most experiments were performed on
medium-heavy nuclei. More recently it has been estimated that
$B_{\mu e}$ may keep increasing with $Z$ \cite{Chi93,Kos97}. The
same calculations predict the coherent fraction to be larger than
80\% for all nuclear systems.

The dependence of $B_{\mu e}$ on the normalized neutron excess
$(N-Z)/(N+Z)$, with $N \equiv A-Z$, depends on the nature of the
LFC-violating propagator. Although no mechanism considered so far
gives rise to a cancellation of the neutron and proton contributions
for any given nucleus, a general model-independent analysis requires
at least two measurements with significantly different values for
$(N-Z)/(N+Z)$ \cite{Sha79,Cza}.

When negative muons stop in matter, they quickly get captured and form
muonic atoms, which mostly reach their ground state before decaying.
The main decay channels are muon decay in orbit
$\mu^- (A,Z) \to e^- \overline {\nu}_e \nu_{\mu} (A,Z)$ and nuclear
muon capture $\mu^- (A,Z) \to \nu_{\mu} (A,Z-1)$.
Experimentally coherent $\mu \to e$ conversion offers a number of
advantages. Since the observation of the process requires the
detection of only one particle in the final state, there are no
problems with accidental coincidences, which constitute one of the
major limitations in searches for
$\mu \to e \gamma$ and $\mu \to 3e$. The electron is emitted with a
momentum $p_e \approx m_{\mu}c$, which coincides with the endpoint
of muon decay in orbit (MIO), which is the only intrinsic background.
Since the momentum distribution of muon decay in orbit falls steeply
above $m_{\mu}c/2$ the experimental set-up may have a large
geometrical acceptance and  the detectors can still  be protected
against the vast majority of decay and capture events. Energy
distributions for MIO electrons have been calculated for a number of
muonic atoms \cite{Her80} and energy resolutions in the order of 1\%
are sufficient to keep this background below $10^{-16}$.

There are several other potential sources of electron background in
the energy region around 100 MeV, involving either beam particles or
cosmic rays. Beam-related background may originate from muons, pions
or electrons in the beam. Apart from MIO, muons may produce
background by  muon decay in flight or radiative muon
capture (RMC). Pions may produce background by radiative pion
capture  (RPC). Capture gammas from RMC and RPC produce electrons
mostly through $e^+ e^-$ pair production inside the target.

Beam-related background can be suppressed by various methods:
\begin {itemize}
\item
Beam pulsing\\
Since muonic atoms have lifetimes of order 100~ns, a pulsed beam with
buckets short compared to this lifetime would allow one  to remove prompt
background by measuring in a delayed time window. As will be
discussed below there are stringent requirements on the beam
suppression during the measuring interval. This is the concept of the
new MECO proposal.
\item
Beam veto\\
In the past, prompt background has been removed with the help of a beam
counter. For the beam intensities proposed here this method can not be
applied. 
\item
Beam purity\\
A low-momentum ($< 70$ MeV/c) $\mu^-$ beam with extremely low pion
contamination ($< 100$ $\pi^-$'s per day) would keep prompt
background at a negligible level. A major advantage of the method is
that heavy targets such as gold with lifetimes around 70~ns can be
studied. This scheme was applied by  SINDRUM II  and will be
discussed in the following.
\end {itemize}

At present, $\mu e$ conversion is being searched for by the {SINDRUM II}
Collaboration at PSI \cite{sndr_proposal} and a new experiment (MECO
\cite{meco_proposal}) is planned at BNL. The sensitivity levels are
$\approx 5 \times 10^{-13}$ and $\approx 5 \times 10^{-17}$,
respectively.

\begin {figure}[htb]
\begin {center}
\includegraphics[width=10cm]{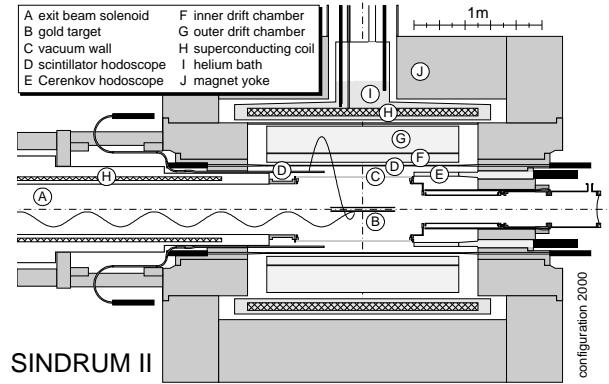}
\end {center}
\caption{SINDRUM II
\label{sndr_setup}}
\end {figure}  
Figure \ref{sndr_setup} shows the {SINDRUM II} spectrometer. Beam
particles are extracted from a carbon production target at
backward angles and momentum-analysed in a conventional quadrupole
channel. The beam is focused into a 9~m long transport solenoid (see
Fig.~\ref{sndr_setup}) called  PMC (pion muon converter).

A pion reaching the gold target has a chance of order 10$^{-5}$ to
produce
an electron in the energy region of interest, so the pion stop rate must
be
below one every ten minutes. At the PMC  entrance the beam contains
similar
amounts of muons and pions. Since the pion range in matter is about half
as large
as the corresponding muon range, the pion contamination can be reduced
strongly
with the help of a moderator at the PMC entrance. Only one out of
$10^6$
pions may cross this moderator. Typically 99.9\% of them would decay
before
reaching the target. The requirement puts strong constraints on the
high-momentum
tail transported by the beam line which could be met after a careful
optimisation
of the beam settings.

 \begin {figure}[htb]
 \parbox{0.54\textwidth}{
 \includegraphics[width=0.5\textwidth]{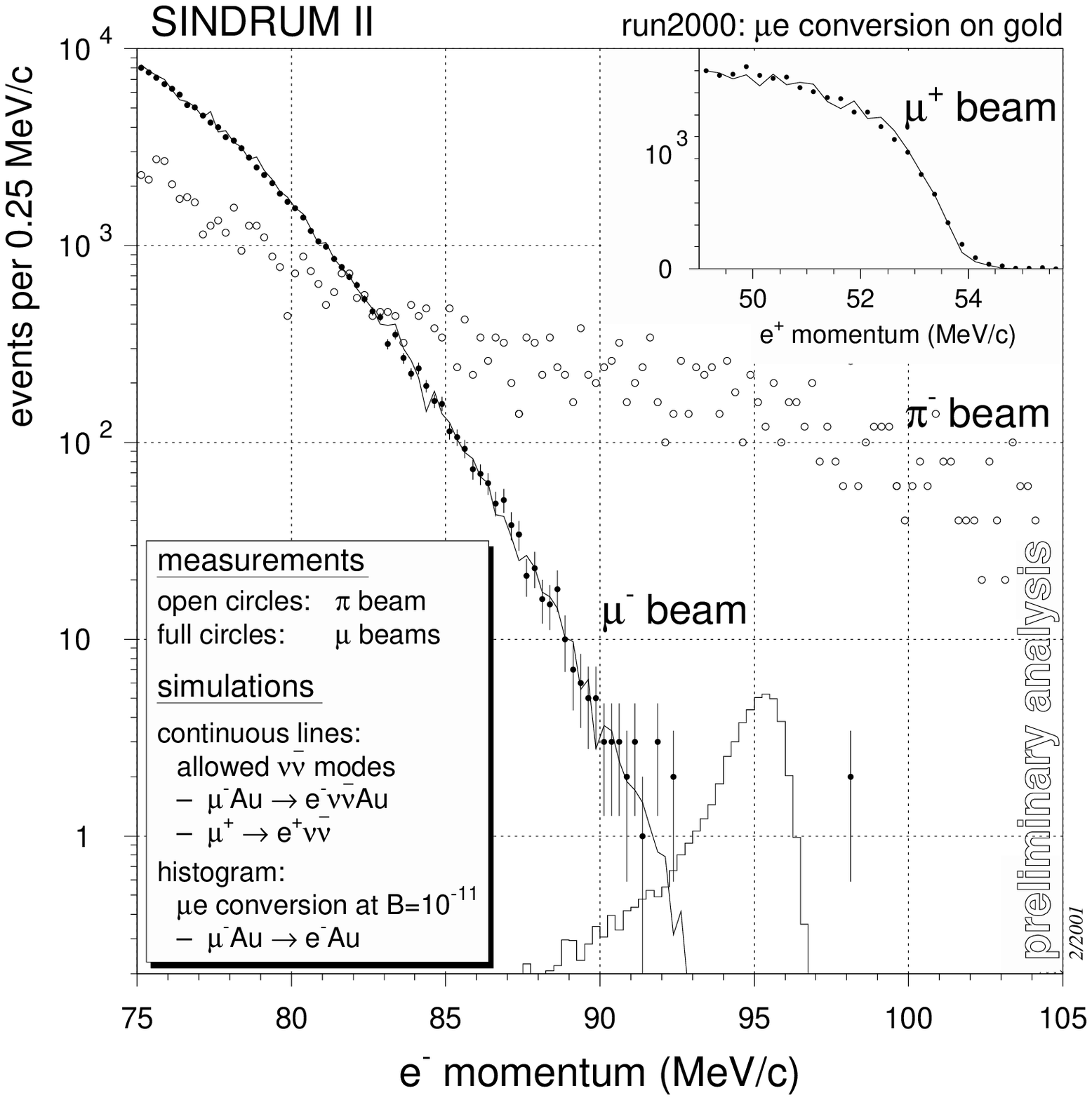}}
 \parbox{0.46\textwidth}{\vspace*{3cm}
 \caption{ 
 \label{sndr_gold2000}
 Recent results by SINDRUM II.
 Momentum distributions for three different beam momenta
 and polarities:
 (i) 53 MeV/$c$ negative, optimised for $\mu^-$ stops, (ii) 63~MeV/$c$
negative,
 optimised for $\pi^-$ stops, and (iii) 48 MeV/$c$ positive, for $\mu^+$
stops.
 The 63~MeV/$c$ data were scaled to the different measuring times. The
$\mu^+$
 data were taken at reduced spectrometer field.}}
 \end {figure}

 During an effective measuring period of 75 days about $4\times 10^{13}$
 muons stopped in the gold target.
 Figure~\ref{sndr_gold2000} shows as a preliminary result various
momentum
 distributions. The main spectrum, taken at 53~MeV/$c$, shows the steeply
falling
 distribution expected from muon decay in orbit. Two events were found at
 higher momenta, but just outside the region of interest. The agreement
between
 measured and simulated positron distributions from $\mu^+$ decay gives
 confidence in the momentum calibration. At present there are no hints
about the
 nature of the two high-momentum events: they might be induced by cosmic
rays
 or RPC, for example. Both processes result in flat momentum
distributions such
 as shown by the data taken at 63 MeV/$c$.

 As a preliminary result the single-event sensitivity was estimated to be 
 $\approx 2 \times 10^{-13}$, i.e.,  an improvement by two orders on
magnitude of the
 previous best result on a heavy target \cite{winfried}.

\begin {figure}[htb]
\centerline{\includegraphics[height=6cm]{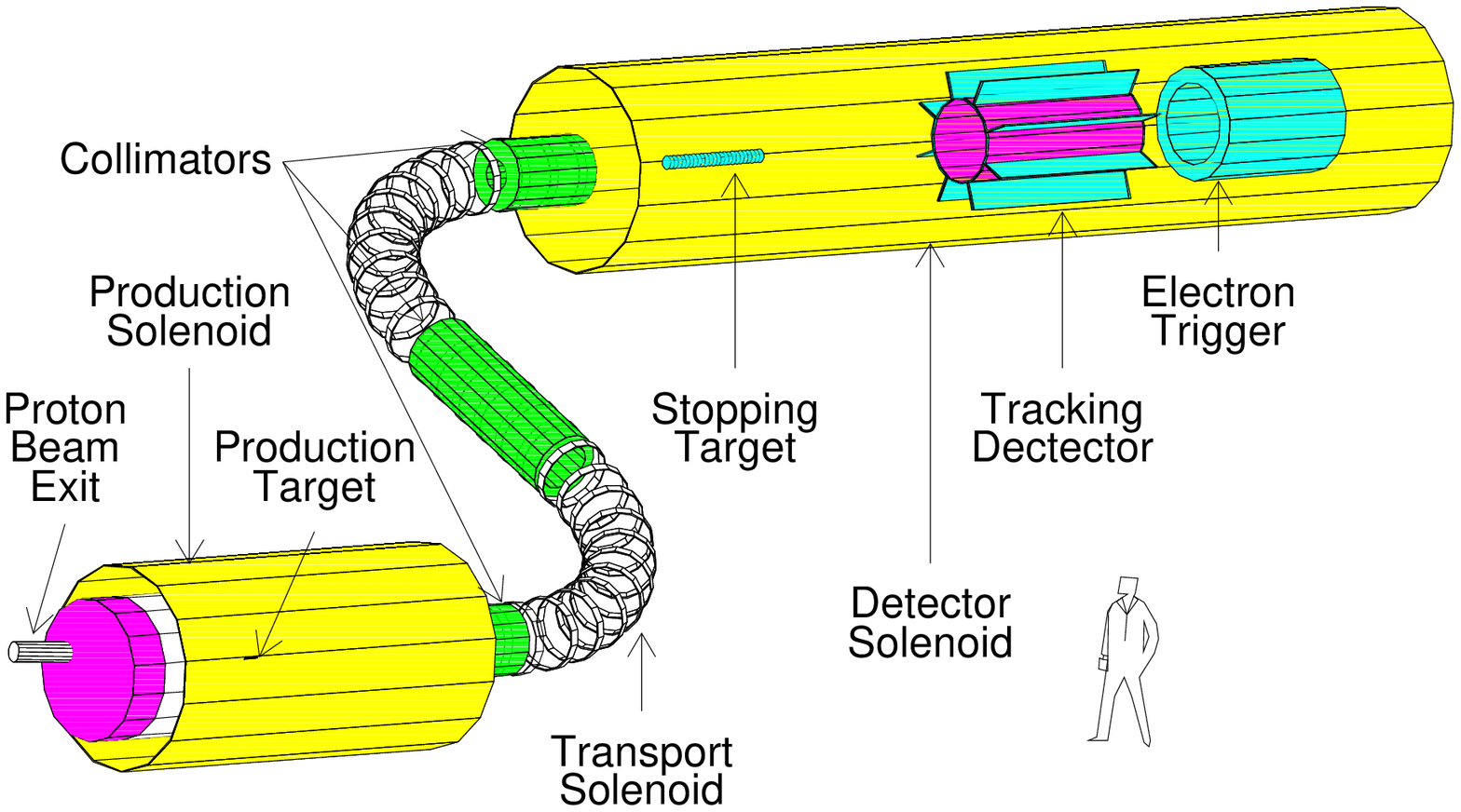}}
\caption{MECO
\label{meco_setup}}
\end {figure}

\begin {figure}[htb]
\centerline{\includegraphics[width=9.5cm]{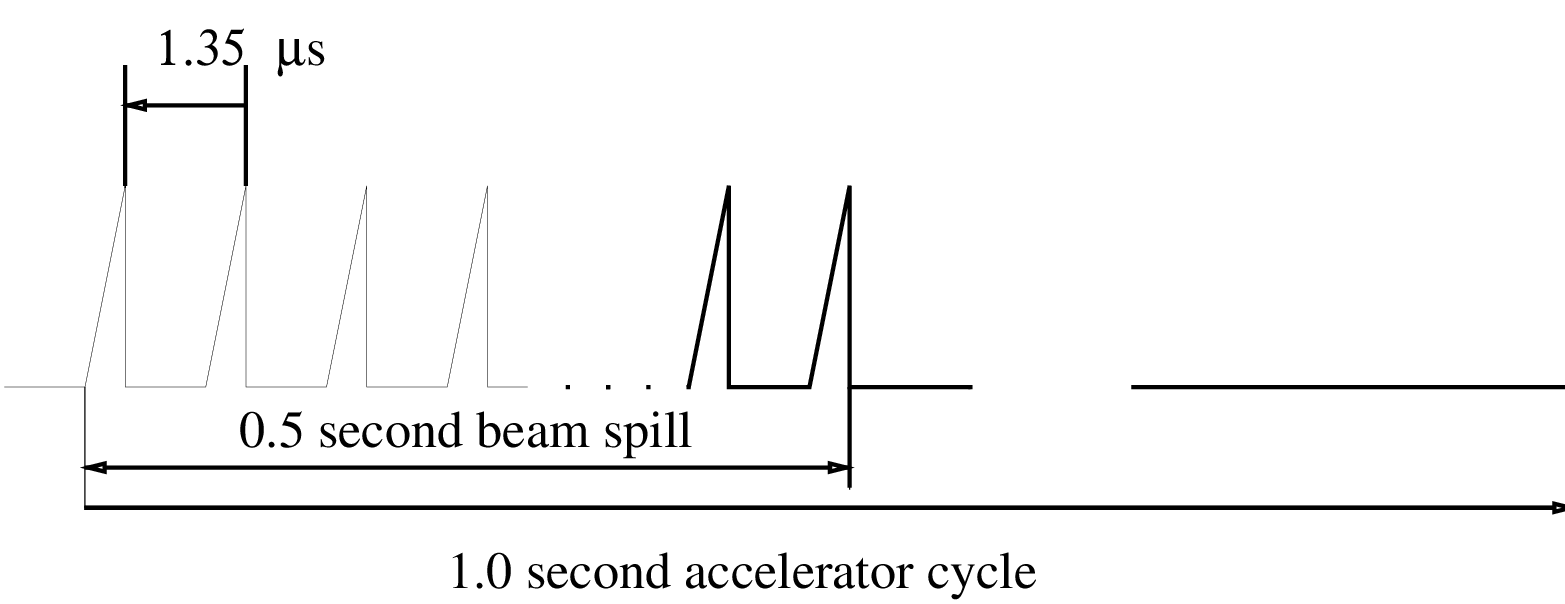}}
\caption{Proposed beam spill
for MECO \label{meco_spill}}
\end {figure}

The MECO experiment plans to combat beam-related background with the
help of a pulsed 8 GeV/$c$ proton beam. Figure~\ref{meco_setup} shows
the proposed layout.
Pions are produced by 8~GeV/$c$ protons crossing a 16~cm long tungsten
target, and muons from their decays are collected efficiently with
the help of a graded magnetic field. Negatively charged particles
with 60--120 MeV/$c$ momenta are transported by a curved solenoid to
the experimental target. In the spectrometer magnet a graded field
is applied as well. Figure~\ref{meco_spill} shows the proposed time
structure of the proton beam.
A major challenge is the requirement for proton suppression in
between the proton bursts. In order to keep the pion stop rate in the
`silent' interval below the aforementioned 100 per day, a beam
extinction factor better than $10^{-8}$--$10^{-9}$ is required.

Table~\ref{mue_table} compares the $\mu ^-$ intensity and single-event
sensitivity obtained by  SINDRUM II  with those expected for
MECO and the $\nu$ factory.
\begin{table}[hbt]
\caption{
\label{mue_table}
Rates in $\mu e$  conversion searches}
\begin{center}
\begin {tabular}{l|c|c|c}
\hline
                         &SINDRUM II         &MECO
&NUFACT      \\
                         &PSI                &BNL                   &CERN
\\
\hline
Status                   &Stopped            &Proposed
&Discussed          \\
Proton momentum [GeV/$c$]  &1.2                &8                     &2
\\
Proton rate [s$^{-1}$]   &$1 \times 10^{16}$ &$4 \times 10^{13}$    &$1
\times 10^{16}$ \\
$\mu ^-$ rate [s$^{-1}$] &$3 \times 10^7$    &$1.7 \times 10^{11}$
&$10^{14}$          \\
$\mu^-$ stops [s$^{-1}$] &$(1\ldots 2) \times 10^7$  &$1.0 \times
10^{11}$  &$2.0 \times 10^{13}$\\
Single event sensitivity &$2\times10^{-13}$         &$2.0 \times
10^{-17}$ &$10^{-19}$         \\
\hline
\end   {tabular}
\end{center}
\end{table}

Two scenarios for a $\mu^-$ beam are under study (see also
Section~\ref{prmb} below):
\begin{itemize}
\item
A {\it bunched proton beam} extracted from the accumulator could be used
in a fashion similar
to the MECO beam line. As explained above, this option requires extinction
factors of ten
orders of magnitude or so to keep prompt background under control.
\item
As an alternative one might use {\it a cooled muon beam}. Such a beam would
probably be free
of pions and  contaminated only by decay electrons. As a second
by-product of the cooling the extinction factor should be much higher at
this stage.
\end {itemize}

\subsubsection{Muonium-antimuonium conversion}
Muonium is the atomic bound state of a positive muon and an electron.
For leptons a spontaneous conversion of muonium ($\mu^+e^-$) into
antimuonium 
($\mu^-e^+$)  would be completely analogous 
to the well known
${\rm K}^0-\overline{{\rm K}^0}$ oscillations in the
quark sector.  A search  already  suggested   in 
1957 by Pontecorvo \cite{Pontecorvo} three years before the atom was
discovered 
by Hughes \cite{Hughes_60}.
The process could proceed
at tree level through bilepton exchange or through various 
loops. Predictions for the process exist in a variety of
speculative models including left--right symmetry, 
$R$-parity-violating supersymmetry, GUT theories and several others
\cite{mmb_theories}. 

\begin {figure}[htb]
\begin {center}
\includegraphics[width=10cm]{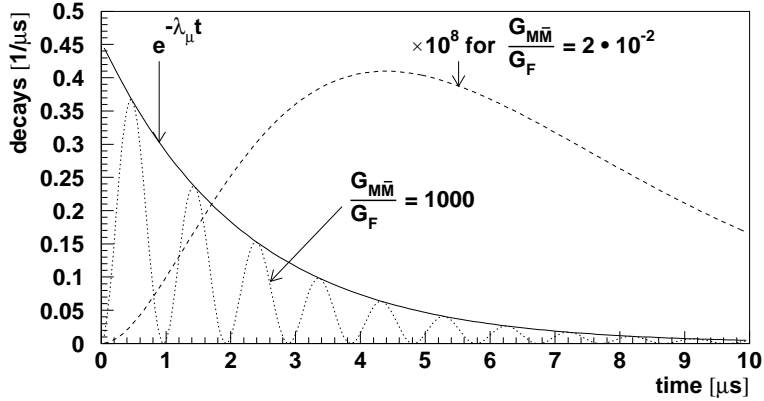}
\end {center}
\vspace*{-0.5cm}
\caption{Time dependence of the probability to observe an antimuonium decay  
for a system which was initially in a pure muonium state.  
The solid line represents the exponential decay of muonium  
in the absence of a finite coupling. 
The decay probability as antimuonium is given for a coupling strength of 
$G_{\rm M\overline{M}}   = 1000 G_F$ 
by the dotted line and for a coupling strength 
small compared to the muon decay rate (dashed line). 
In the latter case the maximum of the probability 
is always at about 2 muon lifetimes.  
Only for strong coupling   could 
several oscillation periods   be observed.
\label{mmbar_oscill}}
\end {figure}

Any possible coupling between muonium and its antiatom  will give rise to
oscillations between them. For atomic s-states with 
principal quantum number $n$
a splitting of their
energy levels
\begin{eqnarray} 
 \delta =
 {\displaystyle \frac{8 G_{\rm F}}{\sqrt{2} n^2 \pi a_0^3}
 \frac{G_{\rm M\overline{M}}}{G_{\rm F}} }
\end{eqnarray}
is caused,
where $a_0$ is the Bohr radius of the atom, $G_{\rm M\overline{M}}$
is the coupling constant in an effective four-fermion interaction  
and ${G}_{\rm F}$ is the weak interaction Fermi 
coupling constant.
For the ground state we have $\delta =   
1.5 \times 10^{-12} \, {\rm eV} \times (G_{\rm M\overline{M}}/G_{\rm F})$
which corresponds to 519 Hz for $G_{\rm M\overline{M}}=G_{\rm F}$.
An atomic system created at time $t = 0$ 
as a pure state of muonium can be expected to be observed 
in the antimuonium state at a later time $t$ with a time dependent 
probability of 
\begin{eqnarray} 
\label{t_oscill} 
{ p_{\rm M\overline{M}}(t)} &= \sin^2\left(\frac{\delta \,
t}{2\,\hbar}\right) 
e^{-\lambda _\mu t}  
&\approx \left(\frac{\delta \, t}{2\,\hbar}\right)^2  e^{-\lambda
_\mu t}\;, 
\end{eqnarray} 
where $ \lambda _\mu = 1/\tau_\mu $ is the muon decay rate
(see Fig. \ref{mmbar_oscill}). 
The approximation is valid for a weak coupling as suggested by the  
known experimental 
limits on $G_{\rm M\overline{M}}$.

\begin {figure}[htb]
\begin {center}
\includegraphics[width=10cm]{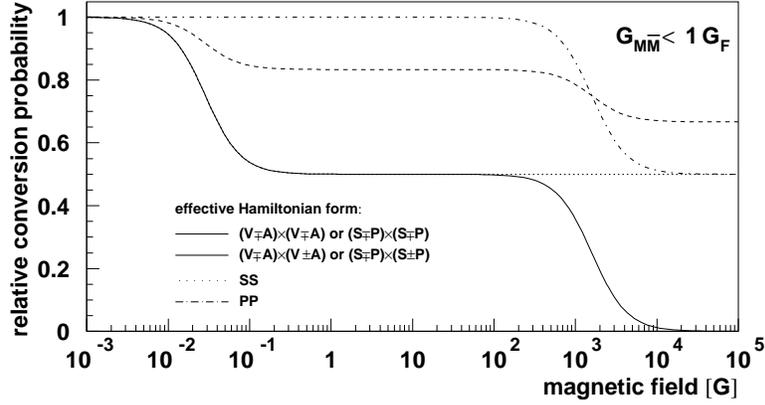}
\end {center}
\vspace*{-0.5cm}
\caption{The muonium to antimuonium conversion probability 
          depends on external magnetic fields 
          and the coupling type. Recent independent 
          calculations were performed  by Wong and Hou 
          {\cite{wong95}} and Horikawa and 
          Sasaki {\cite{hori95}}. 
\label{mbar_in_fields}}
\end{figure}

The degeneracy of corresponding states in the atom and its anti-atom 
is removed by external magnetic fields which can cause a suppression  
of the conversion and a reduction of the probability $p_{\rm
M\overline{M}}$. 
The influence of an external magnetic field depends on the interaction
type of 
the process. The reduction of the conversion 
probability has been calculated for all possible 
interaction types as a function of field strength  
(Fig. \ref{mbar_in_fields}) \cite{wong95,hori95}.  
In the case of an observation of the conversion process,   
the coupling type could be revealed by  measurements of 
the conversion probability at two different  
magnetic field values. 
 
The conversion process is strongly suppressed for muonium 
in contact with matter
since a transfer of the negative  
muon in antimuonium to any other atom is energetically 
favoured and breaks 
up the symmetry between muonium and antimuonium by opening up  
an additional decay channel for the anti-atom only \cite{morg_70} \footnote{In gases
at atmospheric pressures the conversion probability  is approximately
five orders of magnitude smaller than in vacuum 
mainly  due to scattering of the atoms from gas molecules.  
In solids the reduction amounts to even 10 orders of magnitude.}. 
Therefore any new sensitive experiment needs to employ  
muonium atoms in vacuum \cite{Willmann_99}. 

A recent  experiment at PSI utilised 
a powerful signature in which the identification of both
constituents of the anti-tom  and their coincident detection was
requested
after its decay.
In this an energetic electron appears in the $\mu^-$ decay. The 
positron from the atomic shell remains with an 
average kinetic energy of 13.5~eV \cite{Marciano_00}. 
The energetic particle could be observed in a 
magnetic wire chamber spectrometer and a position-sensitive microchannel
plate
(MCP) served as a detector for atomic shell positrons onto which these
particles could be transported in a magnetic guiding field after
post-acceleration in an electrostatic device. A clean vertex
reconstruction and
the observation of annihilation $\gamma$'s in a pure CsI detector
surrounding
the MCP were required in an event signature \cite{Willmann_99}.
Half a year of actual data-taking was carried out  at the currently most
intense
surface muon source, the $\pi$E5 channel at PSI. The previous upper 
bound on the total conversion probability 
per muonium atom 
${  P}_{{\rm M\overline{M}}}= \int  p_{\rm M\overline{M}}(t) {\rm d}t$ 
was improved by more
than three  orders of magnitude and yielded an upper bound of 
${P}_{{\rm M} \overline{{\rm M}}}\leq 8.0 \times 10^{-11}/{ 
S}_{\rm
B}$. Here a magnetic field correction ${S}_{\rm B}$ is
included which accounts for the 0.1 T magnetic
field in the experimental conditions. It is of order unity and 
depends on the type of the ${\rm M\overline{M}}$ interaction. 
For an assumed effective (V--A)$\times $(V--A)-type four-fermion interaction   the
quoted result   corresponds to an upper limit   for the coupling
constant of    
${G}_{{\rm M} \overline{{\rm M}}}\leq 3.0\times 10^{-3} {
G}_{\rm F}$  (90~\%~C.L.).
Several limits on model parameters 
were significantly improved like the mass of the  bileptonic gauge boson 
and some models  were strongly disfavoured
such as a certain $Z_8$ model with radiative mass generation 
and the minimal version of 331 models \cite{Willmann_99}.

\begin{figure}[bt]  
  \begin{minipage}{1.5in}  
  \centering{  
   \hspace*{-1.6cm}  
   \mbox{  
   \includegraphics[width=2in]{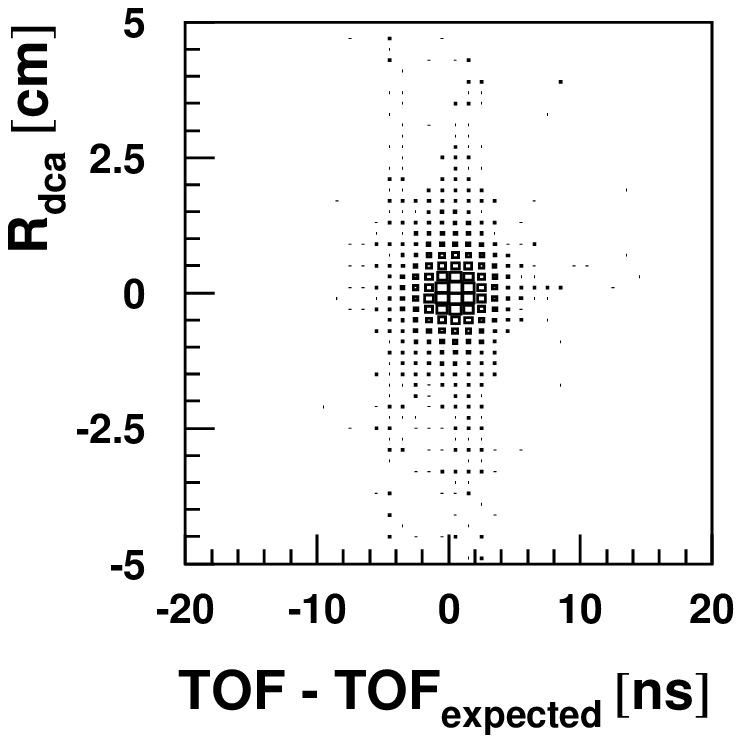}
         }  
             }  
  \end{minipage}     
 \hspace*{0.1cm}  
 \begin{minipage}{1.5in}  
    \centering{  
   \hspace*{.3cm}  
   \mbox{  
   \includegraphics[width=2in]{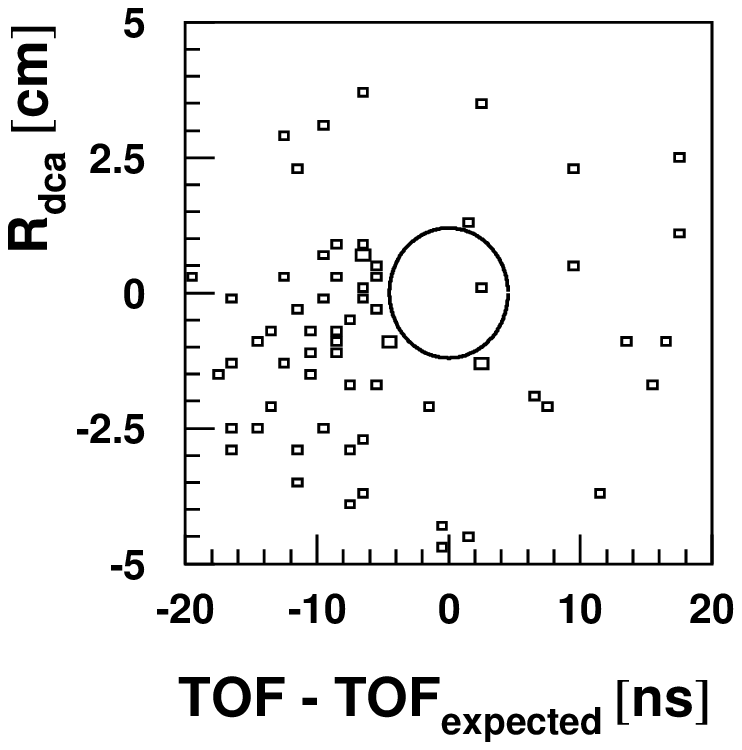}  
         }  
             }  
   \end{minipage}  
 \centering\caption 
        {The distribution of the distance of closest approach ($R_{dca}$)
        between a track from an energetic   
        particle in the magnetic spectrometer and the back projection of
the  
        position on the MCP detector versus the time of flight (TOF)  
        of the atomic shell particle for a muonium measurement (left)  
        and for all data recorded within 1290 h of   
        data taking while searching for   
        antimuonium (right).   
        One single event falls within 3 standard deviations region  
        of the expected TOF and $R_{dca}$ which is indicated by the
ellipse.  
         The events concentrated at early times   
       and low $R_{dca}$correspond   
        to a background signal from the allowed decay  
        $\mu \to 3e2\nu$. In a new experiment such background
could be
        suppressed significantly through the characteristically different
        time evolution of a potential antimuonium signal and the
background.
          }\label{Fmmbar_res}  
   
\end{figure}  
  
With a new and  intense pulsed beam
the characteristic time dependence  of the conversion process
could be exploited, if only the decay of  atoms  that  have survived several muon
lifetimes 
$\tau_{\mu}$ could be observed. Whereas all beam-muon-related background decays 
exponentially, the anti-atom population increases quadratically in time
giving the signal an advantage over background,
which for a 3-fold coincidence signature as in the 
PSI experiment can be expected to decay with a time
constant of $\tau_{\mu}/3$ [compare Eq.~(\ref{t_oscill})].

Some two orders of magnitude improvement can be envisaged 
\cite{Jungmann_00a} with still no principal background arising from
the $\mu \to 3e2\nu$ process or internal Bhabha scattering 
in which the positron from $\mu^+$ decay would transfer its
energy to the electron in the atomic shell and mimic an
event which is searched for (Fig.~\ref{Fmmbar_res}).

The requirements forradiation hardness and rate capability of the
set-up are similar to those of a $\mu \to 3e$ experiment. As before, a common
approach to these two measurements may be found. 

\subsubsection{Conclusions}
All experiments discussed above would benefit from the highest-possible
stop-densities, allowing thin targets with minimal distortion of the
outgoing particles. In the case of $\mu^+$, most promising are subsurface
$\mu$ beams that can be stopped in targets of less than a few mg/cm$^2$.

Better experiments need better detection systems as well. In particular
a search for $\megs$ is limited by detector technology and it
seems unlikely that the sensitivity will rise in proportion to the
beam intensity. Searches for $\mu \to 3e$ and
$\mu^+ e^- \leftrightarrow \mu^- e^+$ may have similar requirements on
the set-up. High granularity will be required for the tracking system
which has to stand the load of the full muon decay spectrum.

From the four tests of lepton flavour conservation discussed here, 
$\mu $-$e$ conversion is expected to exploit best
the increased beam intensity:
\begin {itemize}
\item
Intrinsic background can be limited to the decay in orbit which can be
kept under control by pushing the electron momentum resolution below
$\approx 0.5$~MeV/$c$. Care has to be taken to avoid high-momentum
components in the momentum response function. Ultimately the
resolution would be given by the uncertainty in the energy loss
inside the experimental target. 
\item
Detector rates can be kept under control thanks to the steeply-falling
transverse momentum distribution.
\item
Prompt beam-correlated background, in particular induced by pions
contaminating the beam, can be removed using a pulsed proton beam.
Alternatively, a cooled muon beam might have a sufficiently low
pion contamination such that only electrons would have to be suppressed.
\end {itemize}

In conclusion, all searches would benefit from
the projected increase in muon stop rate. 
In some cases the gain in
sensitivity over ongoing experiments 
might, however, be not more than one order of magnitude.

\section{FUNDAMENTAL MUON PROPERTIES}

Besides measurements on normal muon decay and the 
`classical' searches for the rare decays $\mu \to e \gamma$,
$\mu \to eee$, $\mu^-N \to e^-N$ conversion, and
muonium-antimuonium conversion (see Section~\ref{LNV})
which are strongly motivated by our present thinking towards extensions
of the Standard Model, there are a number of very interesting experiments
on free muons and atomic systems containing muons:
\begin{itemize}
\item 
precise determinations of the properties characterising the muon
(e.g. the  mass $m_{\mu}$, magnetic moment $\mu_{\mu}$, 
magnetic anomaly $a_{\mu}$, charge $q_{\mu}$ and lifetime $\tau_{\mu}$,
see Section~\ref{subsec:2.1.3}), 
\item
CPT tests from a comparison of $\mu^-$ and $\mu^+$ properties, 
\item 
measurements of fundamental constants of nature (e.g. the electromagnetic
fine structure constant $\alpha$ or the weak interaction Fermi constant
$G_{\rm F}$), 
\item
sensitive searches for deviations of muon parameters from Standard Model
predictions,
including novel approaches to search for a permanent  muon
electric dipole moment $edm_{\mu}$,
\item
study of nuclear properties in muonic (radioactive) atoms, and
\item
applications in condensed matter and life sciences.
\end{itemize}

In the past most of these topics have been addressed very successfully 
and in many cases unique information has been extracted. With more
intense
muon sources they can be accessed with novel and much more precise
techniques and substantial progress may be achieved.

\subsection{Anomalous magnetic moment of the muon }

\begin{figure}[htb]
\centerline{
\includegraphics[width=0.65\textwidth]{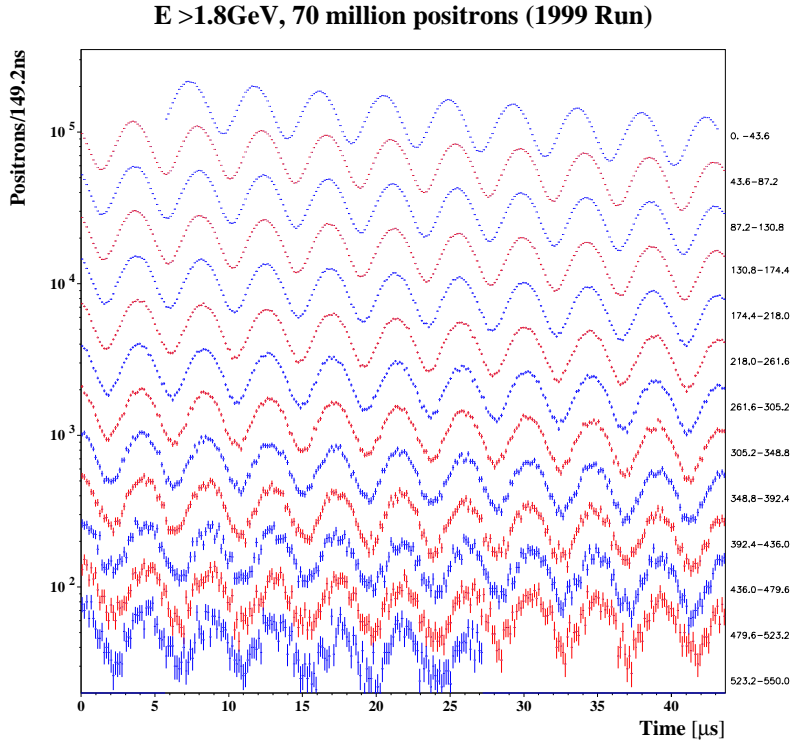}}
\caption{A sample of new muon $g-2$ data. There is no sign of
background after 10 time-dilated
muon lifetimes of 64~$\mu$s. \label{g2wiggles}}
\end {figure}

\begin{figure}[htb]
\centerline{
\includegraphics[width=8cm]{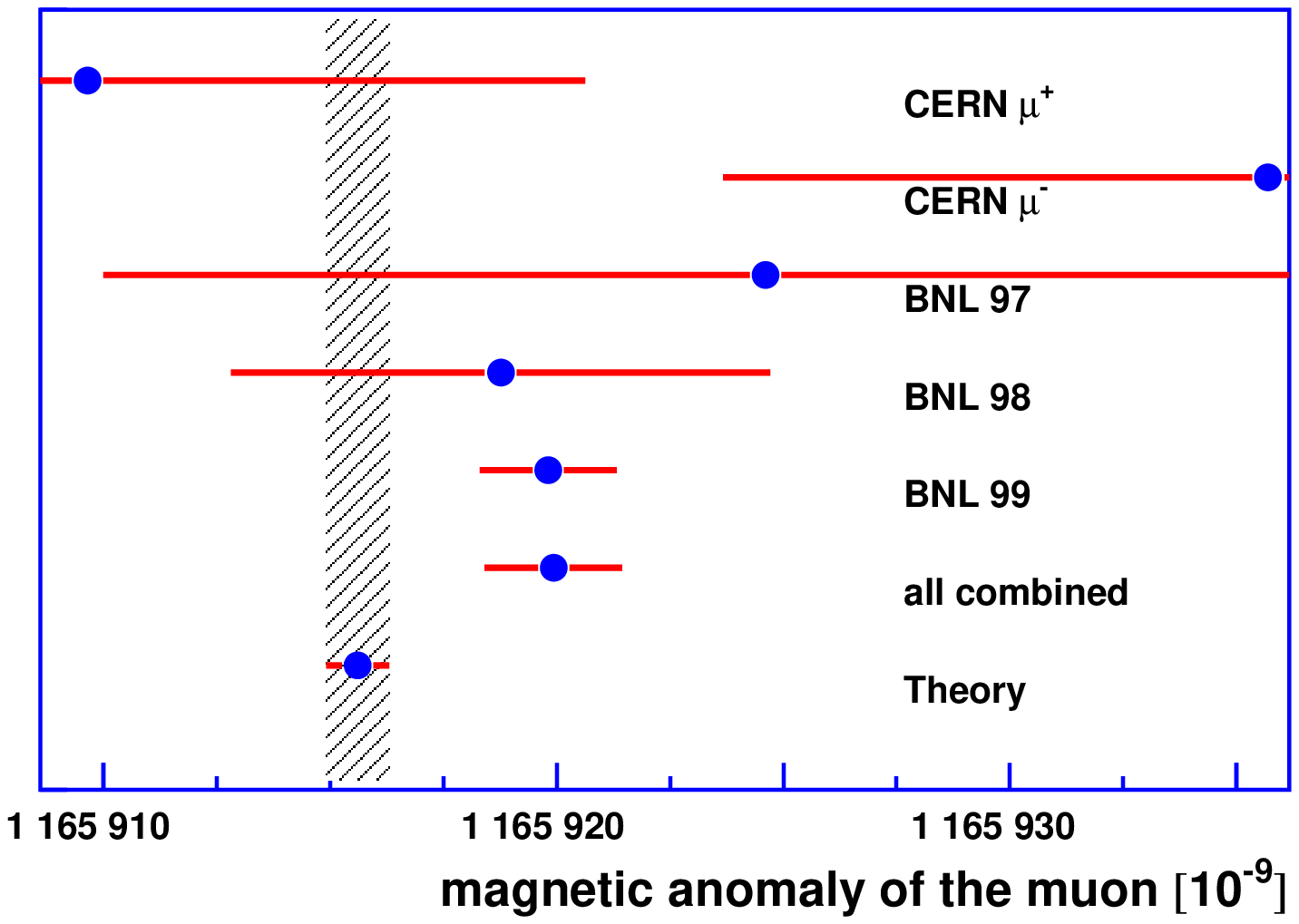}}
\caption{Comparison of recent measurements of the muon magnetic
anomaly with the most accurate and most recent theoretical value \cite{Davier_98}. 
The difference is 2.6 times the combined theoretical and experimental
uncertainty. 
\label{g2result}}
\end {figure} 

The anomalous magnetic moment of the muon, $a_{\mu} \equiv g_\mu - 2$,
has
been measured in three experiments at CERN, and is presently being
measured at BNL~\cite{Farley_90} (Fig.~\ref{g2wiggles}). A superferric precision
magnetic field storage ring is employed in which the difference between the spin
precession and the cyclotron frequencies is determined.  Currently
$a_{\mu}$ is experimentally known to 1.3~ppm: 
$a_{\mu}^{exp}=11\,659\,202(14)(6)\times 10^{-10}$ \cite{Brown_01}. 

The most recent and most accurate theory value in the framework of the
Standard Model $a_{\mu}=11\,659\,159.6(6.7)\times10^{-10}$
(0.57~ppm) appears to differ from the experimental result by $\Delta
a_{\mu} = 43(16)\times 10^{-10} $ which is about 2.6 times the combined
experimental and theoretical uncertainties.  
Although this discrepancy may be just caused by an underestimate of the 
uncertainties in the hadronic contributions to the SM value of $a_{\mu}$,
it is also possible that we are facing
hints of physics beyond the Standard Model.  At this stage,
definitive statements cannot yet be made.

Data already recorded at BNL are expected to provide a statistical error
for
positive muons which is about two times smaller. Data taking with similar
accuracy for negative muons is in progress in 2001, with a potential
continuation in 2002.  Upon completion, this experiment will either
provide more confirmed hints of New Physics beyond the Standard Model, or
at least limit parameters appearing in speculative models. The large
number of models to which $g_\mu - 2$ has some sensitivity includes
supersymmetry, lepton compositeness, radiative muon mass models,
anomalous
$W$ boson parameters, new gauge bosons, leptoquark models, exotic
fermions
and many more \cite{Czarnecki_00_a}. 

As examples of where speculations based on the present difference $\delta
a_{\mu}$ may lead, we note that for the constrained minimal
supersymmetric
extension of the Standard Model (CMSSM) all the preferred parameter space
is within reach of the LHC, but may not be accessible to the Tevatron
collider, nor to a first-generation $e^+$ $e^-$ linear collider with
centre-of-mass energy below 1.2 TeV.  In many of these models there is a
direct connection between a non-standard anomalous muon magnetic moment
and other processes discussed in this article.  For example, in models
with muon substructure, a contribution to $a_{\mu}$ from muon
compositeness would suggest that the process $\mu \to e \gamma$
could have a branching ratio above $10^{-13}$ (see Section~\ref{LNV}). 

From the experimental point of view, the measurements are predominantly
statistics-limited. The systematic uncertainties could still be
significantly reduced, allowing about one order of magnitude higher
accuracy before fundamentally new experimental approaches would be
needed. 
At this point, the present limitations on magnet and field measurement
technology as well as
particle pile-up would start to impose significant systematic influences
on the signal.  In order to reduce detector pile-up, a 3.1 GeV short
pulsed ($<$100 ns) beam of a few $\times 10$ Hz repetition rate would be
most
advantageous for a future experiment.  A considerable improvement in
accuracy could still be envisaged from a new experimental set-up using a
similar approach to the current BNL experiment with a super-ferric storage
ring, where the main advantage arises from exploiting higher particle
fluxes.

There is, however, a severe limitation to the interpretation of a $g_\mu - 2$
measurement, when seeking to extract or limit parameters of speculative
models, namely that due to the hadronic vacuum polarization. Its
influence
restricts the accuracy of calculations within the Standard Model. Until
recently, the hadronic contribution was obtained using a dispersion
relation and data from electron-positron annihilation into hadrons, as
well as QCD theory. The accuracy of earlier
work~\cite{Yndurain_01} was improved significantly by newer approaches
using hadronic $\tau$ decay data~\cite{Davier_98}. It appears that all
newer evaluations agree very well with the newest and most accurate
determination and yield an accuracy of better than
0.6~ppm~\cite{Narison_01,Marciano_Roberts_01}. Newer data for $e^+ e^-$
annihilation into hadrons are now available from experiments at
Novosibirsk
and Beijing and needs to be included. Further improvements can be
expected
from the DAPHNE accelerator at Frascati, and data on $\tau$ decays from
$B$ factories may also contribute. However, even if the accuracy of
the theoretical value for hadronic vacuum polarisation could be further
improved using even higher-quality experimental input, there would still
remain an uncertainty arising from the treatment of hadronic
light-by-light scattering. 

A new and independent $g_\mu - 2$ experiment would undoubtedly be highly
desirable if the experiment--theory discrepancy at BNL survives after the
experiment is completed.

\subsection{Permanent muon electric dipole moment}

The identification of new sources of CP violation
appears to be a crucial requirement for explaining the dominance 
of matter over antimatter in the Universe. 
Permanent electric dipole moments of fundamental particles
would violate time reversal (T) invariance and with the assumption of
CPT conservation also the CP symmetry.

In the muon $g  - 2$ experiment 
a hypothetical electric dipole moment of the muon $edm_{\mu}$
would cause the plane of 
its spin precession in the storage ring  to be
tilted against the plane of its orbit.
This inclination can be measured from an asymmetry of the decay electron
distribution with respect to the orbital plane. 
With this method in the latest CERN muon $g -2$
experiment a limit for $edm_{\mu}$ was established 
at $1.05 \times10^{-18}$ e\,cm 
\cite{Farley_90}.
The ongoing BNL measurements will provide one order of magnitude 
improvement.

 \begin {figure}[htb]
 \hspace*{2cm}
 \includegraphics[width=2in,angle=0]{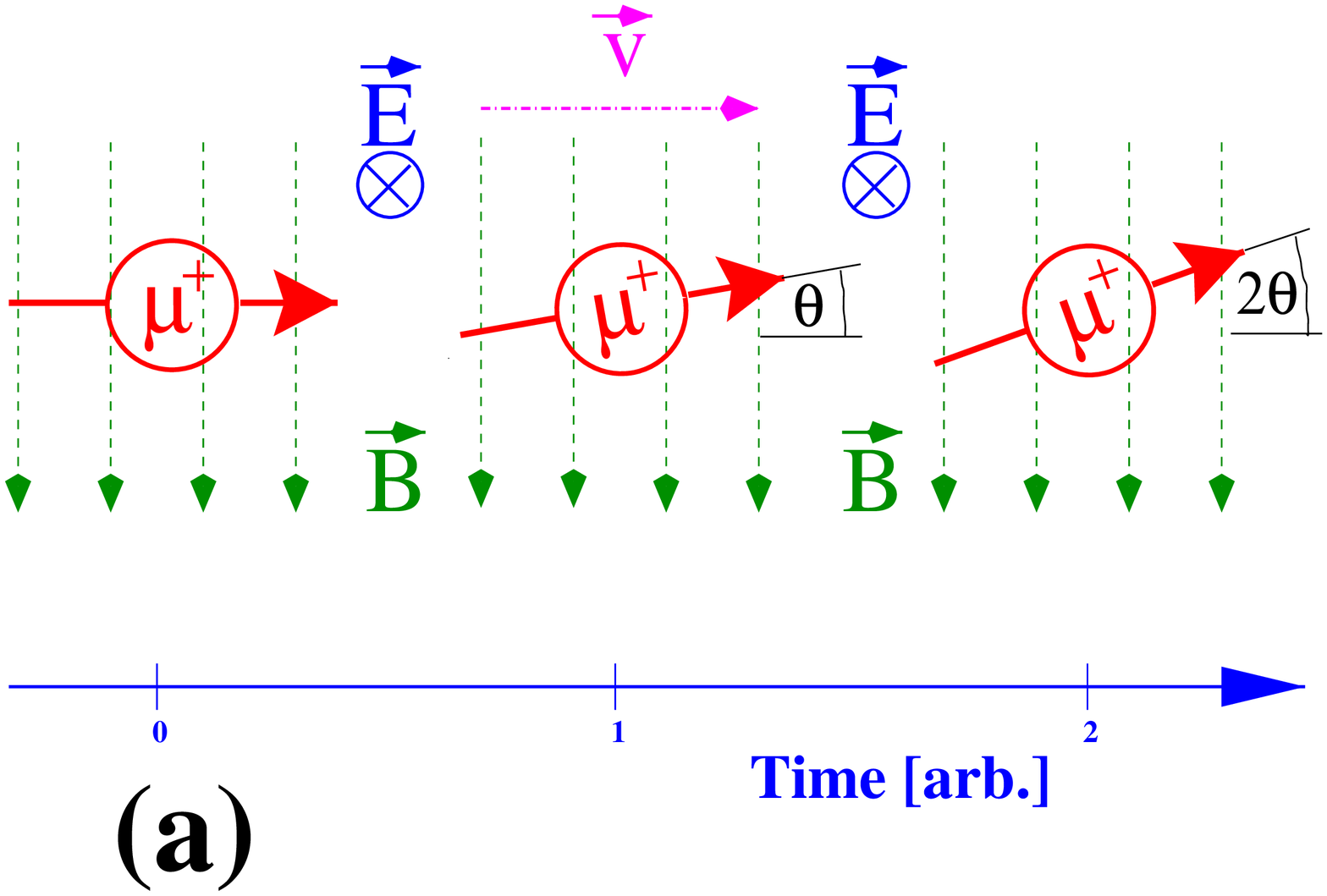}
 \hspace*{2cm}
 \includegraphics[width=2in,angle=0]{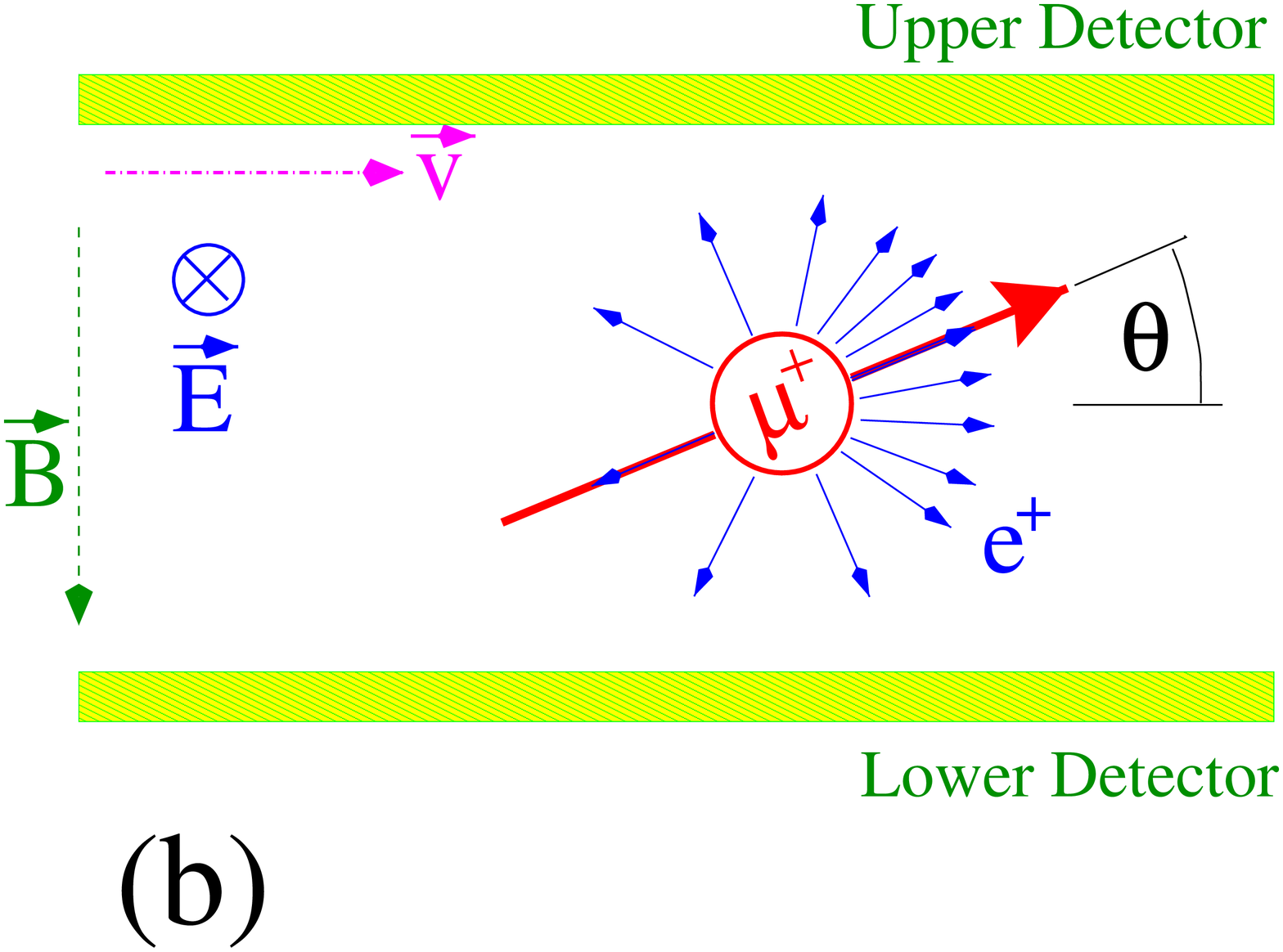}
 \caption{The basic principle of a novel muon edm experiment. 
         $\mu^+$'s of velocity $\vec{v}$ 
         moving in a magnetic storage ring with field $\vec{B}$ find
themselves 
         exposed to a motional electric field $\vec{E} \propto \vec{v}
\times \vec{B}$. 
         In case of a finite EDM the muon spin precesses with
         an angle $\Theta$ which grows linearly in time (a), which can 
         be monitored by observing the time dependence of the angular
distribution of
         the decay positrons (b).
 \label{muedm}}
 \end {figure}

Recently a novel idea was introduced to exploit fully the high
motional electric fields in a magnetic storage ring. With an
additional radial electric field the spin precession due to the
magnetic moment anomaly can be compensated. In this case an  $edm_{\mu}$
would express itself as a spin rotation angle out of the orbital
plane that increases linearly in time (Fig. \ref{muedm}).
This can be observed as a time-dependent increase of the above/below
plane asymmetry. For the possible muon beams at the AGS synchrotron
of 
BNL a sensitivity of $10^{-24}$e\,cm is expected 
\cite{Semertzidis_00}. 
In this experiment the possible muon flux is a major limitation.
For models with nonlinear mass scaling of edm's such an experiment would 
already be more sensitive than present experiments
on neutral atoms \cite{Babu_00}. 
An experiment carried out at a more intense muon source could provide
a significantly more sensitive probe to CP violation in the second 
generation of particles without strangeness.

It should be noted that the very same experimental technique 
(and the same experimental set-up) could also be
utilized for searching for permanent electric dipole moments 
of a large variety of relativistic nuclei \cite{Khriplovich_99}. 
Such possibilities were recently explored in a workshop at GSI \cite
{KJ_99}.
Results that  are superior to those extracted from present experiments
on neutrons or mercury atoms can be expected. Such a novel experiment
could
be well accommodated at a new and intense ISOL facility associated with 
the proton machine discussed here.  

\subsection{CPT tests}
In general, any two measurements of muon parameters for 
both positive and  negative muons constitute a CPT test.
Such tests were performed for the mass, the magnetic moments 
and most sensitively for the muon magnetic anomaly.
In the framework of a rather general ansatz
the past muon g-2 experiments at CERN have provided the best test of 
CPT invariance at a level of $2\cdot10^{-22}$ 
which is a more than 3 orders of magnitude
tighter bound than the mostly quoted ${\rm K}^0-\overline{{\rm K}^0}$
mass difference and the value extracted from measurements of
the electron and positron magnetic anomalies \cite{Kostelecki_00}. 
From the ongoing measurements of $a_{\mu}$ at BNL one can expect 
an improvement by at least one order of magnitude in this figure.

\section{BOUND MUON SYSTEMS}

\subsection{Muonic atoms}

\subsubsection{Muonium}

The hydrogen-like muonium atom (${\rm M}= \mu^+e^-$) is the bound state of a 
positive muon and an electron
\cite{Jungmann_00}. Since both particles may be regarded as 
structure-less objects with dimensions below $10^{-18}$~m, their
electromagnetic 
interactions can be calculated to much higher precision than for
hydrogen
isotopes containing hadrons. In natural hydrogen, for example, the
hyperfine structure has been 
measured six orders of magnitude more accurately than can be
calculated because of uncertainties in the
proton magnetic radius and
polarizability.
For muonium the comparison between theory and experiment of this
transition 
could be made  with two orders of magnitude higher accuracy than for 
natural hydrogen. Therefore the system gives an important input to the
set of adjusted fundamental constants \cite{Mohr_00}.

Muonium can be produced most efficiently in the atomic 1s state.
For this reason, precision experiments can be carried out
on the ground state hyperfine interval and the 
1s--2s frequency interval. 

\begin {figure}[htb]
 \parbox{0.5\linewidth}{
 \includegraphics[width=\linewidth]{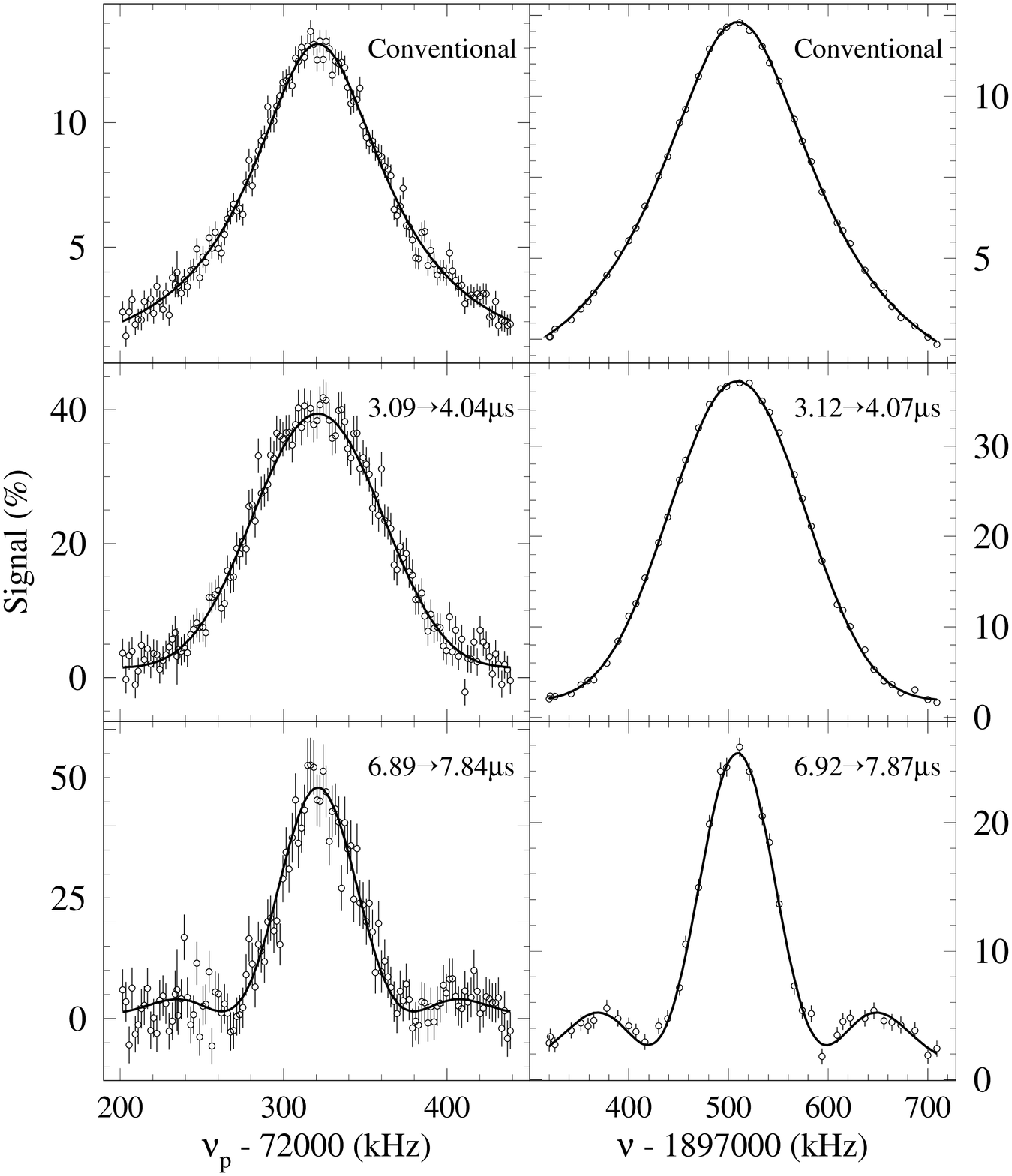}}
 \hspace*{\fill}\parbox{0.45\linewidth}{
 \caption{Muonium hyperfine structure transition 
 signals \cite{Liu_99} obtained with magnetic field and with
 microwave frequency scans using the `old muonium' technique. Lines
 of width significantly below the natural width were obtained.
 \label{M_hfs}}}
 \end {figure}

\begin {figure}[htb]
 \parbox{0.5\linewidth}{
 \includegraphics[width=\linewidth]{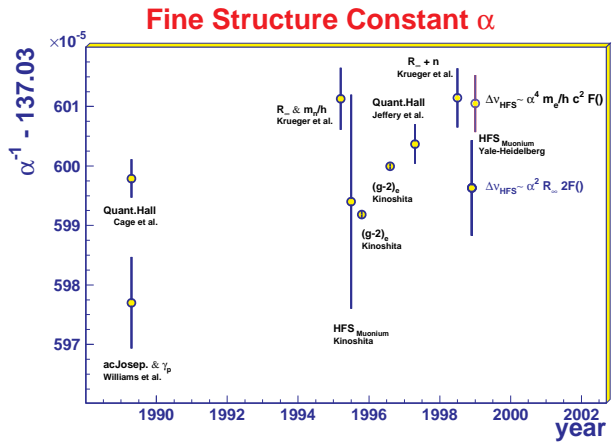}}
 \hspace*{\fill}\parbox{0.45\linewidth}{
 \caption{The fine structure constant $\alpha$ has been determined
with various methods
 (from Ref.~\cite{Jungmann_01}).
  Most precise is the value extracted from the magnetic anomaly of the 
  electron. 
  The muonium atom offers two different routes which use independent
sets of fundamental
  constants. The disagreement (the error bars are mostly statistical)
seems to indicate
  that the uncertainty of $h/m_e$ from neutron de Broglie wavelength
measurements may have been underestimated. 
 The degree of agreement between the $\alpha$ values from electron $g-2$ 
  and muonium hyperfine structure is a measure for the internal
consistency of
  quantum electrodynamics and the set of fundamental constants involved
in the
  evaluation.
 \label{alpha}}}
 \end {figure}

Recently an experiment was completed at LAMPF 
\cite{Liu_99}
in which 
two Zeeman transitions in the ground state were induced
by microwave spectroscopy.
They both involved a muon spin flip and could be detected through 
a change in the spatial anisotropy of the muon decay positrons.
The experiment takes advantage of line narrowing that occurs  
when only such atoms are detected which have coherently interacted
with a microwave field for periods long compared to
the average muon lifetime $\tau_{\mu}$ (`old muonium' technique
\cite{Liu_99}, Fig. \ref{M_hfs}).
The hyperfine interval was determined to 12~ppb as a test of QED 
or alternatively as a determination of the fine structure constant
$\alpha$
to 58 ppb.
The good agreement between this $\alpha$  and the value extracted from
a measurement of the electron magnetic anomaly 
(Fig. \ref{alpha}) is generally interpreted
as
the best test of internal consistency of quantum electrodynamics.
The comparison between theory and experiment is for the muonium hyperfine
structure
some two orders of magnitude more precise than for electronic hydrogen, 
where the accuracy of knowledge of the proton magnetic radius 
and polarizability limit the QED calculations
more than seven orders of magnitude below the experiment.
In addition, in the measurements 
the muon magnetic moment (respectively its mass) were
determined to 120~ppb.
 
At RAL the 1s--2s interval was determined to 4~ppb using Doppler-free
two-photon spectroscopy with pulsed lasers
\cite{Meyer_00}. 
The result may be interpreted as a precise muon mass measurement or
alternatively
as the by far the best test of the charge equality of muon and electron
(2~ppb),
i.e. two particles from different particle generations. It should be
noted that the gauge invariance principle can provide charge quantization
only within
one particle generation.

The muonium spectroscopy 
experiments at LAMPF and at RAL were limited largely by statistics only.
Systematic errors arising mainly from technical imperfections of the pieces of 
  apparatus  involved 
 can be expected to be kept under control for one 
further order of magnitude gain in accuracy for the hyperfine structure
experiment.
The laser experiment can be performed with CW lasers and promises
some 2--3 orders of magnitude improvement without significant
systematics problems, provided 4--5 orders of 
magnitude increased muon flux  is  available. 
In both cases a pulsed time structure 
would be required.  

\subsubsection{Radioactive muonic atoms}

Muonic atoms have been employed both for testing QED vacuum
polarisation and to determine nuclear parameters, 
most importantly nuclear mean square 
charge radii. 

A novel programme is currently  being initiated in a collaboration between
various
universities and  the Paul Scherrer Institute (PSI) in Switzerland  and 
CERN to pursue the possibility to use muonic X-rays to obtain
information on very exotic nuclei far from the valley of stability. The
exact measurements on muonic X-ray spectra can yield the most precise
values for the charge radii of nuclei as well as other ground-state
properties such as moments and even B(E2) transition strengths for
even--even nuclei. In general,
muonic X-rays promise higher accuracy for most of these quantities 
compared to electron scattering.

The experimental approach proposed is to use ion trapping of exotic
isotopes combined with a newly developed muon decelerator concept as
well as other muon manipulation techniques of PSI. In this approach,
radioactive ions or isotopes capture muons in their outer atomic orbits,
subsequently leading to cascade into the inner orbits, resulting
in emission of characteristic X-rays of MeV energies due to the increase
in binding energies for massive muons. Since   the capture
is of atomic nature, very high sensitivities are expected for this
approach which, if proven successful,   would eventually mean one of the
greatest new developments in physics of nuclei under extreme conditions.
In addition, other physics opportunities offered by muon capture in
nuclei will be pursued in this collaboration.  

Among the possible scenarios,
nested traps for radioactive nuclei and for muons can be envisaged.
Large formation rates can be expected from a set-up 
containing a Penning trap 
\cite{Penning_trap}
the magnetic field of which serves
also for a cyclotron muon trap 
\cite{Simons}. 
For muon energies in the
range of electron binding energies 
the muon capture cross-sections grow to atomic values,
efficient atom production can be expected of order 50
systems per second.                                  
CERN could be a unique
place   where such experiments become possible.
The ideal beam time  structure is the shortest possible pulse
with non-specific requirements for repetition rates. 

Further, nuclear muon capture in both stable and radioactive muonic atoms
offers the possibility to study weak interactions. Of particular
interest is the radiative muon capture in hydrogen, where the induced 
pseudoscalar coupling is a long-standing problem \cite{Hasinoff_93}.

\subsubsection{Muonic hydrogen}

In muonic hydrogen \cite{Kars_00} the Bohr radius of the system is only 
100 times larger than the size of the proton. Therefore  electromagnetic
transitions in this system are very sensitive to the size and
inner structure of the proton.
The atom is  currently being investigated in an experiment at PSI where
the 2s--2p Lamb-shift transition will be induced with infrared laser light
\cite{Pohl_00}.
This quantity is mostly determined by QED effects. A significant
contribution arises
from the proton mean square charge radius the determination of which is
the expressed goal of the collaboration at PSI. The proton radius is of
significant
relevance for electronic hydrogen spectroscopy since it gives a large
uncertainty in
the interpretation of the observed 1s--2s energy interval.
It should be noted that several electron scattering experiments have
failed
to produce a reliable value of the proton mean square charge radius. 

Another important quantity is the hyperfine splitting between the $\rm F=0$ and
$\rm F=1$ states
in the  $\mu$p system. A measurement to an accuracy of $10^{-4}$ or better
would yield new
information on the proton polarisability \cite{Bacal_93}.
Such an experiment can be performed with muonic hydrogen atoms produced
by depositing a short pulse of muons in the centre of a small cell in
which molecular
hydrogen gas at high pressure is surrounded by high-$Z$ walls. The average
flight time
of the system can be measured from the timing of muonic X-rays that 
are emitted upon arrival of the muonic hydrogen atoms at the wall
where immediate muon transfer occurs. The flight time shortens after
laser resonance
absorption at 6.8 $\mu$m wavelength due to the immediate collisional
quenching and the
0.12~eV gain in kinetic energy from recoil in this process.
Such  an experiment would require an intense pulsed beam with less than
500~ns bunch length.
If the $\mu^-$ beam were polarised the laser transition could be
observed as a change
in the angular distribution of the decay electrons \cite{Kars_00}.

\subsection{Condensed matter}

Muon Spin Rotation and Muon Spin Resonance ($\mu$SR) are well-established
methods to study
bulk condensed matter \cite{muon_science}. A typical $\mu$SR experiment
requires $10^6$ to
$10^7$ muons. Since any signal is extracted from a spin rotation
frequency, the resolution
of the apparatus is inversely proportional to the duration of the muon
pulse which therefore
should be as short as possible. 

Efforts are currently under way to
produce beams of very slow muons which would make it possible to expand
the $\mu$SR method to
research topics such as surface magnetism, surface diffusion, catalysis. 
Bio-science applications are currently being  explored \cite{Nagamine_99}. 
They include for example electron transfer mechanisms in proteins.
Element analysis at surfaces and in thin layers and membranes through
muonic X-rays
can be envisaged.

\begin {figure}[htb]
 \begin {center}
 \includegraphics[width=4in]{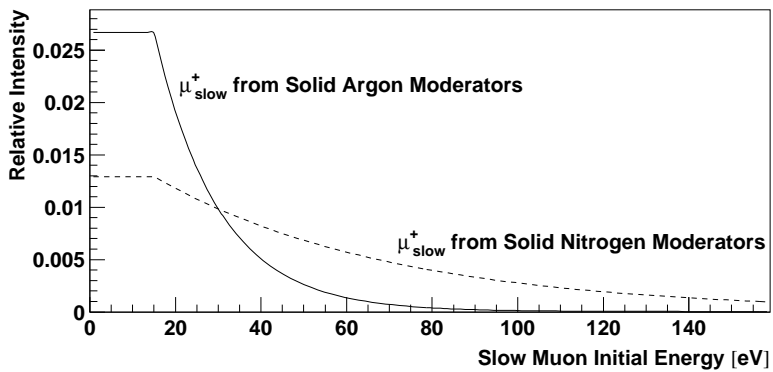} 
 \end {center}
 \caption{The energy distributions for slow muons determined for
frozen Ar and N$_2$ targets.
 \label{slomu_e}}
\end {figure}

For positive muons, moderation to typically 20$\pm$10 eV (Fig.~\ref{slomu_e})
has been demonstrated \cite{Traeger_00}.

In frozen noble gas 
targets the particles have negligible interactions with the 
moderator atoms once they have reached an energy below the 
materials band gap. The process which generates such a
tertiary muon beam has some $10^{-5}$ efficiency and
requires therefore intense muon sources. The highest flux right 
now is available at PSI \cite{Morenzoni_99}.

\section{MUON BEAMS}

\begin{figure}[htb]
 \begin{center}
 \includegraphics[width=8.5cm]{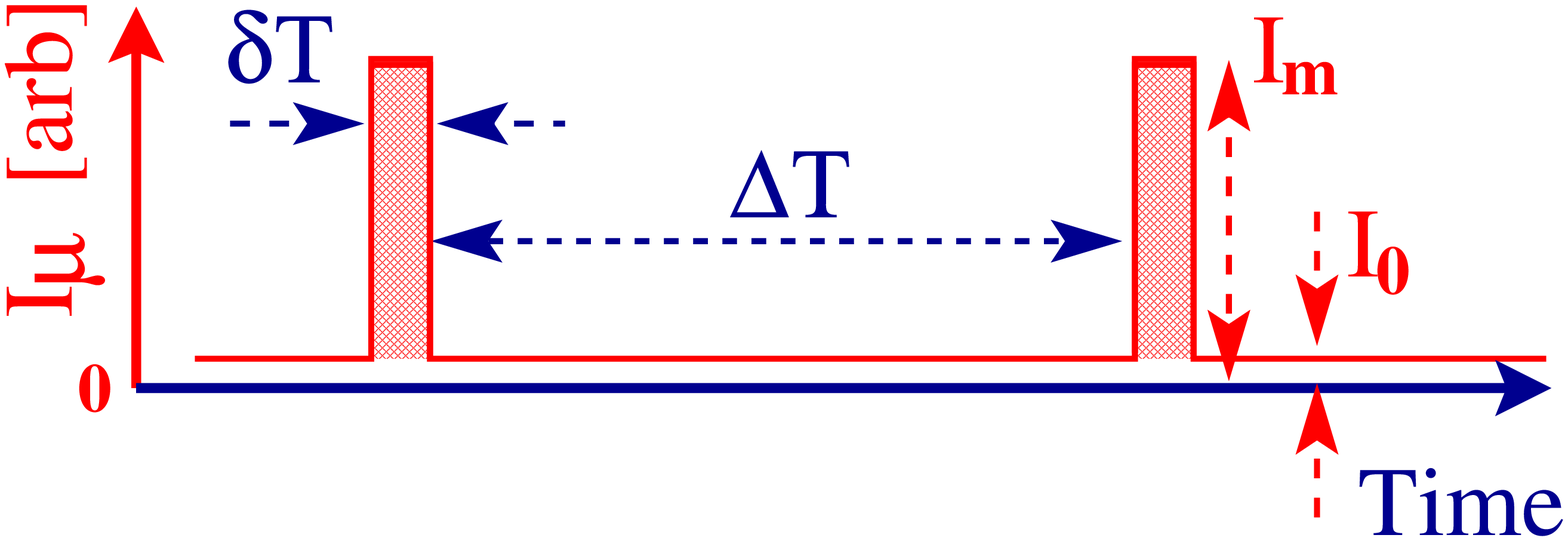}
 \end{center}
 \caption{A pulsed time structure is important for many stopped muon
experiments. Crucial parameters are the muon suppression in between
the bunches $I_0/I_{m}$, the bunch width $\delta T$ and separation 
$\Delta T$.}\label{Fig27}
 \end {figure}  

\begin{table} 
\label{muon_beam_rates}
\caption{Beam requirements for new muon experiments. 
Given are required sign of charge $q_{\mu}$ and the minimum of the total
usable number of muons
$\int I_{\mu}dt$ above which significant progress can be
expected in the physical interpretation. The experiments which require
pulsed beams (see Fig.~\ref{Fig27}) are 
sensitive to the muon suppression $I_0/I_{m}$  
between pulses of length $\delta T$ and separation $\Delta T$.
This does not apply (n/a) for continuous beams.
Most experiments require energies below 4 MeV corresponding to
29 MeV/$c$ momentum. Thin targets and storage ring acceptances,
demand rather small momentum bites $\Delta p_{\mu}/p_{\mu}$.}
\centerline{\begin{tabular}[hbt]{|c|c|c|c|c|c|c|c|}
\hline
&&&&&&&\\
Experiment  & $q_{\mu}$ &$\int I_{\mu}dt$&$I_0/I_{m}$&$\delta
T$&$\Delta
T$&$E_{\mu}$&$\Delta p_{\mu}/p_{\mu}$\\
            &           &                &             & [ns]     & 
[$\mu$s]    & [MeV]   &[\%]                   \\
\hline
$\mu^-N \to e^-N$$^{\dag}  $    &--
&$10^{21}$&$<10^{-10}$&$\leq 100$&$\geq 1$
&$<20$     &$<10$         \\
$\mu^-N \to e^-N$$^{\ddag} $     &--   
&$10^{20}$ &n/a                &n/a            &n/a
&$<20$     &$<10$         \\
$\mu \to e \gamma$     &+    &$10^{17}$& n/a      &n/a &n/a
&1...4     &$<10$         \\
$\mu \to eee$          &+    &$10^{17}$& n/a      &n/a 
&n/a  &1...4     &$<10$         \\
$\mu^+e^- \to \mu^-e^+$&+    &$10^{16}$&$<10^{-4}$&$<1000$   
&$\geq 20$      &1...4     &1...2         \\
\hline
$\tau_{\mu}$                   &+    &$10^{14}$&$<10^{-4}$&$<100   $  
&$\geq 20$  &4  &1...10        \\
transvers. polariz.          &+&$10^{16}$&$<10^{-4}   $
&$<0.5 $ &$ >0.02  $ &30-40     &1...3        \\
\hline
$g_{\mu}-2$                    &$\pm$&$10^{15}$&$<10^{-7}$&$\leq 50  $
&$\geq 10^3$    &3100      &$10^{-2}$     \\
$edm_{\mu}$                    &$\pm$&$10^{16}$&$<10^{-6}$&$\leq 50  $
&$\geq 10^3  $  &$\leq$1000&$\leq 10^{-3}$\\
\hline
$M_{HFS}$                      &+    &$10^{15}$&$<10^{-4}$&$\leq 1000$
&$\geq 20$  &4  &1...3         \\
$M_{1s2s}$                     &+    &$10^{14}$&$<10^{-3}$&$\leq 500 $
&$\geq 10^3$    &1...4     &1...2         \\
\hline
$\mu^- $atoms                   &--   &$10^{14}$&$<10^{-3}$&$\leq 500 
$&$\geq 20$     &1...4     &1...5         \\
\hline
condensed matter            &$\pm$&$10^{14}$&$<10^{-3}$&$< 50    $ 
&$\geq 20$      &1...4     &1...5         \\
(incl. bio sciences)     &&&&&&&\\
\hline
\end{tabular}}
{\footnotesize \flushleft 
\noindent$^{\dag}$ Scenario in which a pulsed beam is utilized.\\
\noindent
{$^{\ddag}$ Scenario in which a continuous beam after the muon cooling
stage is employed.}}
\end{table}

\subsection{Beam requirements}

The experiments discussed in this report require different types of beam.
Measurements with stopped muons require rather low momentum muons with
either    a pulsed or    a continuous time structure. Other
experiments would need   higher energy pulsed beams (see Table
\ref{muon_beam_rates}). Continuous (or quasi continuous) 
beams would be required particularly
for experiments with tight coincidence signatures.

Low-momentum positive beams can be obtained up to momenta of $p_s =29$
MeV/$c$ from so-called `surface muon' channels \cite{surface_beams}. 
These are beams of muons that originate in the decay of pions
that stopped close to the surface of the production target.
Their rate increases with $p^{3.5}$ up to a maximum at $p_s$.
Low-momentum negative beams may be obtained from pion decay in the cloud 
surrounding the production target.

There are important requirements for most experiments 
on the tolerable momentum bite $\Delta p/p$. This arises from
the fact that either only thin targets are possible in which one would
like to stop an as high as possible fraction of the beam, or this is 
determined by the acceptance of subsequent equipment like storage rings.
Typically one should aim for $\Delta p/p \approx 1\%$. Muons outside of 
this band will not be able to contribute to the useful signal, but will 
most likely contribute to the background. 

Muon beams are regularly contaminated by electrons which originate from
$\pi_0$ decays and pair creation.
An essential piece of equipment in a low-energy muon beam line is 
therefore an electromagnetic separator (Wien filter). The efficiency
of such devices is best for low-momentum spread. (They have also 
an advantage over degrading material, which on the  one hand would allow  muons 
to be separated from particles of different masses through different  fractional
energy losses. On the other hand they would increase the momentum bite.)

\subsection{Possible realizations of muon beams}\label{prmb}

The precise design of muon beams at a neutrino factory complex has not
been performed but the following summarises a number of ways by which one
could imagine increasing the muon flux by several orders of magnitude over
present or foreseen facilities.

The great strength of the CERN-based Neutrino Factory complex as 
envisaged in the CERN baseline scenario~\cite{nufact28}, is twofold.
Firstly the high intensity offered by the 2.2~GeV/$c$
Superconducting Proton Linac (SPL)~\cite{SPL}, which is given to
provide 4 MW average power in a 75~Hz pulsed mode or even possibly 24 MW in
continuous wave (CW) mode.
Secondly, the great flexibility in the beam timing offered by the fact
that a proton accumulator and a bunch compressor are foreseen in the
complex. These rings are necessary to match the basically continuous time
structure of the SPL to the pulsed structure needed to operate the muon
acceleration system as well as the storage ring.

\begin {figure}[htb]
\centerline{\includegraphics[width=14cm]{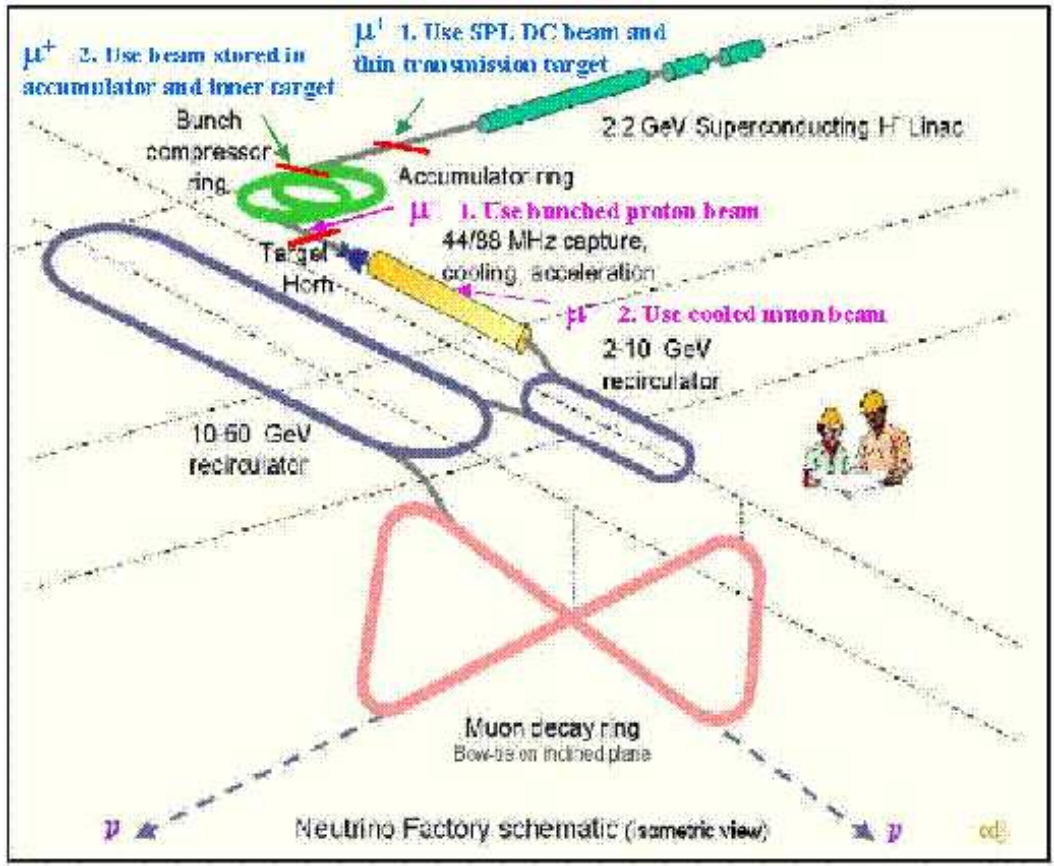}}
 \caption{Possible locations of muon target along the proton path in the CERN neutrino
factory baseline scenario.}
\label{muonbeamlocations}
 \end {figure}

Figure~\ref{muonbeamlocations} shows various locations where targets
could be installed along the proton path.

\subsubsection{Continuous beams}

\begin {figure}[htb]
\begin {center}
\includegraphics[width=10cm]{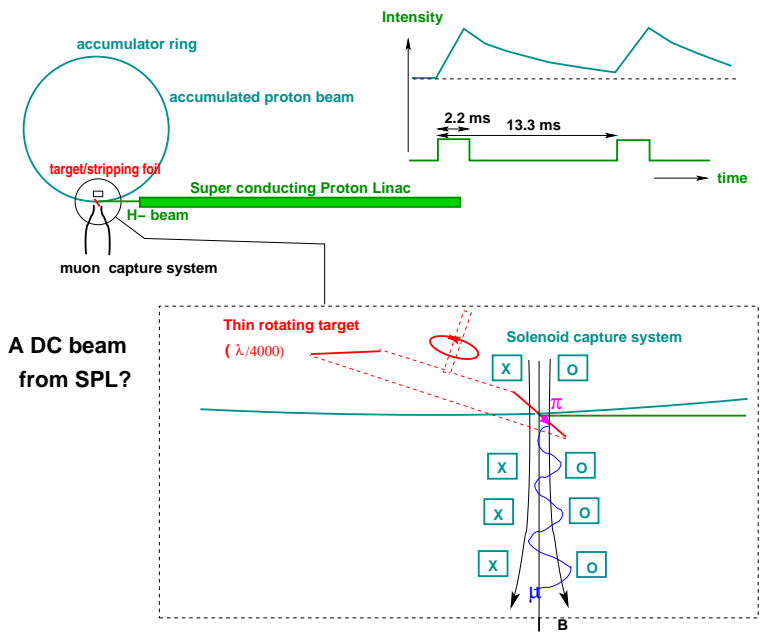}
\end {center}
\vspace*{-0.5cm}
\caption{Sketch of a thin muon production internal target
for a quasi-continuous muon beam.}
\label{internaltarget}
\end {figure}

For experiments requiring a continuous beam ($\mu^+ \to e^+
\gamma$, $\mu^+ \to e^+e^+e^-$) two solutions can be envisaged.
The benchmark beam for these experiments is the PSI  beam  for
the foreseen $\mu \to e \gamma$ experiment~\cite{PSIprop}.
\begin{itemize}
\item An internal target inside the proton accumulator.
This solution, shown in Fig.~\ref{internaltarget}, is 
similar to that sketched in Ref.~\cite{thintarget} and it
would consist in inserting in the proton accumulator a thin
target (typically 1/4000 of an interaction length),
either in place of the H$^-$
stripping foil, or elsewhere in the lattice. With respect to the
benchmark beam, the gain is obtained from the fact that the beam
recirculates many
times through the target so that essentially all protons
-- up to tails generated by
large scatters -- end up interacting. This gives right away a factor 20
more intensity than the benchmark set-up, where a 5\% interaction length
target is placed on the proton beam. Since the benchmark beam
uses a quadrupole capture system, with the typical
aperture of 150~mrad, an additional gain of a factor 40 or so
should be achievable with a more efficient capture system,
using for instance a solenoidal magnetic field.

The time structure of the beam is described
in Fig.~\ref{internaltarget}. Because the target is very thin, the
protons stay in the accumulator for a long time, and the system behaves as an
approximately continuous muon source.

This scheme is currently under evaluation.
A serious technical problem to be solved is the design
of the target, which will be subject to very intense heating.
This can be in principle solved by using a rotating carbon target so
that the
area of material exposed to the beam (about one square centimetre) is
removed and radiation-cooled fast enough as in the PSI E and M
targets~\cite{PSImuontargets} or
in a study for the neutrino factory~\cite{benett}.
Another serious technical problem is the local irradiation around the
target area. Since all particles produced in the target, except the muons, will
have to be absorbed locally, this will necessitate a very careful design of
the area, which receives the same amount of radiation as a beam dump.

\item
Another possibility would be to use directly the proton beam from the
linac operated in high-power, CW mode. This solution is less
efficient than the previous one, but could constitute a back-up
possibility.
\end{itemize}

It is therefore seen that improvements due to the
target design and capture design, could in principle lead to an
increase of the flux by almost three orders of magnitude. This will
constitute a considerable opportunity as well as a tremendous challenge to the
design of the experiments themselves.

\subsubsection{Pulsed beams}
Muon conversion experiments,  $\mu^-N\to e^-N$, and various other precision experiments
could take significant advantage of a pulsed beam structure (see
Table~\ref{muon_beam_rates} and Fig.~\ref{Fig27}).
 A pulsed beam allows a
time delay of many pion lifetimes to reduce the pion contamination. For
this rejection to be efficient, a time separation of a microsecond between
pulses is desirable. Two possibilities come to mind.

One could choose to place the production target in a proton beam line
at the exit of the buncher. Taking advantage of the SPL timing
flexibility, the bunches could in principle be delivered
for this application at a frequency different from the nominal one.

Another interesting possibility would be to take advantage of the main
muon cooling channel, in which the particles travel for typically 300 metres,
ensuring an efficient clean-up from the pions. The muons have a rather
broad energy spectrum (typically 200 MeV with an energy spread of 5\%)
at that level, however, and the stopping power may be limited.

To conclude this section on possible beams, it can be said that the SPL
together with its  accumulator and buncher rings offers attractive
possibilities. A conceptual design and more precise performance
estimates of target and beam lines for muon physics
at the SPL is now necessary.

\section{CONCLUSIONS}

The main conclusion of this study is that the physics potential of
a new slow muon facility,
such as the one that will become available as a necessary step on the
way to building a muon storage ring (Neutrino Factory),
is very rich and compelling, with a large variety of applications in many 
fields of basic research.  Indeed, 
muon physics, that has already played an important
role in establishing the Standard Model of particle physics, 
may provide us with crucial information regarding the theory that lies beyond, 
proving itself to be still far from having exhausted its potential.

This new low-energy muon source will have unprecedented intensity, three to
four orders of magnitude larger than presently available. It can have the large
degree of flexibility necessary to satisfy the requirements of very different 
experiments, providing muon beams with a wide variety of momenta and time 
structures. Both continuous and pulsed beams are possible. In addition, 
it is capable of producing physics results at the very early stages of the 
Muon Complex, well before the completion of muon cooling, 
acceleration, and storage sections.

Only preliminary ideas on the design of this facility are introduced
here, suggesting ways by which the muon flux could be boosted orders of
magnitude above present or foreseen facilities. The tasks of detailed
conceptual design of target and capture systems and of quantitative 
estimates of beam
performances are still entirely ahead of us.
It will be now enthusiastically tackled in view of the physics rewards
that are to be expected.

A major interest in muon physics lies in the search for
rare processes that violate muon number conservation. In many extensions of
the Standard Model, such as supersymmetry, they may occur
at rates close to the current experimental bounds.
Their discovery would have far-reaching consequences.
The most interesting processes are $\mu^+\to e^+\gamma$, $\mu^+\to
e^+e^-e^+$, and $\mu^-$--$e^-$ conversion in nuclei. 
We emphasise that all the different processes should be pursued. Indeed,
the relative rates of the different modes provide a powerful tool for discriminating
different manifestations of new physics. The muon
facility discussed here has enough flexibility to allow
the study of different muon processes, and 
promises to be more sensitive 
by at least a few orders of magnitude, when compared with current experiments.  

Experimental prospects on these muon number-violating processes
and on the related process of muonium--antimuonium conversions can be
 summarised as follows:
\begin{itemize}
\item The present upper limit on $\mu^+\to e^+\gamma$ ($1.2 \times
  10^{-11}$) is expected to be reduced (or the process observed), by 2005, to $\sim
  10^{-14}$. Exploiting the higher intensities available at
  a neutrino factory complex is a nontrivial challenge here,
  as the main limitation to improving the sensitivity to $\mu^+\to e^+\gamma$
  comes from the severe background of accidental
  coincidences. A number of paths in detector design, aiming at
  a sensitivity below $\sim
  10^{-15}$, are suggested for further exploration.
\item 
The reduction in background necessary to improve the sensitivity 
to $\mu \to 3e$ beyond 
the current upper limit
$B(\mu \to 3e) < 1 \times 10^{-12}$ appears to be 
an easier task, if detectors working at the high beam rates proposed 
can be developed. Sensitivity to
$B(\mu \to 3e) > 1 \times 10^{-16}$ or better seems to be attainable.
\item
The limit on neutrinoless $\mu^- \to e^-$ conversion nuclei 
is expected to be pushed from the present 
$6.1 \times 10^{-13}$ down to $1 \times 10^{-13}$ soon, while 
future experiments may be sensitive to conversion probabilities down to
$1 \times 10^{-16}$. As the experimental 
sensitivity is completely dominated by the performance of the 
(pulsed) muon beam, this mode should continue to benefit
from increased beam intensity, and conversion probabilities of 
$1 \times 10^{-18}$ or smaller 
could be within reach. 
\item
Muonium--antimuonium conversion could be further probed by using a pulsed beam
and taking advantage of the
time  evolution of the conversion process. One may reach 
sensitivities to conversion probabilities as small as $10^{-13}$, almost 
three orders of magnitude beyond the present limit of $8.1 \times 10^{-11}$. 
Detector requirements closely resemble  those
for $\mu \to 3e$ searches. 
\end{itemize}

Normal muon decay can also be studied with increased precision. 
The muon lifetime could be measured with a precision which is ten
times higher than achieved in the ongoing PSI experiments.
Improvements on the measurements of decay parameters, 
via a number of other observables related to the muon and electron spin 
variables, are also possible but are not limited by muon rates, and appear
less promising in view of the current efforts at TRIUMF and PSI. 

{
\begin{table}[bth] \centering
\label{muon_experiments}
\caption{Experiments which could beneficially take advantage of the intense future
stopped-muon sources  at the CERN neutrino factory (NUFACT)}
{\tiny \begin{tabular}[b]{|c|c||c|c|c||c|}
\hline
Type of   & Physics Issues & Possible   & Previously established &Present
activities &Projected for \\
experiment&                & experiments&accuracy&(proposed accuracy)&
NUFACT @ CERN \\
\hline \hline
`Classical' & Lepton Number Violation;&$\mu^-N \to e^-N$     
&$6.1 \times
10^{-13}$       & PSI, proposed BNL       ($5 \times 10^{-17}$) &  $ <
10^{-18}$     \\
rare \&       & Searches for New Physics:&$\mu \to e \gamma$
&$1.2 \times 10^{-11}$
                & proposed PSI           ($2 \times 10^{-14}$) &  $ <
10^{-15}$      \\
forbidden     & SUSY, L-R Symmetry,&$\mu \to eee$     &  $1.0\times 10^{-12}$ 
                & completed 1985 PSI     & $ < 10^{-16}$ \\
decays        & R-parity violation,.....&$\mu^+e^- \to
\mu^-e^+$&$8.1 \times 10^{-11}$
                & completed 1999 PSI &  $ < 10^{-13}$ \\
\hline
Muon          & $G_F$; Searches for New
Physics;&$\tau_{\mu}$                   &$18
\times 10^{-6}$         & PSI (2x), RAL         ($1 \times 10^{-6}$) &
$ < 10^{-7}$ \\
decays         & Michel Parameters& transv. Polariz.                   
&$2\times 10^{-2}$& PSI, TRIUMF           ($5 \times 10^{-3}$) &   $ <
10^{-3}$ \\
\hline
&Standard Model Tests;&&&&\\
Muon           &   New Physics; CPT Tests
&$g_{\mu}-2$                    &$1.3 \times 10^{-6} $         & BNL
($3.5\times10^{-7}$) &  $ <10^{-7}$        \\
moments        &T- resp. CP-Violation &$edm_{\mu}$    &$3.4 \times
10^{-19} e$ cm  
            & proposed BNL           ($10^{-24} e$ cm) &  $ < 5 \times
10^{-26} e$ cm  \\
&in 2nd lepton generation&&&&\\
\hline
Muonium        & Fundamental Constants,
$\mu_{\mu}$,$m_{\mu}$,$\alpha$;&$M_{HFS}$ &$12 \times 10^{-9}$          
              & completed 1999 LAMPF  &   $ 5 \times 10^{-9}$   \\
spectroscopy   &  Weak Interactions; Muon Charge
&$M_{1s2s}$                     &$1 \times 10^{-9}$           
                     &  completed 2000 RAL  &  $ < 10^{-11}$    \\
\hline
Muonic atoms  & Nuclear Charge Radii;&$\mu^-$ atoms  & depends 
& PSI, possible CERN    &  new nuclear\\
&Weak Interactions&&&($<r_p>$to $10^{-3}$)& structure           \\
\hline
Condensed      & surfaces, catalysis & surface $\mu$SR             &n/a
& PSI, RAL             (n/a)& high rate                \\
matter&bio sciences ... &&&&\\
\hline
\end{tabular}}
\end{table}}

Several other experiments on free muons and atomic systems containing 
muons also promise substantial progress:
\begin{itemize}
\item
A new dedicated muon $g-2$ experiment could improve the experimental
precision  by almost a factor 15 compared with present experiments, and by almost a
factor of  5 compared with the precision expected to be reached by the ongoing BNL
experiment  (0.35~ppm). This effort may become particularly 
worthwhile if the recently 
reported discrepancy is confirmed, although
hadronic contributions limit the accuracy of the theoretical prediction. 
It should also be noted that 
systematic effects on the magnetic field become important at the level of 0.1~ppm.
\item 
A search for a permanent muon electric dipole moment (edm) would be a
fundamentally novel experiment. It could reach a sensitivity similar to or better than
present neutron edm searches, down to $5 \times 10^{-26}$ e\,cm. 
\item
Muonium microwave and laser spectroscopy could provide more precise measurements 
of important fundamental constants like the muon mass, the muon magnetic moment, 
and the electromagnetic fine structure constant. Improvements by
one to two orders of magnitude are possible. 
\item
The combination of intense muon beams
and a new ISOL facility offer new physics opportunities. 
 Measurements on muonic atoms of radioactive isotopes would reveal
properties of nuclei with half-lifes down to milliseconds.
One can envisage unknown effects in nuclear structure, comparable to 
the recent discovery of halo nuclei.
\item 
Muonic hydrogen spectroscopy allows measurements of
the proton charge radius with much better accuracies
than what is reached in electron scattering experiments.
\end{itemize}

There are also applications in other fields, such as
condensed matter physics and life sciences. 
Slow muon beams with four orders of magnitude more flux than the
most intense sources at PSI are expected to yield new insight
in a variety of areas, including surface science, thin films, the 
chemistry of catalysis and dynamical processes in biological molecules.

We find the discussed experiments theoretically well motivated and
capable of providing answers to urgent questions in fundamental physics.
The described experiments appear technically feasible at the projected
accuracies (Table~\ref{muon_beam_rates},\ref{muon_experiments}).  In due time, new
international collaborations should be encouraged to work on detailed
proposals for the possible experiments with forefront detector
technology.  We enthusiastically recommend  to continue efforts towards
designing the new stopped muon facility, and we are ready to fully support
the CERN management in its efforts to achieve such a unique
physics opportunity.

\end{document}